\documentclass{aa}

\usepackage[dvipsnames]{xcolor}
\usepackage{multirow}
\usepackage{subcaption}
\usepackage{rotating}
\usepackage{natbib}

\bibpunct{(}{)}{;}{a}{}{,} % to follow the A&A style

\usepackage{graphicx}
\usepackage[varg]{txfonts}
\usepackage{amsmath}
\usepackage[mathscr]{euscript}

\begin{document} 

\title{3D simulations of magnetospheric accretion in T Tauri stars\\
  I. Disk truncation, stellar torques, and application to
  observations}
\titlerunning{3D simulations of accretion in T Tauri stars. I. Disk truncation, stellar torques, and application to observations}
%\subtitle{}

\author{G. Pantolmos\inst{1,2},
  C. Zanni\inst{3},
  \and J. Bouvier\inst{2}
}
\authorrunning{Pantolmos et al.}

   \institute{Department of Physics, National and Kapodistrian University of Athens, University Campus,
     Zografos GR-157 84 Athens, Greece\\
     \email{gpantolmos@phys.uoa.gr}
     \and
     Univ. Grenoble Alpes, CNRS, IPAG, 38000 Grenoble, France
     \and
     INAF – Osservatorio Astrofisico di Torino, Strada Osservatorio
     20, 10025 Pino Torinese, Italy
     % \thanks{}
   }

   %\date{}

% \abstract{}{}{}{}{} 
% 5 {} token are mandatory
 
   \abstract
% context heading (optional)
{Young stars accrete material from their circumstellar disk through their magnetosphere and undergo contraction, these two processes impact their rotational evolution.}
% aims heading (mandatory)
{We investigate stable and unstable magnetospheric accretion regimes attributed to the development 
of an interchange instability in the disk truncation region. We examine the
associated stellar torques to assess the spin evolution of young stars.}
% methods heading (mandatory)
{We performed 3D magnetohydrodynamic (MHD) simulations of disk accretion onto
an inclined stellar dipole. We ran 21 simulations with varying
stellar rotation rates, dipole field strengths and obliquities, and
mass accretion rates. For each simulation, we estimated the disk
truncation radius, a critical quantity to determine the accretion regime
and the star-disk angular momentum transfer. We also
computed the torques that dictate the stellar spin evolution.}
% results heading (mandatory)
{We find that stars with a ratio of truncation-to-corotation radius of $R_t/R_{co}\gtrsim0.80-0.85$ accrete through a stable regime, whereas accretion becomes unstable otherwise. 
In addition, our $R_t/R_{\star}$ parametrization 
weakly depends on the mass accretion rate and the dipolar intensity,  
while strongly depending on the stellar rotation rate. We derived torque formulae for
each flow component affecting the stellar rotation; namely, accretion, 
magnetospheric ejections, and stellar winds. 
Finally, we applied our results to a sample of young stars with measured magnetic 
fields, mass accretion rates, and rotational periods. We find that most of
them should currently accrete in an unstable regime and undergo spin-up torques.}
% conclusions heading (optional)
{Our study confirms and expands upon previous results. Unstable accretion 
should lead to a net spin-up torque on the central star, while stable accretion 
can lead to stellar spin-down. When applying our truncation radius and torque prescriptions 
to observational data, we find that most young stars in our sample should be in a
spin-up state. Thus, for young stars, which are slow rotators in spite of accretion and contraction, the angular momentum problem remains.} 


   \maketitle
   \nolinenumbers
%
%________________________________________________________________

\section{Introduction}

The interaction of a stellar magnetosphere with a surrounding accretion disk is the process responsible for mass accretion onto magnetized stellar objects such as classical T Tauri stars (CTTs) in the pre-main sequence phase of low-mass star formation. These stars exhibit clear signatures of both strong ordered large-scale magnetic fields reaching kilogauss strengths \citep[see, e.g.,][]{Johnstone:2014ab} and active accretion \citep[see the review by][]{Hartmann:2016aa}. 
The stellar magnetic field can truncate the accretion disk at a few stellar radii, force the radially incoming gas to be lifted up from the disk midplane, and funnel it along the magnetic field lines. The gas then falls onto the stellar photosphere at near-free-fall velocities, thus forming accretion shocks and hotspots near the magnetic poles. \citep[see the review by][]{Romanova:2015aa}.

This dynamical picture is supported by several observations. Accretion shock models can explain the optical/UV excess observed in CTTs \citep{Calvet:1998aa,Gullbring:2000aa}. Atomic lines, often displaying classical or inverse P-Cygni profiles, are characterized by red-shifted features at free-fall speeds, probing the gas within the accretion columns, and blue-shifted components at escape speeds, highlighting the presence of magnetospheric outflows \citep[e.g.,][]{Hartmann:1994aa,Muzerolle:2001aa,Edwards:2006aa,Edwards:2013aa,Sousa:2021aa,Gravity:2023ab}.  

The interaction of the stellar magnetosphere with the surrounding environment is also considered to play a fundamental role in the angular momentum evolution of CTTs. These are slow rotators, with rotation periods between 1 and 10 days, which corresponds to $\lesssim 10\%$ of their break-up limit \citep[e.g.,][]{Herbst:2007aa,Bouvier:2014aa}. Besides, their rotation distribution in open clusters of different ages seems to remain roughly constant as long as they are actively accreting \citep{Gallet:2013aa,Smith:2023aa}, despite the fact that they are still contracting and acquiring angular momentum from the disk. This fact suggests that CTTs do not spin-up as long as they are surrounded by an accretion disk and the magnetospheric star-disk interaction has been supposed to provide an efficient process to extract stellar angular momentum to keep the stellar rotation period approximately constant. It had been first proposed that the stellar rotation was ``locked'' to the disk itself, with the disk able to remove the excess stellar angular momentum \citep{Koenigl:1991aa,Collier-Cameron:1993aa,Armitage:1996aa}, but this mechanism turned out to be too inefficient \citep{Agapitou:2000aa,Uzdensky:2002ab,Matt:2005ab,Zanni:2009aa}. The current consensus is that the excess stellar angular momentum could be expelled by different types of magnetospheric outflows, possibly operating at the same time: (1) outflows efficiently extracting angular momentum from the disk before it is accreted, such as conical \citep{Romanova:2009ab,Takasao:2022aa} or X-winds \citep{Shu:1994aa} to reduce or even cancel the accretion torque; (2) stellar winds, possibly enhanced by the accretion power \citep[Accretion Powered Stellar Winds,][]{Matt:2005aa}, extracting angular momentum directly from the star along open magnetic field lines; (3) nonstationary or variable outflows such as magnetospheric ejections \citep[MEs,][]{Zanni:2013aa} or ReX-winds \citep{Ferreira:2000aa} that exploit inflating and/or reconnecting magnetic field lines connected both to the star and the disk to exchange angular momentum with both of them. 

The magnetospheric star-disk interaction is an inherently 3D process, as stellar magnetic fields are clearly nonaxisymmetric \citep{Johnstone:2014ab}. For example, a misalignment between the stellar rotation axis and the magnetic dipolar moment is expected to favor the development of two opposite, approximately point-symmetric accretion funnels and shocks so that the stellar rotation determines a periodic modulation of the aforementioned spectroscopic, photometric, and interferometric signatures \citep[see, e.g.,][]{Robinson:2021aa,Tessore:2023aa}. A dynamical 3D model obviously requires full 3D numerical magneto-hydrodynamic (MHD) simulations, based on either laminar \citet{Shakura:1973aa} ``alpha'' disk models \citep{Romanova:2008aa,Kulkarni:2008aa,Blinova:2016aa} or turbulent disks driven by the magnetorotational instability (MRI; \citealt{Romanova:2012vx,Takasao:2022aa,Zhu:2024aa,Zhu:2025aa}. Both types of numerical experiments have clearly shown that the magnetospheric boundary (i.e., the accretion disk's inner rim) can become unstable due to the development of an interchange instability, the magnetic counterpart of the Rayleigh-Taylor instability. As a consequence, instead of forming two main ordered accretion funnels (stable accretion regime), the magnetospheric accretion flow fragments into multiple streams sneaking in the magnetosphere and impacting the stellar surface at different azimuths and latitudes with variable patterns (unstable accretion regime). It has been suggested that this instability could conceal the stellar periodicity, producing more stochastic light curves \citep{Romanova:2008aa,Kulkarni:2008aa,Kulkarni:2009aa,Kurosawa:2013aa} or triggering periods different from the stellar rotation
\citep{Blinova:2016aa,Armeni:2023aa,Armeni:2024aa,Romanova:2025aa}.

The threshold between stable and unstable magnetospheric accretion regimes was expressed as a function of the ratio between the truncation radius, $R_\mathrm{t}$, the radial distance where the disk accretion is diverted by the stellar magnetosphere into forming accretion funnels, and the corotation radius, $R_\mathrm{co}$, where the disk Keplerian rotation period equals the stellar one $P_\star$, $R_\mathrm{co} = \left(GM_\star P_\star^2/4\pi^2\right)^{1/3}$, where $G$ is the gravitational constant and $M_\star$ is the stellar mass \citep{Blinova:2016aa}. Analogously, the magnetospheric star-disk interaction torque and the threshold between the stellar spin-up and spin-down regimes have been often parametrized as a function of the same $R_\mathrm{t}/R_\mathrm{co}$ ratio \citep{Ghosh:1979ab,Koenigl:1991aa,Ostriker:1995aa,Matt:2005ab,Long:2005aa,Zanni:2013aa,Zhu:2025aa}. While these appear to be two distinct phenomena, this fact suggests that it could be possible to find a correlation between stable and unstable magnetospheric accretion and stellar spin-down and spin-up regimes, respectively.

The aims of this work are as follows: (1) investigate and characterize stable and unstable regimes of magnetospheric accretion with a particular focus on the disk truncation dynamics to establish a definition of the truncation radius and its dependence on observable quantities (i.e., dipolar field strength, mass accretion rate, stellar period) that clearly identifies the threshold between different accretion regimes; (2) compute and parametrize the stellar torques associated with the star-disk interaction and assess their correlation with stable and unstable accretion regimes; (3) apply the outcome of our models, the position of the truncation radius, and the stellar torque prescriptions to a sample of CTTs observed in spectropolarimetry as well as (when possible) in interferometry so that we can provide all the relevant stellar parameters and infer their accretion regime and spin state.

Our results are based on 3D MHD time-dependent numerical simulations of an ``alpha'' accretion disk interacting with a tilted magnetosphere of a rotating star. To explore a wide parameter space, we performed 21 numerical simulations, with varying stellar rotation rates, dipolar field strengths and obliquities, and mass accretion rates. In this work, we focus on the time-averaged dynamical properties of our solutions to characterize in a simple way the accretion regimes and the stellar torques. In a companion paper, we will investigate the dynamical and photometric variability of our solutions to better discern different accretion regimes and spin states.

In Sect. \ref{sec_num_set}, we describe the numerical method employed to develop our simulations, the initial and boundary conditions of our numerical models, their
normalization, and the parameter space explored here. In Sect. \ref{sec_bndry}, we characterize stable and unstable magnetospheric accretion regimes by looking at how the disk truncation and the magnetospheric accretion occur in both cases. In Sect. \ref{sec_rtrunc}, we provide a parametrization for our definition of the truncation radius as a function of the main stellar and disk parameters. In Sect. \ref{sec_torques}, we compute and parametrize the torques associated with the star-disk interaction that impact the angular momentum evolution of the young forming star, looking for a correlation with the accretion regimes. In Sect. \ref{sec_disc}, we compare our results with recent theoretical related works, and, above all, we test our findings on a sample of CTTs observed in spectropolarimetry and interferometry. In Sect. \ref{sec_concl}, we summarize our results.

\section{Numerical setup}
\label{sec_num_set}

\subsection{MHD equations and numerical method}
\label{sec_mhd_num}

The models presented in this paper are obtained by numerically solving
the MHD equations including viscous and resistive effects. In
cgs-Gaussian units, these equations are expressed as
\begin{equation}
\begin{aligned}
\label{eq_MHD}
\frac{\partial \rho}{\partial t} & + \nabla\cdot\left(\rho \vec {v}\right) = 0 \\
\frac{\partial \rho\vec{v}}{\partial t} & + \nabla \cdot \left[ 
\rho \vec{v}\vec{v} +
\left( P + \frac{\vec{B}\cdot\vec{B}}{8\pi} \right)\vec{I}-
\frac{\vec{B}\vec{B}}{4\pi} - \mathcal{T}
\right]  = \rho \vec{g} \\
\frac{\partial E}{\partial t} & + \nabla\cdot\left[
\left(E + P + \frac{\vec{B}\cdot\vec{B}}{8\pi}\right)\vec{v}-
\frac{\left(\vec{v}\cdot\vec{B}\right)\vec{B}}{4\pi} \right] =  \\
& = \rho \vec{g} \cdot \vec{v} + 
\left( \nabla \cdot \mathcal{T} \right) \cdot \vec{v}
- \frac{\vec{B}}{4\pi} \cdot \left(\nabla \times \eta_{\mathrm{m}} \vec{J}\right) \\
\frac{\partial \vec{B}}{\partial t} & + \nabla \times \left(\vec{B}\times\vec{v} + \eta_{\mathrm{m}} \vec{J} \right)= 0 \; . 
\end{aligned}
\end{equation}
The system of Eqs. (\ref{eq_MHD}) conveys the mass, momentum, and
energy conservation coupled to the induction equation to
follow the evolution of the magnetic field. Here, $\rho$ is the mass
density, $P$ is the thermal pressure, $\vec{v}$ and $\vec{B}$ are,
respectively, the velocity and magnetic field vectors, $\vec{I}$ is the
identity tensor, $\mathcal{T}$ is the viscous stress tensor, $\vec{g}
= -(GM_{\star}/R^2) \hat{R}$ is the gravitational acceleration (where
$G$ is the gravitational constant, $M_\star$ the stellar mass and $R$
the spherical radius), $\vec{J} = \nabla \times \vec{B}/4 \pi$ is the
electric current and $\eta_{\mathrm{m}}$ is the magnetic
resistivity. The magnetic diffusivity is defined as $\nu_{\mathrm{m}}
= \eta_{\mathrm{m}}/4 \pi$. The total energy $E$, given by the sum of
internal, kinetic, and magnetic energy, is defined as  
\begin{equation}
  E = \rho e +\rho \frac{\vec{v} \cdot \vec{v}}{2}
  +\frac{\vec{B} \cdot \vec{B}}{8 \pi} \; ,
  \label{eq_totenrg}
\end{equation}
where $e(T)$ is the specific internal energy as a function of
temperature $T$. As described in detail in Appendix A of
\citet{Pantolmos:2020aa}, we employed an equation of state for a
calorically imperfect gas, where the specific heats at constant
pressure and volume and their ratio $\gamma$ (the polytropic index)
are temperature-dependent. In particular, the plasma in our models
behaves almost isothermally ($\gamma = 1.05$) at high temperatures
and adiabatically ($\gamma = 5/3$) at low temperatures, offering the
ability to simulate  hot stellar winds and cold adiabatic
accretion disks at the same time. The equation of state is set so that $\gamma=1.05$
for $P/\rho > 0.1 GM_\star/R_\star$ and $\gamma=5/3$ for $P/\rho <
0.01 GM_\star/R_\star$, where $R_\star$ is the stellar radius. Only
the $\mathcal{T}_{R\phi}$ component of the viscous stress tensor is
included 
\begin{equation}
\mathcal{T}_{R\phi} = \eta_\mathrm{v}R \frac{\partial}{\partial R} \left( \frac{v_\phi}{R}\right) \; ,
\end{equation}
where $\eta_\mathrm{v}$ and $\nu_\mathrm{v} = \eta_\mathrm{v}/\rho$
are the dynamic and the kinematic viscosities, respectively. The
viscous and resistive terms were included, so they did not
provide any dissipative viscous and Ohmic heating. We also solved two
passive scalar equations: one for the entropy, as defined in Appendix
A in \citet{Pantolmos:2020aa}, to control the numerical dissipation
and the possible negative pressures associated with the conservation of
the total energy, along with a second one for a tracer used to distinguish
the disk material from the coronal and stellar wind plasma.  

The system of Eqs. (\ref{eq_MHD}) was numerically solved using a
second-order Godunov method provided by the PLUTO code
\citep{Mignone:2007aa}. Primitive variables were spatially
reconstructed using a mix of linear and parabolic limiters. The inter-cell
fluxes were computed using the HLLD Riemann solver by
\citet{Miyoshi:2010aa}, which allowed us to subtract the contribution of
potential force-free magnetic fields (i.e., the initial stellar
magnetosphere) from the estimate of the Laplace force. A second-order
Runge-Kutta method was used to advance the equations in time. The
divergence-cleaning method \citep{Dedner:2002aa} was employed to
control the solenoidal $\nabla\cdot\vec{B} = 0$ condition. Resistive
and viscous terms were explicitly integrated in time. The system
of Eqs. (\ref{eq_MHD}) was solved in a frame of reference
co-rotating with the star. In this frame of reference, the velocity
field $\vec{u}$ is defined as 
\begin{equation}
\vec{u} = \vec{v} - r\Omega_\star \hat{\phi} \; ,
\end{equation}
where $r$ is the cylindrical radius and $\Omega_\star$ is the stellar
angular speed. We assume that in the rotating frame of reference,
the electric field at the stellar surface is equal to zero, so the magnetic flux through the stellar surface is frozen and does not
change over time. 

We solved the MHD equations in three spatial dimensions using a
spherical system of coordinates ($R$, $\theta$, $\phi$), where we use $R$ to denote the spherical radius, $\theta$ the colatitude, $r=R
\sin \theta$ the cylindrical radius, and $\phi$ the longitude. The
computational domain covers a radial distance $R \in [1,34.28]
R_\star$ with 144 point logarithmically spaced so that the grid
spacing is proportional to the distance, $\Delta R \propto R$. The
colatitude $\theta \in [0,\pi]$ and the longitude $\phi \in [0,2\pi]$
are both resolved with 128 points. In addition, we employed the adaptive mesh
refinement (AMR) tool provided with PLUTO 4.3 \citep{Mignone:2012ac} to
lower the spatial resolution around the polar axis. 
This approach increases the integration time step, 
which leads to a decrease in the computational cost of the simulations.
The spatial resolution employed in this work is similar
to that of the \citet{Blinova:2016aa} models, but it is about a factor of 2 lower
than in \citet{Takasao:2022aa}. This resolution is therefore adequate
to resolve the structure of an $\alpha$ viscous and resistive disk,
while the lower computational cost enables us to perform a much
larger number of simulations compared to the turbulent models of
\citet{Takasao:2022aa} or \citet{Zhu:2025aa} to provide better
statistics and scaling relations. 

\subsection{Initial and boundary conditions}
\label{sec_ic_bc}

For the simulation initial conditions, we needed to set up an accretion disk, a stellar
corona, and the stellar magnetic field. The disk density,
$\rho_\mathrm{d}$, thermal pressure, $P_\mathrm{d}$ and toroidal speed,
$v_{\phi\mathrm{d}}$, are determined by the vertical and radial
hydrostatic equilibrium, assuming a polytropic relation, $P_\mathrm{d}
\propto \rho_\mathrm{d}^\gamma$ with $\gamma = 5/3$, via
\begin{equation}
\label{eq_disk}
\begin{aligned}
\rho_\mathrm{d} &= \rho_\mathrm{d0} \left \{ \frac{2}{5 \epsilon^2}
\left[\frac{R_\star}{R} - \left( 1 - \frac{5
        \epsilon^2}{2}\right) \frac{R_\star}{r} \right] \right \}^{3/2}, \\
P_\mathrm{d} &= \epsilon^2 \rho_\mathrm{\mathrm{d}0} v_{\mathrm{K} \star}^2
\left(\frac{\rho_d}{\rho_{d0}} \right)^{5/3}, \\
v_{\phi \mathrm{d}} &= \sqrt{\left(1-\frac{5}{2}\epsilon^2\right)\frac{GM_\star}{r}} \, ,
\end{aligned}
\end{equation}  
where $\epsilon = c_\mathrm{sd}/v_\mathrm{K}\rvert_{\theta=\pi/2}$ is the disk aspect ratio defined by the ratio between the disk isothermal
sound speed, $c_\mathrm{sd}=\sqrt{P_\mathrm{d}/\rho_\mathrm{d}}$, and
the Keplerian speed, $v_\mathrm{K} = \sqrt{GM_\star/r}$, evaluated at
the disk midplane; $\rho_{\mathrm{d}0}$ and $v_{\mathrm{K} \star}$ are
the disk density and Keplerian speed at the disk midplane at
$R_\star$. We assume that MRI-driven turbulence leads to the
development in the disk of an anomalous $\alpha$ viscosity
\citep{Shakura:1973aa} and resistivity \citep{Ferreira:1993aa} that we can
parametrize as
\begin{equation}
\label{eq_nu}
\nu_\mathrm{v} = \alpha_\mathrm{v}\frac{c_\mathrm{sd}^2}{\Omega_\mathrm{K}}, \qquad \qquad
\nu_\mathrm{m} = \alpha_\mathrm{m}\frac{c_\mathrm{sd}^2}{\Omega_\mathrm{K}} \; ,
\end{equation}
where $\Omega_\star = \sqrt{GM_\star/r^3}$ is the disk Keplerian
angular speed. We refer to Sect. 2.2 in
\citet{Pantolmos:2020aa} for the complete spatial and time-dependent
definition of the disk isothermal sound speed, $c_\mathrm{sd}$, and the
expression for the transport coefficients, $\nu_\mathrm{v}$ and
$\nu_\mathrm{m}$, used to restrict the viscous and resistive effects to
the low-magnetization ($\beta = 8\pi P/B^2 > 1$) regions of the
disk. With the viscosity defined by Eq. (\ref{eq_nu}), the initial
disk accretion speed is given by 
\begin{equation}
\label{eq_vacc}
v_{R\mathrm{d}} = -\frac{3}{2}\alpha_{\mathrm{v}} \frac{c_\mathrm{sd}^2}{v_\mathrm{K}} \sin\theta \, .
\end{equation}
This equation shows that the initial accretion inertia is of the order of $\mathcal{O}(\alpha_\mathrm{v}^2\epsilon^4)$ and therefore does not strongly affect the disk hydrostatic equilibrium.

Above the disk, we set a stellar corona using the thermal pressure and
density profiles of a 1D, spherically symmetric,
isentropic transonic Parker-like wind model. This solution is defined
by its density, $\rho_\star$, and sound speed, $c_{\mathrm{s}\star}$, at
$R=R_\star$. Its speed is set to zero to avoid initial supersonic
motions. With suitable boundary conditions, this initial condition
will drive an almost isothermal wind emerging with a subsonic velocity
from the inner boundary, accelerating to supersonic speeds to fill the
polar regions. The initial vertical boundary between the disk and the
corona is set by the thermal pressure equilibrium, while the disk is
initially truncated at a radius where the star-disk interaction torque
should prevail over the internal viscous one. We refer to Sect. 2.2 in
\citet{Pantolmos:2020aa} for more details.  

The initial stellar magnetosphere is a dipolar magnetic field with the
magnetic moment misaligned with respect to the stellar rotation axis
by an angle $\Theta$. The vector components are 
\begin{equation}
\begin{aligned}
B_R & = B_\star \left(\frac{R_\star}{R}\right)^3\left(\cos\Theta\cos\theta + \sin\Theta\sin\theta\cos\phi \right), \\
B_\theta & = \frac{B_\star}{2} \left(\frac{R_\star}{R}\right)^3\left(\cos\Theta\sin\theta - \sin\Theta\cos\theta\cos\phi \right), \\
B_\phi & = \frac{B_\star}{2} \left(\frac{R_\star}{R}\right)^3\sin\Theta\sin\phi \; ,
\end{aligned}
\end{equation}
where $B_\star$ is the field intensity at the magnetic pole.

Along the polar axis, we assume $\pi$-periodic boundary conditions that
allow the flow to smoothly cross the polar regions. At the stellar
boundary, we must consider two different boundary conditions: one for a
subsonic inflow into the computational domain
for the stellar wind, which is also suitable for the almost
hydrostatic magnetospheric cavity, and a second one for a supersonic outflow 
leaving the computational domain (i.e., the accretion funnels). 
For the subsonic inflow condition, we imposed the pressure and density profiles
used to initialize the stellar corona. Since this condition is close
to an hydrostatic equilibrium, it is also suitable for the
magnetospheric cavity. For the supersonic outflow condition, the
density was extrapolated along the magnetic field lines with an
adjustable power-law, while the thermal pressure was determined
assuming a constant entropy. We used the values of the sonic Mach
number and the passive tracer in the first layer of cells of the
domain to interpolate between the two boundary conditions. The
boundary conditions on the other primitive variables were the same in
the two cases. The radial component of the magnetic field was kept
constant to its initial value to conserve the distribution of stellar
magnetic flux in the rotating frame of reference. The $\theta$
component was linearly extrapolated into the ghost zones. Following
\citet{Zanni:2009aa,Zanni:2013aa,Pantolmos:2020aa}, the toroidal field
was linearly extrapolated to apply a torque onto the first layer
of cells above the inner boundary that forces the footpoints of the
magnetic field lines to co-rotate with the star. This condition
imposes a toroidal speed of $v_\phi =
r\Omega_\star-v_\mathrm{p}B_\phi/B_\mathrm{p}$ or $u_\phi =
-v_\mathrm{p}B_\phi/B_\mathrm{p}$, where $v_\mathrm{p}$ and
$B_\mathrm{p}$ are the poloidal speed and magnetic field; thus, in
the rotating frame of reference, the stellar surface electric field is
zero and, therefore, the magnetic flux is frozen.  
In the rotating frame of reference, the velocity is set to be parallel
to the magnetic field using the conservation of the ideal MHD
invariant $k = \rho u/B$ along magnetic field lines; namely,
$u_{(R,\theta,\phi)} = k B_{(R,\theta,\phi)}/\rho$. This condition
guarantees a smooth inflow and outflow avoiding the formation of
shocks at the stellar surface. 
At the outer radial boundary, the density and thermal pressure were
extrapolated using adjustable power-laws to avoid negative values,
while all the other variables were extrapolated linearly. Particular
attention has been devoted to the boundary condition on the toroidal
magnetic field component in the region where the stellar wind exits
the computational domain. Using an approach similar to the one
employed at the stellar boundary, we imposed a boundary condition that
tends to force the open stellar wind magnetic field lines to co-rotate
with the star. This condition allowed us to obtain a correct stellar wind
torque even if the outflow has not become super-Alfv\'{e}nic at the
outer boundary due to the flow geometry \citep{Pantolmos:2020aa} and
the limited size of the domain. The effectiveness of this boundary 
condition  is confirmed by the fact that we obtain an Alfv\'{e}n radius
scaling very similar to trans-Alfv\'{e}nic stellar wind models 
\citep[see, e.g.,][]{Reville:2015aa}, as we show in Sect. \ref{sec_sw}. 

\subsection{Units and normalization}
\label{sec_units}

We performed the simulations in dimensionless units. Here, the density is
expressed in units of the stellar wind density at the stellar surface,
$\rho_\star$, lengths in units of the stellar radius, $R_\star$,
velocities in units of the Keplerian speed at the stellar surface,
$v_{\mathrm{K}\star} = \sqrt{GM_\star/R_\star}$. 
The unit time is therefore $t_0 = R_\star/v_{\mathrm{K}\star}$ while
the magnetic field is expressed in units of $B_0 =
\sqrt{4\pi\rho_\star v_{\mathrm{K}\star}^2}$ , 
\begin{equation}
B_0 = 109.5 \, \left(\frac{\rho_\star}{\scriptstyle 10^{-12}\ \mathrm{g}\
        \mathrm{cm}^{-3}}  \right)^{1/2} 
\left(\frac{M_\star}{M_\sun}\right)^{1/2} 
\left(\frac{R_\star}{2R_\sun}\right)^{-1/2}\ \ \mathrm{G} \; ,
\end{equation} 
the mass accretion/ejection rates in units of $\dot{M}_0 = \rho_\star R_\star^2 v_{\mathrm{K}\star}$ ,
\begin{equation}
\dot{M}_0 = 9.48 \times 10^{-9} \,
\left(\frac{\rho_\star}{\scriptstyle 10^{-12}\ \mathrm{g}\
        \mathrm{cm}^{-3}}  \right)
\left(\frac{M_\star}{M_{\sun}}\right)^{1/2}
\left(\frac{R_\star}{2R_{\sun}}\right)^{3/2}\ \
M_{\sun}\ \mathrm{yr^{-1}} \; ,
\end{equation}
and torques in units of $\dot{J}_0 = \rho_\star R_\star^3
v_{\mathrm{K}\star}^2$. Since the density at the stellar surface,
$\rho_\star$, can be difficult to constrain, it is possible to express
the normalization in terms of the dipolar field intensity at the
magnetic pole, $B_\star$. For example, the mass accretion rate
normalization, $\dot{M}_0$, can be rewritten as 
\begin{equation}
\dot{M}_0 = 1.98 \times 10^{-9} \,
\left(\frac{B_\star}{\mathrm{kG}}\right)^2
\left(\frac{B_\star/B_0}{20}\right)^{-2}
\left(\frac{M_\star}{M_{\sun}}\right)^{-1/2}
\left(\frac{R_\star}{2R_{\sun}}\right)^{5/2}\ \
M_{\sun}\ \mathrm{yr^{-1}} \; ,
\end{equation}
where the dimensionless dipolar intensity $B_\star/B_0$ is a free
parameter of the simulations; see Sect. \ref{sec_params}. In
Sects. \ref{sec_sditorques} and \ref{sec_sw} we present the
torques exerted onto the star dividing them by the stellar angular
momentum $J_\star = k^2 R_\star^2 M_\star \Omega_\star$, to
directly provide the inverse of the spin-up and spin-down timescales. We show them in units of 
\begin{equation}
\left. \frac{\dot{J}}{J_\star} \right\vert_0 = 10^{-6} \, \left(\frac{k^2}{0.2}\right)^{-1}
\left(\frac{B_\star}{\mathrm{kG}}\right)^2
\left(\frac{M_\star}{M_\sun}\right)^{-3/2}\left(\frac{R_\star}{2R_\sun}\right)^{5/2} \; \mathrm{yr}^{-1} \; ,
\label{eq_jnorm}
\end{equation}
where $k^2=0.2$ is the square of the normalized radius of gyration for
a fully convective star modeled with an $n=3/2$ polytrope
\citep{Rucinski:1988aa}.

  \begin{figure*}[!ht]
        \centering
        \begin{subfigure}{.5\textwidth}
                \centering
                \includegraphics[width=0.95\linewidth]{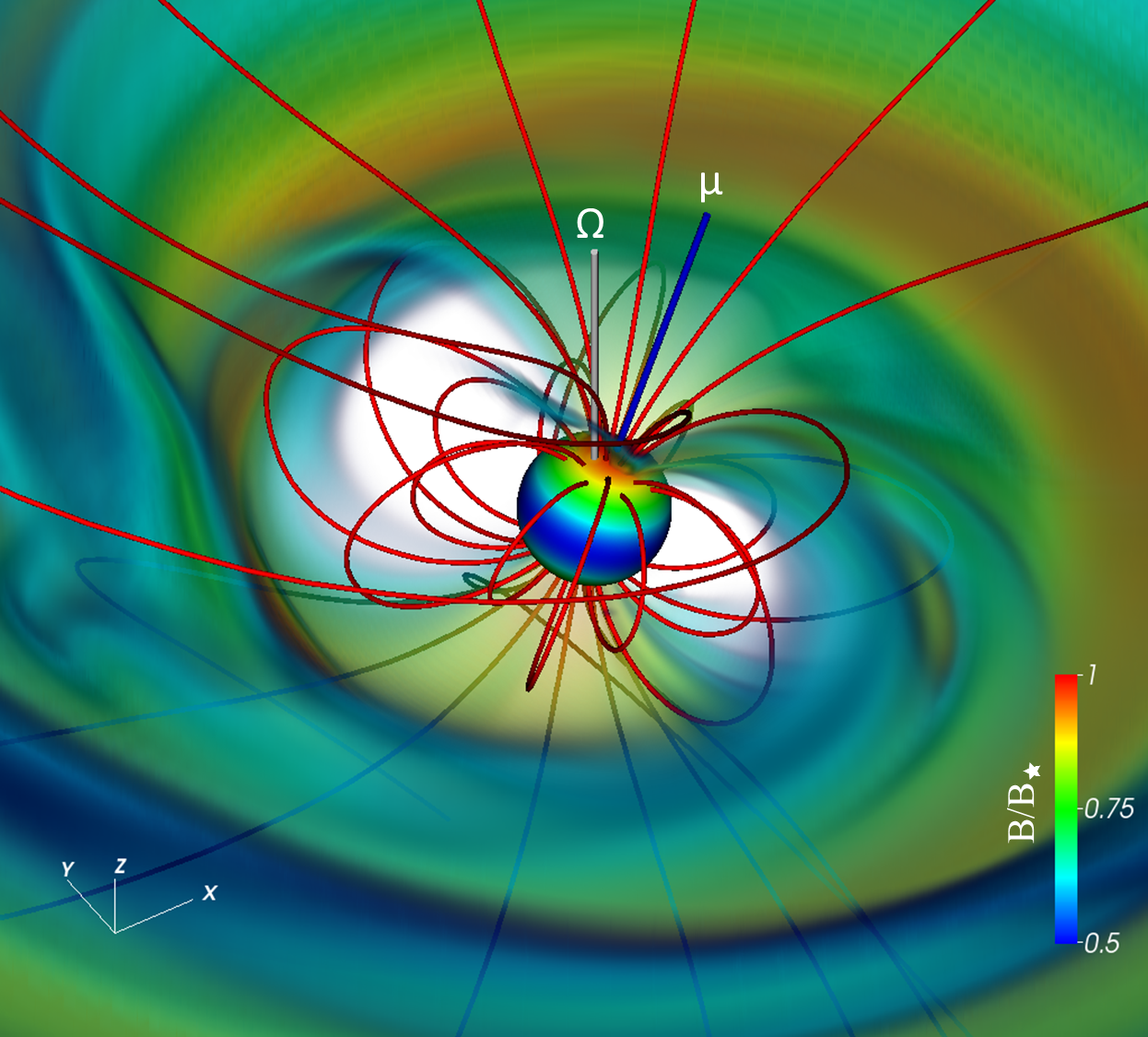}
        \end{subfigure}%
        \begin{subfigure}{.5\textwidth}
                \centering
                \includegraphics[width=0.95\linewidth]{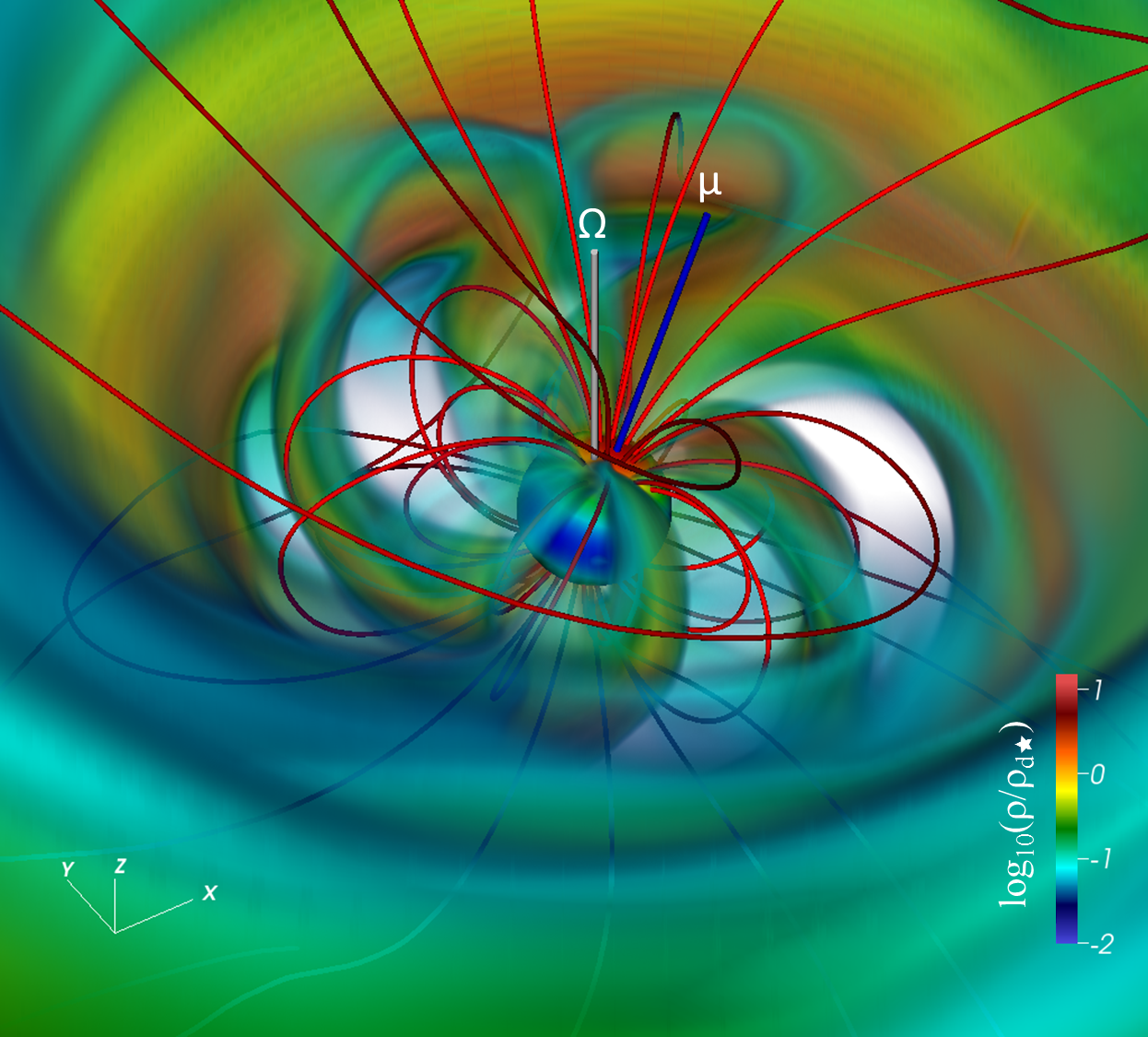}
        \end{subfigure}
        \caption{Volume rendering of the normalized density,
                $\rho/\rho_\star$, in the inner part of the computational
                domain in a stable (\textit{left}) and an unstable (\textit{right}) regime. The snapshots were taken from case 14 after
                21.75 stellar periods (left) and case 16 after 18
                stellar periods (right). The gray line is the stellar
                rotation axis. The blue axis indicate the direction of the
                dipolar moment $\vec{\mu}$ that is inclined by $20^{\circ}$
                from the rotation axis. The magnetic field is displayed with
                red lines. The stellar surface shows the magnetic field
                intensity.}
        \label{fig_rho_3d}
  \end{figure*}
   
\subsection{Simulation parameters}
\label{sec_params}

Given the normalization presented in Sect. \ref{sec_units}, the initial conditions of our simulations described in Sect. \ref{sec_ic_bc} depend on eight dimensionless parameters. We varied four of them: 
\begin{enumerate}
\item the stellar rotation rate, $f$, expressed as the fraction of the stellar break-up speed $f = R_\star\Omega_\star/v_{\mathrm{K}\star}$.
By defining the Keplerian (break-up) rotation period as $P_0=2\pi R_\star/v_{\mathrm{K}\star}$, which can also be written as
\begin{equation}
P_0 = 0.328 \, \left(\frac{M_\star}{M_\sun}\right)^{-1/2}
\left(\frac{R_\star}{2R_\sun}\right)^{3/2}\ \ \mathrm{days} \; ,
\label{eq_P0}
\end{equation} 
the stellar period of rotation is given by $P_\star = P_0 f^{-1}$. For a solar-mass star with
a radius $R_\star = 2R_\sun$ and the three $f$ values considered in this paper, it corresponds
to stellar periods $P_\star = 2.19, 4.69$ and $8.63$ days.
The rotation parameter $f$ also provides the position of the Keplerian corotation radius $R_\mathrm{co}/R_\star = f^{-2/3}$;

\medskip

\item the intensity of the dipolar magnetic field of the star measured at the magnetic pole, $B_\star/B_0$ ;

\medskip

\item the dipolar field misalignment with respect to the rotation axis, $\Theta$ ; 

\medskip

\item the density contrast between the disk and the stellar corona,
  $\rho_\mathrm{d0}/\rho_\star$, measured at the inner radius,
  $R_\star$. 

\end{enumerate}

The parameters that were varied across the 21 simulations performed for this study
are listed in columns 2–7 of Table \ref{tab_data_sim}.
Other four parameters were kept fixed for all the simulations: the disk
thermal aspect ratio $\epsilon = 0.075$, the stellar wind temperature
and sound speed at the stellar surface $c_{\mathrm{s}\star} = 0.37 \
v_{\mathrm{K}\star}$ (corresponding to a specific enthalpy of $h_\star=
1.67 \ v_{\mathrm{K}\star}^2$) and the turbulent viscosity and resistivity parameters, $\alpha_\mathrm{v} = \alpha_\mathrm{m} =
0.2$. These relatively high values of the transport coefficients are
consistent with the scaling of $\alpha_\mathrm{v} \approx 5\beta^{-1/2}$
found at least in shearing box simulations \citep[see,
e.g.,][]{Salvesen:2016aa}, which predicts a strong turbulence in a
situation around equipartition, the typical condition found in the
magnetospheric interaction region of the disk.  

All the simulations were integrated in time for 30 stellar periods.
The choice was made to ensure we had comparable statistics for 
all our simulations regarding the system variability induced by the stellar 
rotation.
In addition, the disk dynamics around the corotation region, where we
we show that important phenomena as the onset of the interchange 
instability and the formation of accretion patterns can take place, 
was followed for the same number of orbits.
In our analysis, we discarded the first ten periods to
avoid initial transients. In addition, in this work, we only present
quantities time-averaged over the final 20 stellar periods and we
defer the analysis of the time-dependent properties of our solutions
to a second companion paper.
Because of these choices, simulations characterized by different
stellar periods end up averaged on different integration timescales. 
In Appendix \ref{sec_variability}, where the temporal evolution of key 
quantities such as the mass accretion rate and the disk truncation radius are 
presented, we show that our assumptions have no major effect on 
the analysis presented in this work.
  
\section{Stable and unstable accretion regimes}
\label{sec_bndry}

\begin{figure*}[!t]
        \centering
        \includegraphics[width=\linewidth]{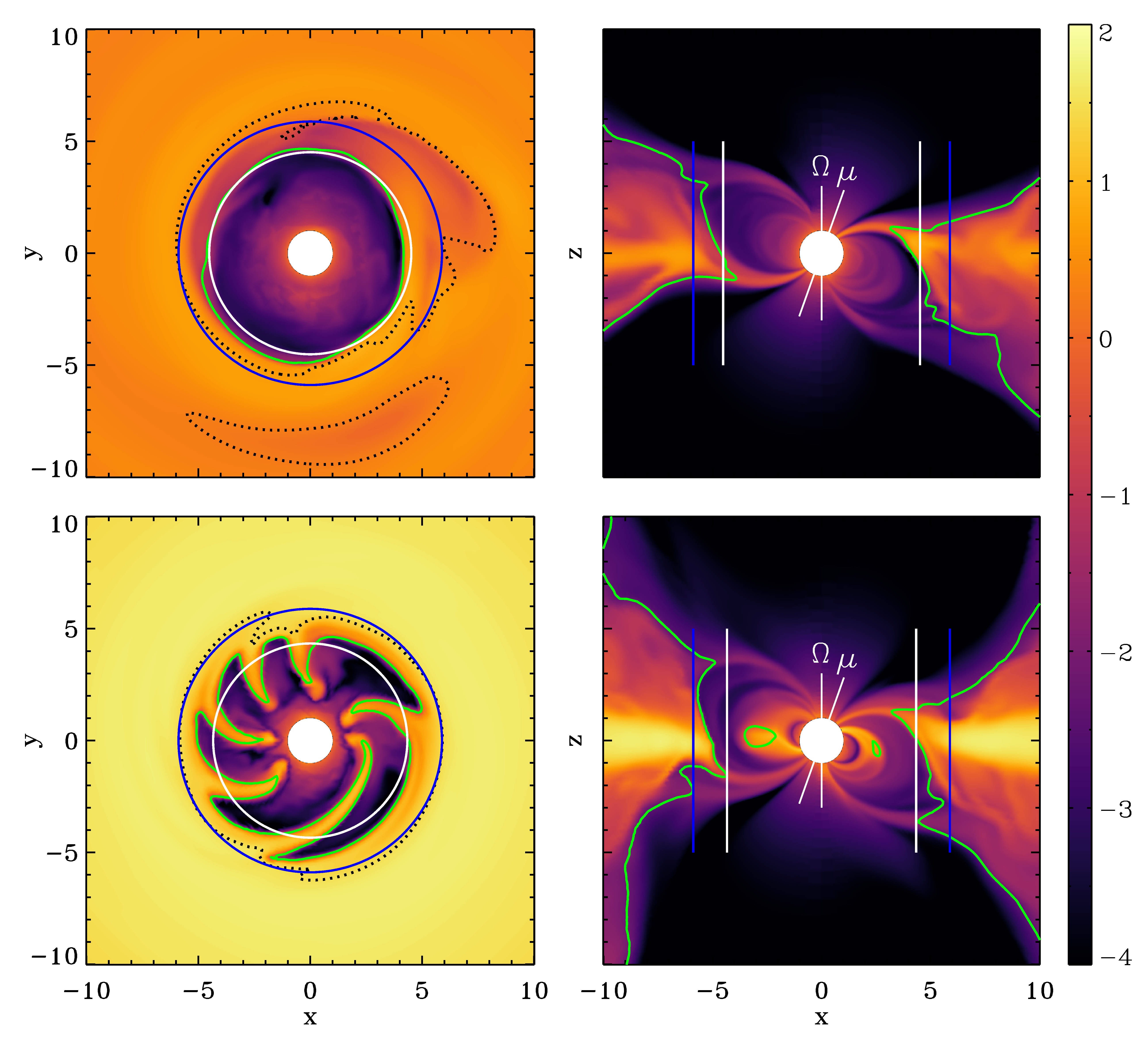}
        \caption{Slices of the logarithmic mass density in the
                equatorial $xy$ plane (\textit{left}) and vertical $xz$ plane
                (\textit{right}). The snapshots correspond to the stable case
                14 after 25.3 stellar periods (\textit{top}) and to the
                unstable case 16 after 29.6 stellar periods (\textit{bottom}). The dotted black lines correspond to the $\beta =
                8\pi P/B^2 = 1$ contour while the green lines indicate the
                $\beta_\mathrm{T}=1$ (Eq. \ref{beta_param}) contour. The
                blue lines indicate the position of the corotation radius
                $R_\mathrm{co}$. The white lines show the average truncation
                radius $R_\mathrm{t}$ defined by Eq. (\ref{eq_rt_avg}). The
                two white axes in the $xz$ slices, labeled as $\Omega$ and
                $\mu$, are the stellar rotation and magnetic moment axes,
                respectively. } 
        \label{fig_rho_2d}
\end{figure*}

The average radial distance of the region where the magnetospheric
star-disk interaction disrupts the disk accretion flow and channels it
towards the stellar surface is commonly referred to as the
truncation or magnetospheric radius, $R_\mathrm{t}$. It has been
proposed and widely accepted 
\citep[e.g.,][]{Kulkarni:2008aa,Blinova:2016aa,Takasao:2022aa,Zhu:2025aa} 
that accretion from the truncation radius towards the stellar surface 
can proceed in two different ways: (1) the disk truncation region has
an approximately circular shape and the accretion flow is channeled 
into orderly curtains that flow along the stellar magnetic field. 
If the stellar magnetic moment is misaligned with its rotation axis, 
two point-symmetric accretion funnels should form that impact the stellar
surface to produce two main opposite accretion spots. This is usually
referred to as a stable accretion regime; (2) the disk is
fragmented in the truncation region and can penetrate the stellar
magnetosphere forming different accretion tongues producing several
accretion spots distributed at different azimuths and latitudes. 
This is usually referred to as an unstable accretion regime. 

\begin{figure*}[!t]
        \centering
        \includegraphics[width=\linewidth]{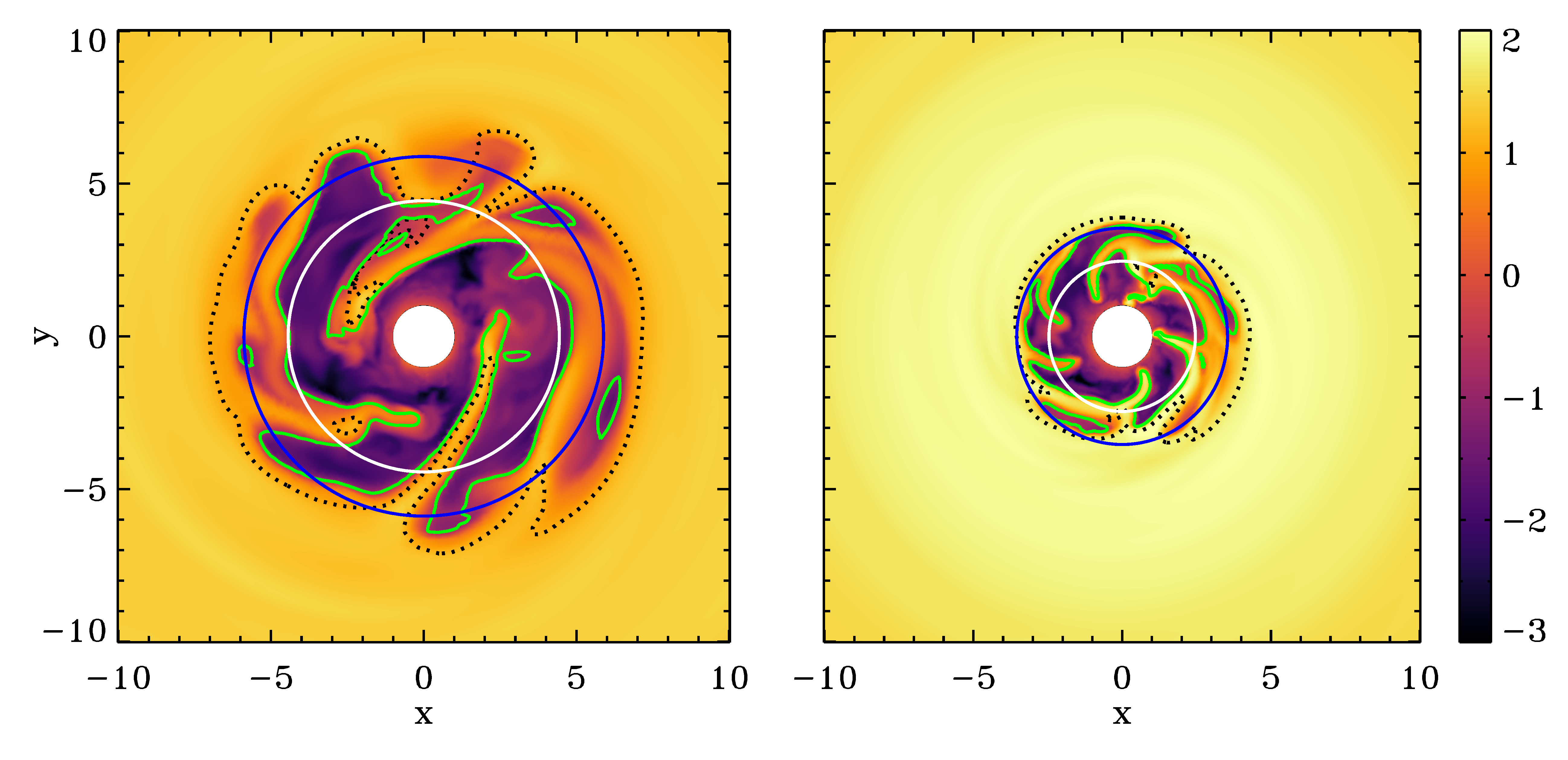}
        \caption{Slices of the logarithmic mass density in the
                equatorial plane $xy$ corresponding to the two unstable cases
                17 at 21 stellar periods (\textit{left}) and 11 
                taken at 14.875 stellar periods (\textit{right}). As in the left panels of Fig.
                \ref{fig_rho_2d}, the dotted black lines correspond to the $\beta =
                8\pi P/B^2 = 1$ contour while the green lines indicate the
                $\beta_\mathrm{T}=1$ (Eq. \ref{beta_param}) contour. The
                blue circles indicate the position of the corotation radius,
                $R_\mathrm{co}$, while the white rings show the average truncation
                radius, $R_\mathrm{t}$, defined by Eq. (\ref{eq_rt_avg}).} 
        \label{fig_rho_2d_unst}
\end{figure*}

This general picture is confirmed by our simulations. In the left
panel of Fig. \ref{fig_rho_3d}, we show a 3D volume
rendering of the density in a stable situation, where only two broad
and almost point-symmetric accretion funnels are formed. A similar
stable configuration is shown in the upper panels of
Fig. \ref{fig_rho_2d}, where we show cuts of the logarithmic density
in the equatorial ($xy$, left panel) and vertical ($xz$, right panel)
planes. Clearly the magnetic cavity has an almost circular shape
(upper-left panel) while two main accretion funnels are formed
preferably in the direction towards which the dipole is tilted
(upper-right panel). On the other hand, the right panel of
Fig. \ref{fig_rho_3d} shows that, in an unstable regime,
magnetospheric accretion is fragmented into multiple tongues accreting
at different azimuthal positions and latitudes. The density slices in
the lower panels of Fig. \ref{fig_rho_2d} show how the equatorial
accretion flow is fragmented into multiple spikes (lower-left panel)
while the vertical cut (lower-right panel) shows how the equatorial
tongues can be truncated closer to the star forming lower-latitude
accretion spots coexisting with higher-latitude but possibly weaker
ones coming from disk regions that are truncated at larger
distances. Since the formation of accretion tongues and their pattern
is time-dependent, unstable configurations are expected to be more
variable than stable ones.     

The formation of equatorial accretion tongues that characterize the
unstable regime is due to the development of an interchange
instability. The interchange is the MHD counterpart of the
hydrodynamic Rayleigh-Taylor (RT) instability, where a higher density
fluid is sitting on top of a lighter one and, due to the action of
gravity, the higher density fluid is pulled into the lighter one
forming plunging spikes and rising plumes \citep{Stone:2007aa,Carlyle:2017aa}. 
These works considered a simplified configuration taking into account, 
for example, uniform magnetic fields. The geometry of the star-disk interaction 
is far more complex and
in this case the heavier fluid is represented by the disk while the lighter 
one is the magnetospheric cavity. Therefore, the truncation region is expected to
be naturally prone to be interchange unstable \citep{Arons:1976aa,Wang:1985aa}. 
\citet{Spruit:1995aa} proposed the following criterion for the development of 
an interchange instability in magnetized accretion disks, 
\begin{equation}
\left(\frac{GM_\star}{r^2}-\Omega^2
  r\right)\frac{\mathrm{d}}{\mathrm{d}r}\left[\mathrm{ln}
  \left(\frac{\Sigma}{B} \right)\right] > 2 \left(r
  \frac{\mathrm{d}\Omega}{\mathrm{d}r}\right)^2\; , 
\label{eq_spruit}
\end{equation}   
where $\Omega$ is the disk angular velocity, $\Sigma$ is its column
density, and $B$ the local poloidal magnetic field. The radial
derivative in the left-hand side is the MHD equivalent of the Atwood
number for the RT instability and is expected to be positive in the
truncation region. This instability criterion shows that the disk
rotation $\Omega$ can have a stabilizing effect: first by reducing the
effective gravity, $g_\mathrm{eff} = -GM_\star/r^2+\Omega^2 r$, thanks
to the centrifugal acceleration and second by increasing the
instability threshold, thanks to the disk differential rotation. It
was shown \citep[e.g.,][]{Kluzniak:2007aa,Zanni:2009aa,Zanni:2013aa} 
that inside the corotation radius the magnetospheric star-disk interaction
attempts to force stellar corotation to yield a sub-Keplerian
rotation and an increase in the effective gravity, and, at the same
time, to flatten the disk rotation profile to decrease the
instability threshold. Therefore we expect the interchange instability
to start developing for $R_\mathrm{t} \le R_\mathrm{co}$, with the
ratio $R_\mathrm{t}/R_\mathrm{co}$ being the main parameter that
determines the transition between stable and unstable regimes. This
trend was first confirmed by
\citet{Blinova:2016aa}. Large magnetic field obliquities, with $\Theta
> 30^{\circ}$, are also expected to suppress the development of the
instability \citep{Kulkarni:2008aa}, since the gravito-centrifugal
potential barrier to overcome to form ordered accretion funnels is
reduced.

\begin{figure}
  \centering
  \includegraphics[width=\linewidth]{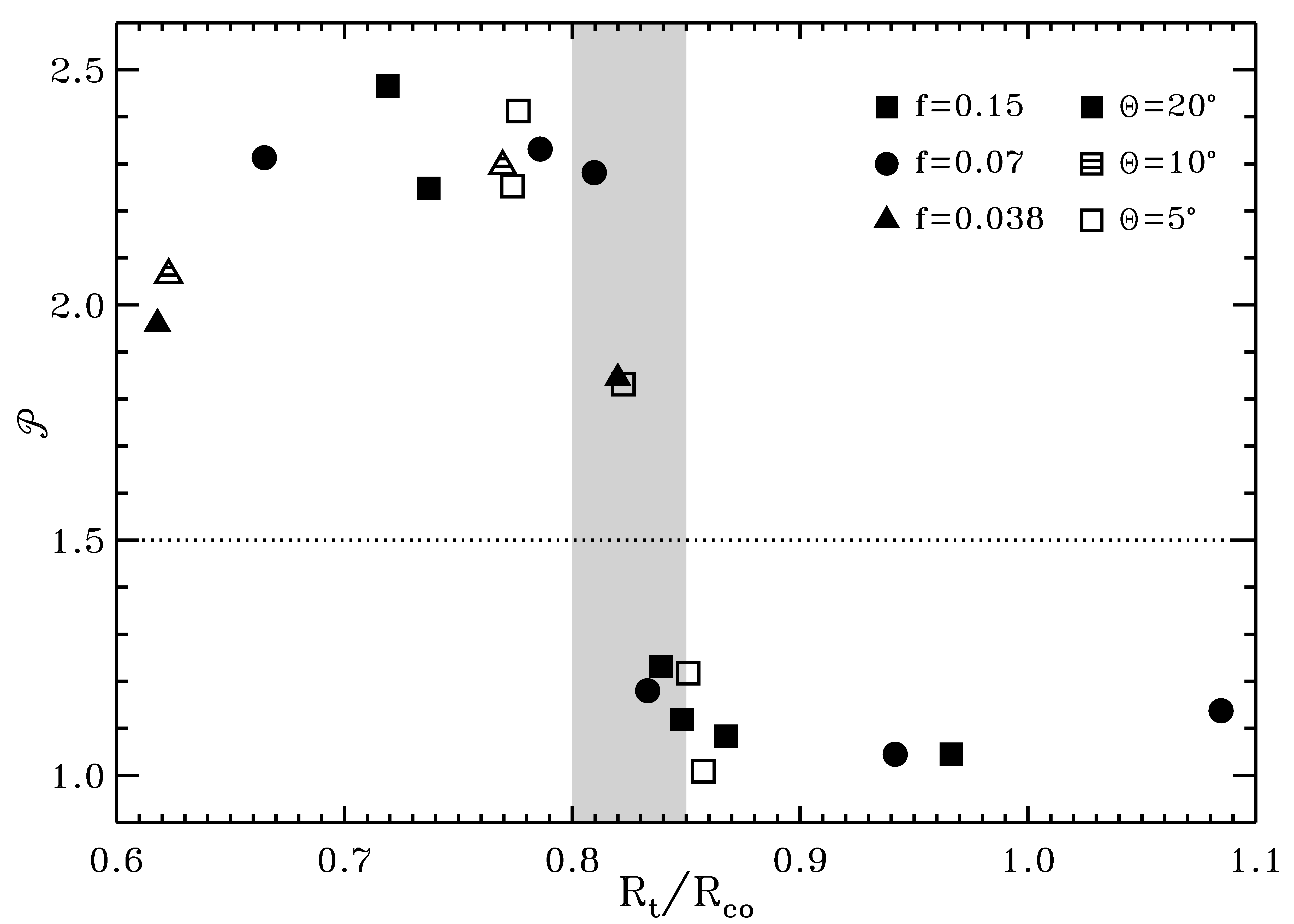}
  \caption{Instability parameter, $\mathcal{P}$, defined by
    Eq. (\ref{p_param}) versus the $R_\mathrm{t}/R_\mathrm{co}$
    ratio. Symbol shapes correspond to a different stellar rotation
    rate ($f=0.15$, squares, $f=0.07$, circles, $f=0.038$,
    triangles). Filled symbols correspond to a  
  $\Theta = 20^\circ$ magnetic field misalignment, striped symbols to
  $\Theta = 10^\circ$ and empty symbols to $\Theta = 5^\circ$. The
  gray area corresponding to $0.8<R_\mathrm{t}/R_\mathrm{co}<0.85$
  roughly marks the transition between unstable ($\mathcal{P}>1.5$)
  and stable ($\mathcal{P}<1.5$) cases. The horizontal dotted line
  corresponds to the arbitrary $\mathcal{P} = 1.5$ instability
  threshold. }
  \label{fig_p_rtrco}
\end{figure}

While Eq. (\ref{eq_spruit}) has often 
and convincingly been tested on the outcome of numerical experiments 
\citep[e.g.,][]{Blinova:2016aa, Takasao:2022aa,Zhu:2025aa}, 
it represents a linear instability criterion derived for a thin disk in a shearing-sheet 
approximation. We therefore prefer to look for another way to estimate the 
amplitude of the interchange instability and identify the boundary between 
stable and unstable regimes.
Since it appears to be the main parameter that controls the instability
development, we need first to estimate the position of the disk
truncation radius, $R_\mathrm{t}$, with respect to corotation,
$R_\mathrm{co}$. 
Customarily, in numerical experiments the truncation radius is
evaluated by looking at the dynamical balance between the magnetic
stellar pressure and the disk mechanical/thermal one. For example, a
common approach is to look at the region where the disk thermal
pressure equals the stellar magnetic pressure, or $\beta = 8\pi P/B^2
= 1$ \citep{Pringle:1972aa,Bessolaz:2008aa,Takasao:2022aa}. This
expression marks a position still located in the accretion disk inside
which the magnetic torque exerted by the large-scale stellar field
becomes dominant and the accretion curtains start to form. 
A similar location can be determined by looking at the geometrical shape
of the outermost accreting magnetic field lines \citep[see,
e.g.,][]{Pantolmos:2020aa}. If instead, the stellar magnetospheric
pressure is larger than the total (thermal plus ram) disk pressure
$P+\rho v^2$, the disk is completely disrupted and the position where   
\begin{equation}
\label{beta_param}
\beta_\mathrm{T} = \frac{P + \rho v^2}{B^2/8\pi} = 1 \; ,
\end{equation}
quite precisely tracks the shape of the magnetospheric cavity. 
  The $\beta_\mathrm{T} = 1$ location has been customarily used in many previous numerical works
 \citep[see, e.g.,][and references therein]{Romanova:2002aa, Romanova:2025aa} to estimate the position of the truncation radius, which they defined as the 
innermost circular orbit unperturbed by the formation of the unstable tongues,
i.e., the maximum radial size of the magnetospheric cavity delimited by the
$\beta_\mathrm{T} = 1$ curve.
Clearly, a truncation definition based on the $\beta = 1$ location tends to provide 
a slightly larger truncation radius than the $\beta_\mathrm{T} = 1$ criterion. 
Since the location determined by Eq. (\ref{beta_param}) seems to be more sensitive 
to the development of the interchange instability and the shape of the
magnetospheric cavity, we use this expression to define an average
truncation radius $R_\mathrm{t}$. Obviously, the shape of the magnetic
cavity can be quite irregular, particularly in an unstable regime. We
therefore define an average $R_\mathrm{t}$ as the radius of a circle
that has the same area $Ar_{\beta_\mathrm{T}=1}$ of the region
encircled by the $\beta_\mathrm{T} = 1$ curve, including the stellar
disk, via 
\begin{equation}
R_\mathrm{t} = \left(\frac{Ar_{\beta_\mathrm{T}=1}}{\pi}\right)^{1/2} \; .
\label{eq_rt_avg}
\end{equation}
In the case of a perfectly circular magnetic cavity this expression
equals the radius of the circle while, in the case of a more irregular
shape, it provides an average between the maximum and the minimum
radial width of the magnetospheric boundary. 
If the truncation radius marks the base of the accretion columns,
our definition takes into account the fact that in an unstable regime accretion 
funnels do not form only at the outer edge of the magnetospheric cavity,
but also from the truncation of the accreting fingers; see the lower panels 
in Fig. \ref{fig_rho_2d}. 
As the accretion tongues become wider and/or more numerous, likely indicating
that accretion through the unstable spikes is becoming more important than 
accretion from the funnels at the outer edge of the magnetospheric cavity, our 
estimate of the truncation radius moves inward with respect to the maximum radial 
extent of the cavity.
The time-averaged values of the 
$R_\mathrm{t}/R_{\star}$ ratio extracted from our simulations are given 
in Table \ref{tab_data_sim}. To estimate the amplitude of the interchange
instability, we can define an instability parameter, $\mathcal{P}$, via
\begin{equation}
\label{p_param}
\mathcal{P} = \frac{Pr_{\beta_\mathrm{T} = 1}}{2 \pi R_\mathrm{t}} \; ,
\end{equation}
where $Pr_{\beta_\mathrm{T} = 1}$ is the perimeter of the $\beta_\mathrm{T} = 1$ curve. 
Since for the same area$-$or an equal $R_\mathrm{t}$ as defined by
Eq. (\ref{eq_rt_avg})$-$a circle has the minimum perimeter compared to
any other closed curve, we expect the $\mathcal{P}$ parameter to be
close to unity in a stable regime, where the magnetic cavity has
an approximately circular shape, and $\mathcal{P}$ values larger than
one in unstable regimes where the magnetic cavity has an
irregular shape. In Fig. \ref{fig_p_rtrco}, we plot a time-averaged
value of the $\mathcal{P}$ parameter (also listed in Table
\ref{tab_data_sim}) as a function of the
$R_\mathrm{t}/R_\mathrm{co}$ ratio. This plot shows quite a sharp
transition around $R_\mathrm{t}/R_\mathrm{co} \approx 0.8-0.85$
between $\mathcal{P}$ values larger than the arbitrary threshold
$\mathcal{P} = 1.5$ for $R_\mathrm{t}/R_\mathrm{co} < 0.8$, that we
associate with an unstable accretion regime, and $\mathcal{P} < 1.5$
values for $R_\mathrm{t}/R_\mathrm{co} > 0.85$, that we associate with
a stable regime.  
Interestingly, Fig. \ref{fig_p_rtrco} shows a decrease in the
instability parameter $\mathcal{P}$ for $R_\mathrm{t}/R_\mathrm{co}
\lesssim 0.7$. This is likely due to the development of fewer but
azimuthally wider unstable accretion tongues that reduce the perimeter
of the magnetospheric cavity compared to an unstable case with more
and thinner spikes. This behavior is likely analogous to the
ordered unstable accretion regime identified by
\citet{Blinova:2016aa}.  

\section{Magnetospheric radius parametrization}
\label{sec_rtrunc}

In the previous section, we show that the transition between stable
and unstable regimes is mainly determined by the value of the
$R_\mathrm{t}/R_\mathrm{co}$ ratio, where we use
Eq. (\ref{eq_rt_avg}) to estimate the truncation radius in our simulations. 
From a practical or observational point of view, this expression 
is not particularly useful as it is not possible to observationally determine
the size or the shape of the $\beta_\mathrm{T} = 1$ (or the $\beta = 1$)
region. Thus, in this section, we derive a parametrization for our
definition of $R_\mathrm{t}/R_\star$ that depends on global or
observable quantities such as the stellar mass, radius and rotation rate, the
magnetic field strength (a dipole in our case), and the mass accretion
rate. 

Customarily, the disk truncation radius has been parametrized as a
function of the accretion parameter, $\Upsilon_\mathrm{acc}$
\citep[e.g.,][]{Pantolmos:2020aa,Ireland:2021aa},  
\begin{equation}
\label{eq_yacc}
\Upsilon_\mathrm{acc} = \frac{B_\star^2 R_\star^2}{4 \pi
        \dot{M}_\mathrm{acc}v_\mathrm{esc}} \; , 
\end{equation}
where $B_\star$ is the stellar magnetic field strength, the polar
intensity of the dipole in our case, $\dot{M}_\mathrm{acc}$ is the
mass accretion rate, and $v_\mathrm{esc} =
(2GM_{\star}/R_{\star})^{1/2}$ is the escape speed from the stellar
gravitational potential well. Using typical stellar and accretion parameters, we obtain
\begin{equation}
\Upsilon_\mathrm{acc} = 559 \left(\frac{B_\star}{\mathrm{kG}}\right)^2
\left(\frac{M_\star}{M_\sun}\right)^{-1/2}\left(\frac{R_\star}{2R_\sun}\right)^{5/2}
\left(\frac{\dot{M}_\mathrm{acc}}{10^{-9} M_\sun\ \mathrm{yr^{-1}}}\right)^{-1}   \; .
\end{equation}
This dimensionless parameter quantifies the dynamical balance between
the magnetospheric outward push and the accretion inward pull. Both
analytical and numerical models have often expressed the position of
the truncation radius as a power law, $R_\mathrm{t}/R_\star =
K_\mathrm{t} \Upsilon_\mathrm{acc}^{m_\Upsilon}$, with $K_\mathrm{t}
\approx 0.5 - 1$. While seminal analytical models estimated an
exponent of $m_\Upsilon = 2/7$ \citep{Pringle:1972aa, Ghosh:1979aa},
assuming a perfectly dipolar magnetic field radial distribution, $B
\propto R^{-3}$, axisymmetric numerical works have reported a somewhat
larger exponent \citep[$m_\Upsilon \approx 0.35$,
][]{Pantolmos:2020aa,Ireland:2021aa}. This result is most likely due to the
compression exerted by the accretion flow on the magnetosphere,
leading to a magnetic field profile in the cavity flatter than
$R^{-3}$, as first proposed and discussed in \citet{Kulkarni:2013aa}.  

\begin{figure}
        \centering
        \includegraphics[width=\linewidth]{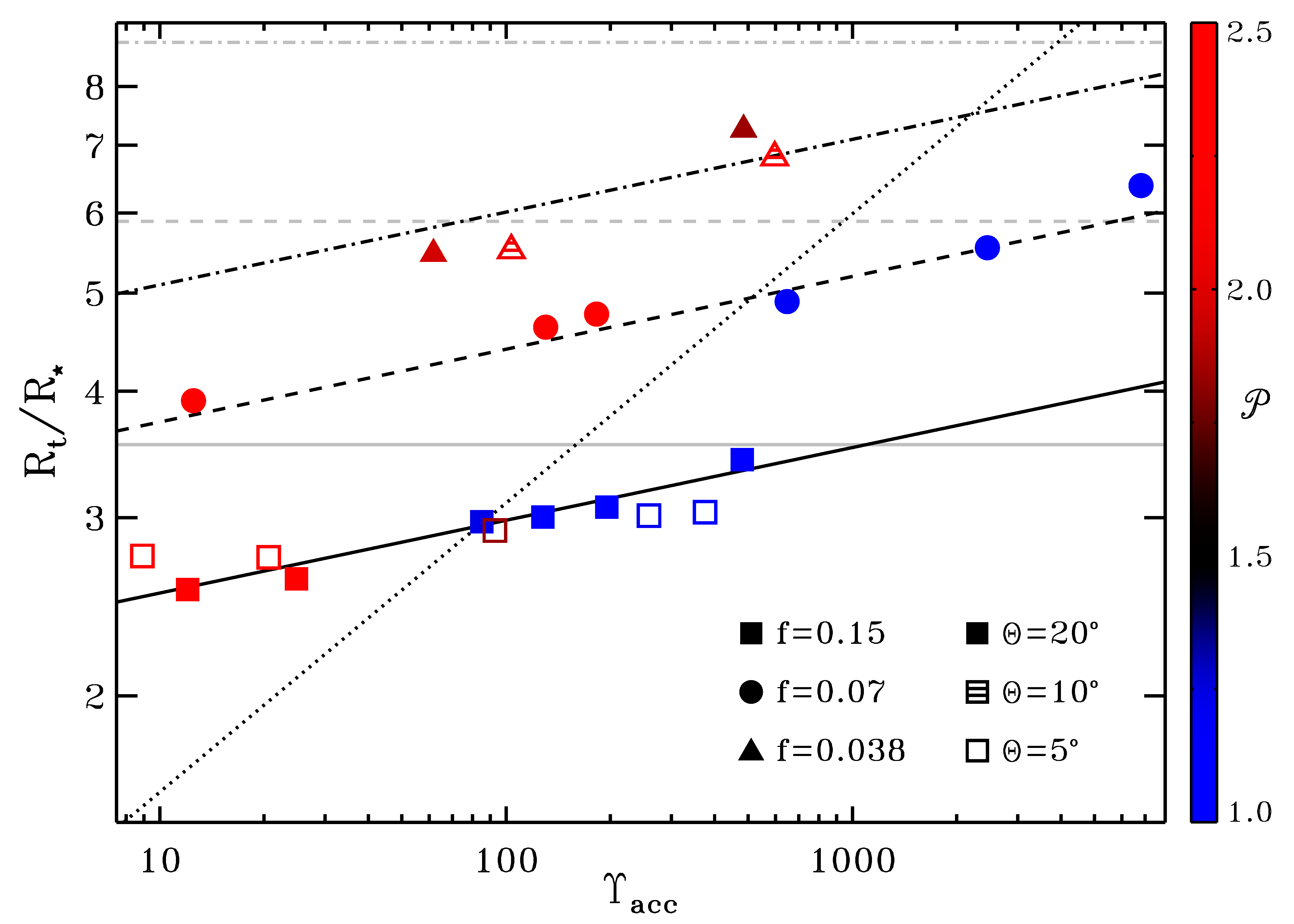}
        \caption{Average truncation radius, $R_\mathrm{t}/R_\star$,
          defined by Eq. (\ref{eq_rt_avg}) as a function of the
          accretion parameter, $\Upsilon_\mathrm{acc}$,
          Eq. (\ref{eq_yacc}). Symbol shapes and fillings have the same
          meaning as in Fig. \ref{fig_p_rtrco}. The color of the
          symbols corresponds to the value of their instability
          parameter, $\mathcal{P}$, with blue and red points indicating
          stable and unstable accretion regimes, respectively. The
          fitting function Eq. (\ref{eq_rt_yaccf}) is plotted for
          different values of the stellar rotation rate, $f=0.15$
          (solid~black lines), $f=0.07$ (dashed~black line), and
          $f=0.038$ (dot-dashed black line). The corresponding
          corotation radii, $R_\mathrm{co}/R_\star$, are plotted as gray
          horizontal lines using the same line style. The dotted black
          line corresponds to a classical \citet{Ghosh:1979aa} scaling of
          $R_\mathrm{t}/R_\star = K_\mathrm{t}
          \Upsilon_\mathrm{acc}^{2/7}$ with $K_\mathrm{t} = 0.8$.} 
        \label{fig_rt_y}
\end{figure}

In Fig. \ref{fig_rt_y}, we plot the time-averaged values obtained from
our simulations of the normalized truncation radius,
$R_\mathrm{t}/R_\star$, as a function of the accretion parameter,
$\Upsilon_\mathrm{acc}$, using different symbols to identify the
stellar rotation rate, $f$. We determined the mass accretion rate,
$\dot{M}_\mathrm{acc}$, to define the accretion parameter,
$\Upsilon_\mathrm{acc}$, by integrating the mass flux via
\begin{equation}
\dot{M} = \left| \int_{S} \rho \vec{u} \cdot d\vec{S} \, \right| \; ,
\label{eq_mdot}
\end{equation}
on the section of the spherical stellar surface that is accreting
inside the closed part of the magnetosphere. The time-averaged
$\dot{M}_\mathrm{acc}$ and $\Upsilon_\mathrm{acc}$ values for our
simulations are listed in Table \ref{tab_data_sim}. We realized that a simple expression 
\begin{equation}
\label{eq_rt_yaccf}
\frac{R_\mathrm{t}}{R_\star} = K_\mathrm{t}\ \Upsilon_\mathrm{acc}^{m_\Upsilon}\ f^{m_f} \; ,
\end{equation}
offers a satisfactory fit to the data points both for stable (blue points,
where the color corresponds to the value of the instability parameter
$\mathcal{P}$) and unstable cases (red points). As reported in Table
\ref{tab_fit}, the fitted exponents $m_{\Upsilon} = 0.072$ and $m_f =
-0.51$ show that, in both accretion regimes, our estimate for the
magnetospheric radius is linked more to the position of the corotation
radius than to the accretion parameter,
$\Upsilon_\mathrm{acc}$. Besides, the $m_\Upsilon$ exponent is much
smaller than the ``classical'' \citeauthor{Ghosh:1979aa} scaling
$m_\Upsilon = 2/7$ plotted in Fig. \ref{fig_rt_y}.
Actually, both analytical (\citealt{Wang:1987aa}; see the informative
discussion in \citealt{Bozzo:2009aa}; \citealt{Matt:2005ab}) and
numerical \citep{Ireland:2022aa} models proposed that, as the disk
truncation approaches corotation, the magnetospheric radius starts to
depend more on the stellar rotation rate, $f$, and the position of
$R_\mathrm{co}$ than on the accretion parameter,
$\Upsilon_\mathrm{acc}$. In particular, \citet{Zanni:2013aa} or
\citet{Ireland:2022aa} showed that, as the truncation radius gets
close to corotation, the system enters a weak propeller regime with a
variable accretion rate, alternating phases of stronger accretion and
weaker ejection, with the truncation radius pushed inside corotation
towards the star, and phases of weaker accretion and stronger
ejection, with the magnetospheric radius pushed by the stellar
centrifugal barrier towards and even possibly beyond corotation. The
magnetospheric radius therefore starts to oscillate around corotation,
staying roughly ``locked'' in that position. This behavior
approximately corresponds  
also to our stable simulations (blue points). On the other hand, the
same axisymmetric models recovered a more usual scaling independent of
$f$ with a higher $m_\Upsilon \approx 0.3$ value for 
stronger accretion and/or weaker magnetic fields, i.e., a smaller
$\Upsilon_\mathrm{acc}$ value
\citep{Ireland:2022aa}. Instead, our
unstable cases characterized by smaller $\Upsilon_\mathrm{acc}$ values
(red points)
appear to follow the same trend, with a strong dependence on $f$ and a weak
one on $\Upsilon_\mathrm{acc}$. Instead of axisymmetrically
compressing more and more the stellar magnetosphere,
3D accretion flows start to fragment into accretion
tongues due the interchange instability which, as shown in
Figs. \ref{fig_rho_2d} and \ref{fig_rho_2d_unst},  
can start to develop close to the corotation radius. 
Since our definition of the average truncation radius
Eq. (\ref{eq_rt_avg}) is quite sensitive to the development of the
interchange instability and therefore to the position of the
corotation radius, our $R_\mathrm{t}/R_\star$ parametrization has a
strong $f$ and a weak $\Upsilon_\mathrm{acc}$ dependence even in
unstable cases. 

\begin{table}[tbp]
        \caption{Best-fit coefficients of scaling laws.}
        \label{tab_fit}
        \centering
        \resizebox{\columnwidth}{!}{
                \begin{tabular}{c c c c}
                        \hline\hline
                        Formulation & Parameter & Value & Equation \\
                        \hline
                        \multirow{3}*{$R_\mathrm{t}/R_\star$} & $K_\mathrm{t}$ & $0.81 \pm  0.03$ & \multirow{3}*{(\ref{eq_rt_yaccf})}\\ 
                        & $m_\Upsilon$ & $0.072 \pm 0.007$ & \\
                        & $m_f$ & $-0.51 \pm 0.02$ \\
                        \hline
                        \multirow{2}*{$\dot{J}_\mathrm{SDI}$} & $K_\mathrm{acc}$ & $0.71 \pm 0.03$ & \multirow{2}*{(\ref{eq_tsdi})}  \\
                        & $K_\mathrm{MEs}$ & $-0.036 \pm 0.002$ & \\ 
                        \hline
                        \multirow{2}*{$\Phi_\mathrm{SW}/\Phi_\star$} & $K_\Phi$ & $0.43 \pm 0.01$ & \multirow{2}*{(\ref{eq_swflux})}  \\
                        & $m_\Phi$ & $-0.69 \pm 0.05$ & \\ 
                        \hline
                        \multirow{2}*{$\langle r_\mathrm{A} \rangle/R_\star$} & $K_\mathrm{A}$ & $0.82 \pm 0.18$ & \multirow{2}*{(\ref{eq_ra})}  \\
                        & $m_\mathrm{A}$ & $0.43 \pm 0.03$ & \\ 
                        \hline
                \end{tabular}
        }
\end{table}

\begin{figure}[!t]
        \centering
    \vspace{\fill}
        \includegraphics[width=0.813\linewidth]{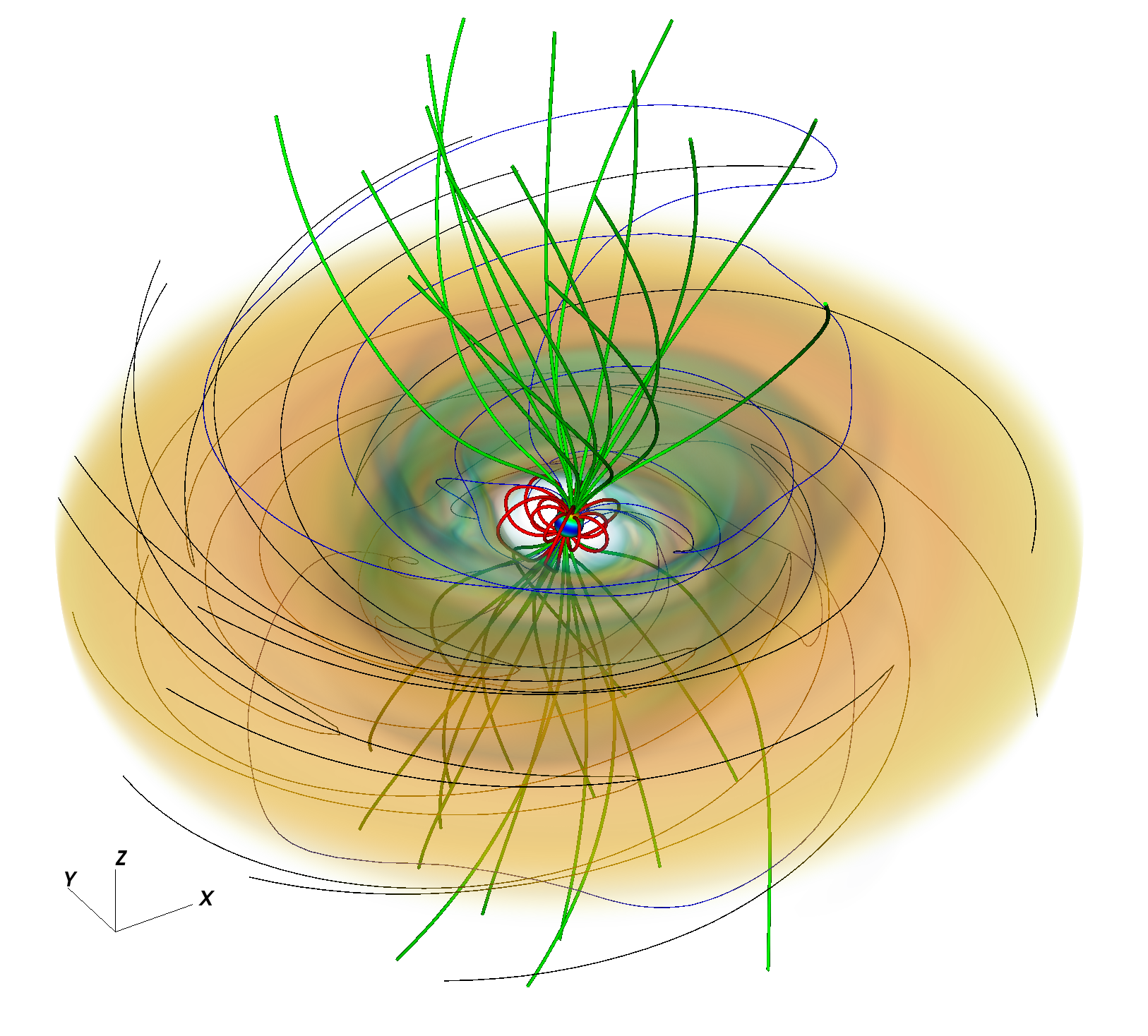}
        \includegraphics[width=0.813\linewidth]{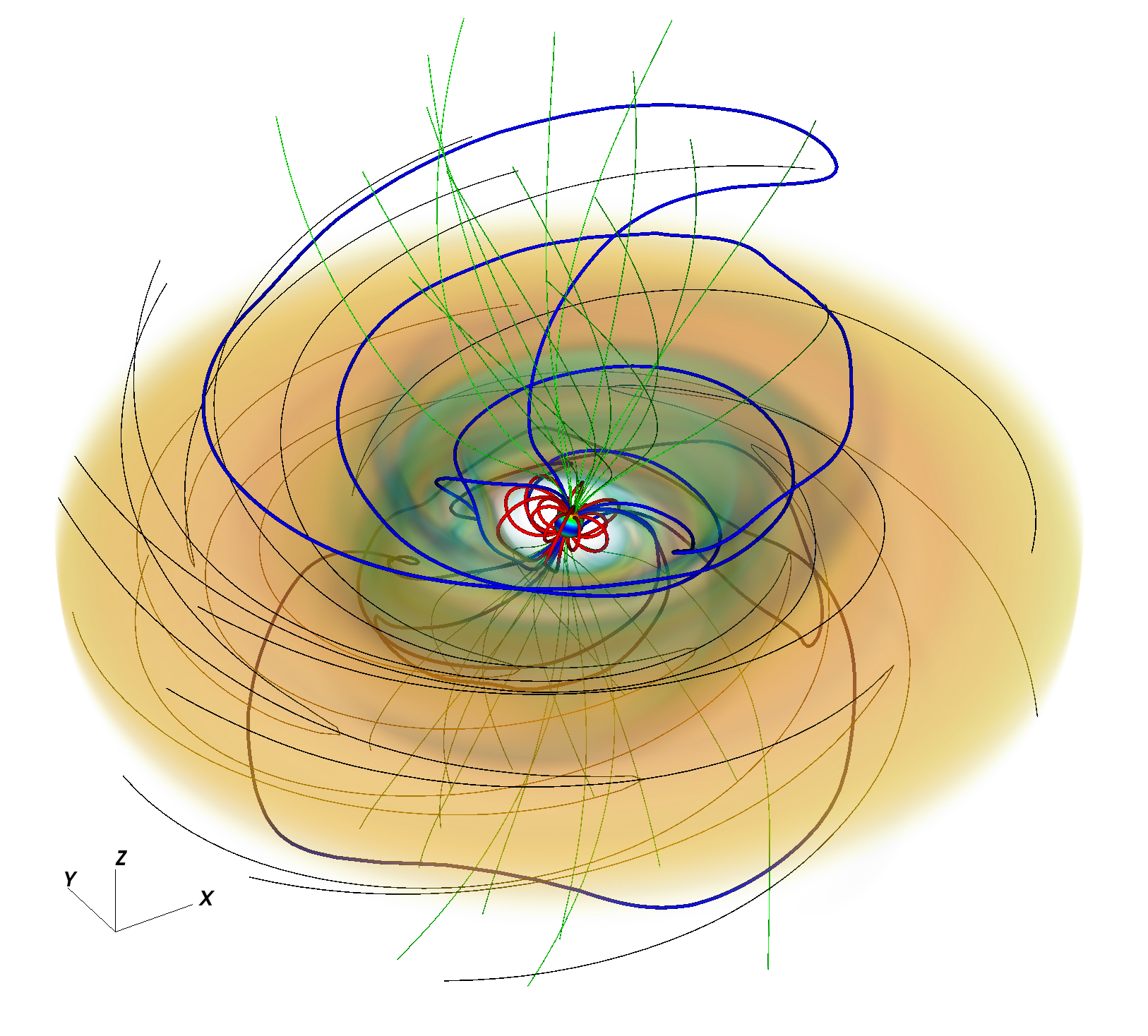}
        \includegraphics[width=0.813\linewidth]{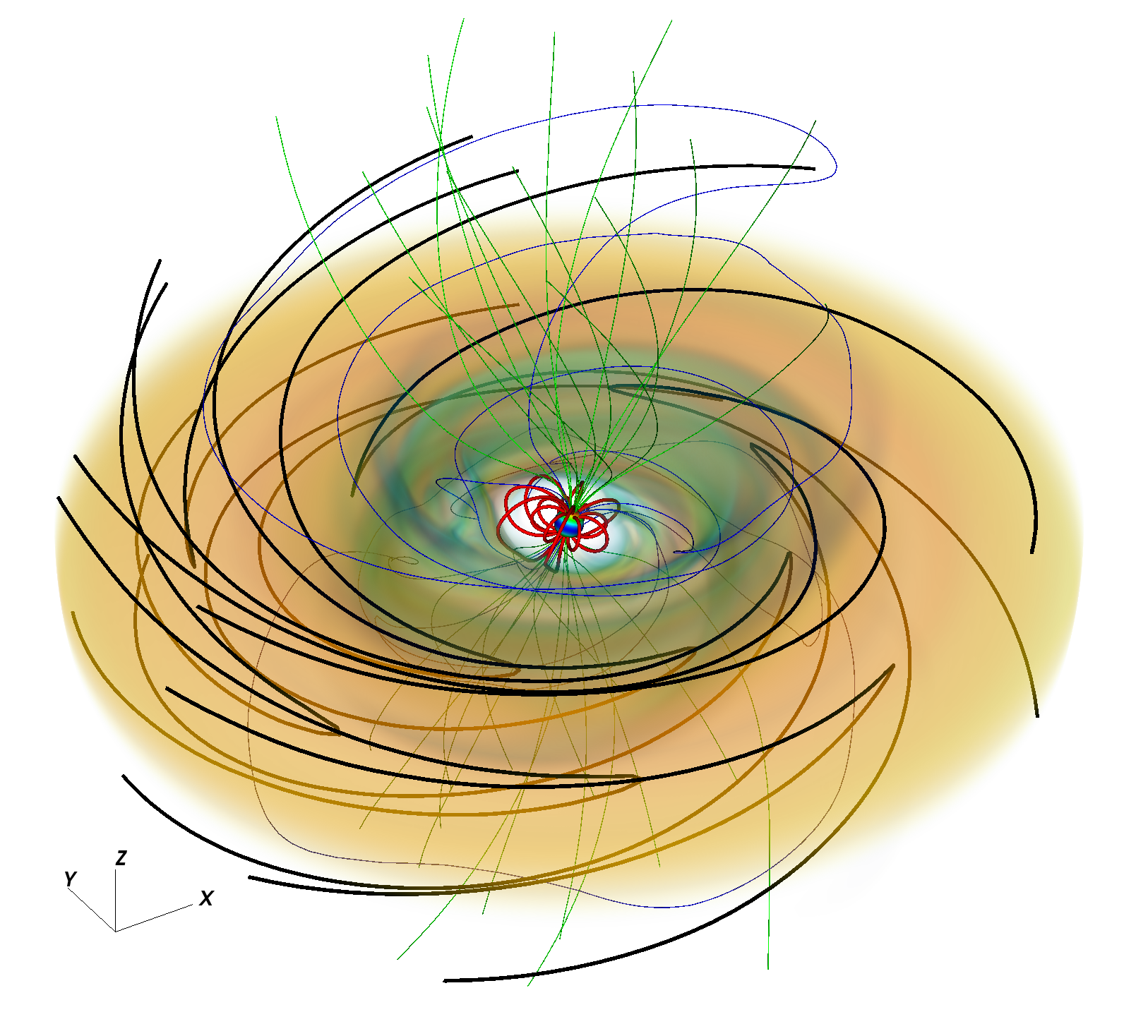}
        \caption{Volume rendering of the normalized logarithmic
                density in the full computational domain of the stable case
                14 after 21.75 stellar periods, corresponding to the
                zoomed-in picture in the left panel of
                Fig. \ref{fig_rho_3d}. Magnetic field lines corresponding to
                different flow components are plotted using different
                colors. Red lines indicate the inner closed magnetic field
                lines that are steadily accreting from the disk. The stellar
                wind is highlighted in green in the upper panel. The
                inflated magnetic field lines of magnetospheric ejections
                are plotted in blue in the central panel. The black lines in
                the lower panel correspond to an outer disk wind.} 
        \label{fig_flow_decomp}
\end{figure}

This behavior is also confirmed by Fig. \ref{fig_rho_2d} and \ref{fig_rho_2d_unst}. 
The simulations in Fig. \ref{fig_rho_2d} and the left panel in Fig. \ref{fig_rho_2d_unst}
have the same stellar rotation period but an accretion parameter 5 times (the case in the 
second line of Fig. \ref{fig_rho_2d}) and 52 times (the left panel in Fig. \ref{fig_rho_2d_unst})
smaller than the stable simulation in the first line of Fig. \ref{fig_rho_2d}; see Table \ref{tab_data_sim}. 
Clearly, both our definition of $R_\mathrm{t}$ and the maximum radial size of the 
magnetospheric cavity do not change significantly despite the
fact that $\Upsilon_\mathrm{acc}$ has varied by a large factor\footnote{Actually, 
the maximum radial size of the magnetospheric cavity appears to be larger in the
unstable cases than in the stable simulation with the same stellar spin. 
If the truncation radius in the first line of Fig. \ref{fig_rho_2d} approximately
marks a stable configuration, this could be due to rising plumes of lighter 
material that typically characterize a Rayleigh-Taylor instability.}. 
If $R_\mathrm{t}$ had followed a \citeauthor{Ghosh:1979aa} scaling, the truncation radius in the 
left panel of Fig. \ref{fig_rho_2d_unst} should have been around three times smaller than the stable case in Fig. \ref{fig_rho_2d}. This result clearly suggests that in our unstable simulations the 
accretion disk starts to fragment close to the corotation radius due to the interchange 
instability before reaching the ``classical'' \citeauthor{Ghosh:1979aa} truncation radius.  
Coherently, Fig. \ref{fig_rt_y} shows that our estimate of $R_\mathrm{t}$ for unstable cases 
(red points) is larger than the \citeauthor{Ghosh:1979aa} radius (dotted line). 
Furthermore, we show in Fig. \ref{fig_rho_2d_unst} two simulations that have approximately the same
accretion parameter while the right case has a smaller corotation radius than the left one; see Table \ref{tab_data_sim}. Both our definition of $R_\mathrm{t}$ and the maximum radial 
size of the magnetospheric cavity move inward following the corotation radius, 
despite the fact that $\Upsilon_\mathrm{acc}$ has hardly changed.

It is important to notice, however, that our model-fitting procedure is limited
to the parameter space that we explored, i.e., $10 \lesssim \Upsilon_\mathrm{acc} \lesssim 7000$
and $0.038\leq f \leq 0.15$, corresponding to $0.6 \lesssim R_\mathrm{t}/R_\mathrm{co} \lesssim 1.1$,
a range which nevertheless covers the majority of the sample of CTTs observed in spectropolarimetry (see Sect. \ref{sec_spec} and Appendix \ref{sec_spol}). For the set of parameters that we
have investigated, the magnetosphere connecting the star to the disk tends to extend and
trigger the interchange instability close to the corotation radius. 
For an $R_\mathrm{t}/R_\mathrm{co}$ ratio smaller than $\approx 0.6$, it becomes more and more
difficult to maintain this connection, so that the instability and the magnetic cavity can start 
to develop increasingly within the corotation radius (see the extreme case in \cite{Zhu:2024aa}, 
with $f=0$ or $R_\mathrm{co} \rightarrow \infty$, which, however, does not correspond
to any observed CTTs), weakening the $f$ and strengthening the $\Upsilon_\mathrm{acc}$ 
dependence of the $R_\mathrm{t}/R_\star$ scaling.
As a consequence, our best fit Eq. (\ref{eq_rt_yaccf}) likely provides only an upper 
limit for small $R_\mathrm{t}/R_\mathrm{co} \lesssim 0.6$ values, particularly 
in the ordered unstable regime.
We also note that the magnetic resistivity, parametrized in this work using a \citet{Shakura:1973aa} prescription with a fixed $\alpha_\mathrm{m}$ value         
(see Sects. \ref{sec_params} and \ref{sec_ic_bc}), can affect both the radial extent of the
magnetospheric star-disk interaction and the disk rotational profile 
\citep{Zanni:2009aa,Zanni:2013aa}, two quantities that have an important impact on the
onset of the interchange instability (see Eq. (\ref{eq_spruit}) and the subsequent discussion). 
As a consequence, a different choice for the resistivity parametrization and the 
$\alpha_\mathrm{m}$ value could modify the scaling of the truncation radius, but a systematic 
investigation of the impact of the magnetic diffusivity is beyond the scope of this work.

Using the fitting constants of Table \ref{tab_fit}, the truncation
radius Eq. (\ref{eq_rt_yaccf}) can be expressed as function of the
stellar parameters in terms of the fractional rotation rate $f$ 
\begin{equation}
  \label{eq_rt_scalingf}
\begin{split}
\frac{R_\mathrm{t}}{R_\star} = 4.15 \, & \left(\frac{B_\star}{\mathrm{kG}}\right)^{0.14}
\left(\frac{M_\star}{M_\sun}\right)^{0.25}
\left(\frac{R_\star}{2R_\sun}\right)^{0.18} \times \\
& \times \left(\frac{\dot{M}_\mathrm{acc}}{10^{-9} M_\sun\ \mathrm{yr^{-1}}}\right)^{-0.072}
\left(\frac{f}{0.1}\right)^{-0.51} \; ,
\end{split}
\end{equation}
or the stellar rotation period $P_\star$
\begin{equation}
  \label{eq_rt_scalingp}
\begin{split}
\frac{R_\mathrm{t}}{R_\star} = 5.15 \, & \left(\frac{B_\star}{\mathrm{kG}}\right)^{0.14}
\left(\frac{M_\star}{M_\sun}\right)^{0.22}
\left(\frac{R_\star}{2R_\sun}\right)^{-0.59} \times \\
& \times \left(\frac{\dot{M}_\mathrm{acc}}{10^{-9} M_\sun\ \mathrm{yr^{-1}}}\right)^{-0.072}
\left(\frac{P_\star}{5 \, \mathrm{days}}\right)^{0.51} \; .
\end{split}
\end{equation}

\section{Stellar torques}
\label{sec_torques}

In this section we examine the torques exerted on the star by the
magnetic interaction with the disk and by stellar winds. As in
\citet{Pantolmos:2020aa} and \citet{Ireland:2021aa, Ireland:2022aa} we
separate the contribution of the different flow components to the mass
flux Eq.(\ref{eq_mdot}) and the angular momentum flux (i.e., the torque)  
\begin{equation}
\dot{J} = -\int_S r \left( \rho v_{\phi}
\vec{u} - \frac{B_{\phi} \vec{B}}{4 \pi}+P_\mathrm{t}\hat{\phi}\right) \cdot d\vec{S} \; ,
\label{eq_jdot}
\end{equation}
where $\vec{u} = \vec{v}-r\Omega_\star \hat{\phi}$ is the speed in the
rotating frame of reference and $P_\mathrm{t} = P +\vec{B}\cdot
\vec{B}/{8\pi}$ is the total pressure. Since these integrals are
performed on spherical surfaces, only the radial components of the
mass and angular momentum fluxes contribute to the result.  

We compute the integrals Eqs. (\ref{eq_mdot}) and (\ref{eq_jdot}) on
the sections of the stellar surface threaded by: (1) open magnetic
flux, to evaluate the contribution of stellar winds (green magnetic
field lines in Fig. \ref{fig_flow_decomp}); (2) steadily closed
accreting field lines for accretion (red lines in
Fig. \ref{fig_flow_decomp}); (3) magnetic field lines connecting the
star to the disk that undergo cycles of inflation and reconnection for
magnetospheric ejections (blue lines in
Fig. \ref{fig_flow_decomp}). The contribution coming from the magnetic
field lines anchored inside the magnetic cavity is included in the
accretion integral, but this term should be equal to zero, at least in
a time-averaged sense. In this way, the sum of the three contributions
covers the entire stellar surface and they can be identified by the
subscripts ``SW,'' ``acc,'' and ``MEs,'' respectively, while we
employ ``SDI'' for the sum of the accretion and MEs
torques\footnote{Since we compute the integral Eq. (\ref{eq_jdot}) by
  separating the closed stellar surface into open areas corresponding
  to different flow components, some caution must be employed to avoid
  the stellar potential field contribution to this integral when
  evaluated on open spherical sectors. See Appendix
  \ref{sec_potential} for a torque definition that takes into account
  this problem.}. 
We adopt the convention that a positive value of the
integral Eq. (\ref{eq_jdot}) corresponds to a spin-up torque, while a
negative one to stellar spin-down. Since we compute the absolute value
of the mass flux integral Eq. (\ref{eq_mdot}), both mass-loss rates
(e.g., the stellar wind) and mass accretion rates will have a positive
value.  

For completeness, to show all the flow components present in our
numerical experiments, we plotted in black in the lower panel of
Fig. \ref{fig_flow_decomp} the magnetic field lines of an outer
disk wind. Clearly this outflow does not directly contribute to the
stellar angular momentum evolution, but it provides an important
torque to accrete the disk material towards the magnetospheric
region. Besides, it is interesting to notice that the magnetic
configuration of our $\alpha$ disk wind solution is very different
from the turbulent disk-wind structure identified in, for example,
\citet{Jacq:2021aa} or \citet[][where it was recognized as a
``failed'' disk wind]{Takasao:2022aa} that show a layered disk
structure with a strong vertical support due to the turbulent magnetic
pressure and an important accretion at the disk surface distorting the
magnetic surfaces and contributing to the magnetic flux
transport. Besides, the MRI turbulent pressure that characterizes
these solutions tends to fill the star-disk interaction region with
high $\beta$ plasma at high latitudes
\citep[see][]{Takasao:2022aa,Zhu:2024aa}, while in our models a low
$\beta$ condition favors the development of coherent large-scale
magnetic structures as those exploited by magnetospheric ejections. 

\subsection{Star-disk interaction torques}
\label{sec_sditorques}

We start by examining the torques directly associated with the
star-disk interaction (SDI), the accretion (Sect.
\ref{sec_accretion}) and the MEs (Sect. \ref{sec_mes}) torques. In
Sect. \ref{sec_sdi} we show how the total SDI torque depends on
the accretion regime, and in Sect. \ref{sec_sw} we analyze the
contribution of the stellar wind torque. 

\begin{figure}[!t]
        \centering
         \includegraphics[width=\linewidth]{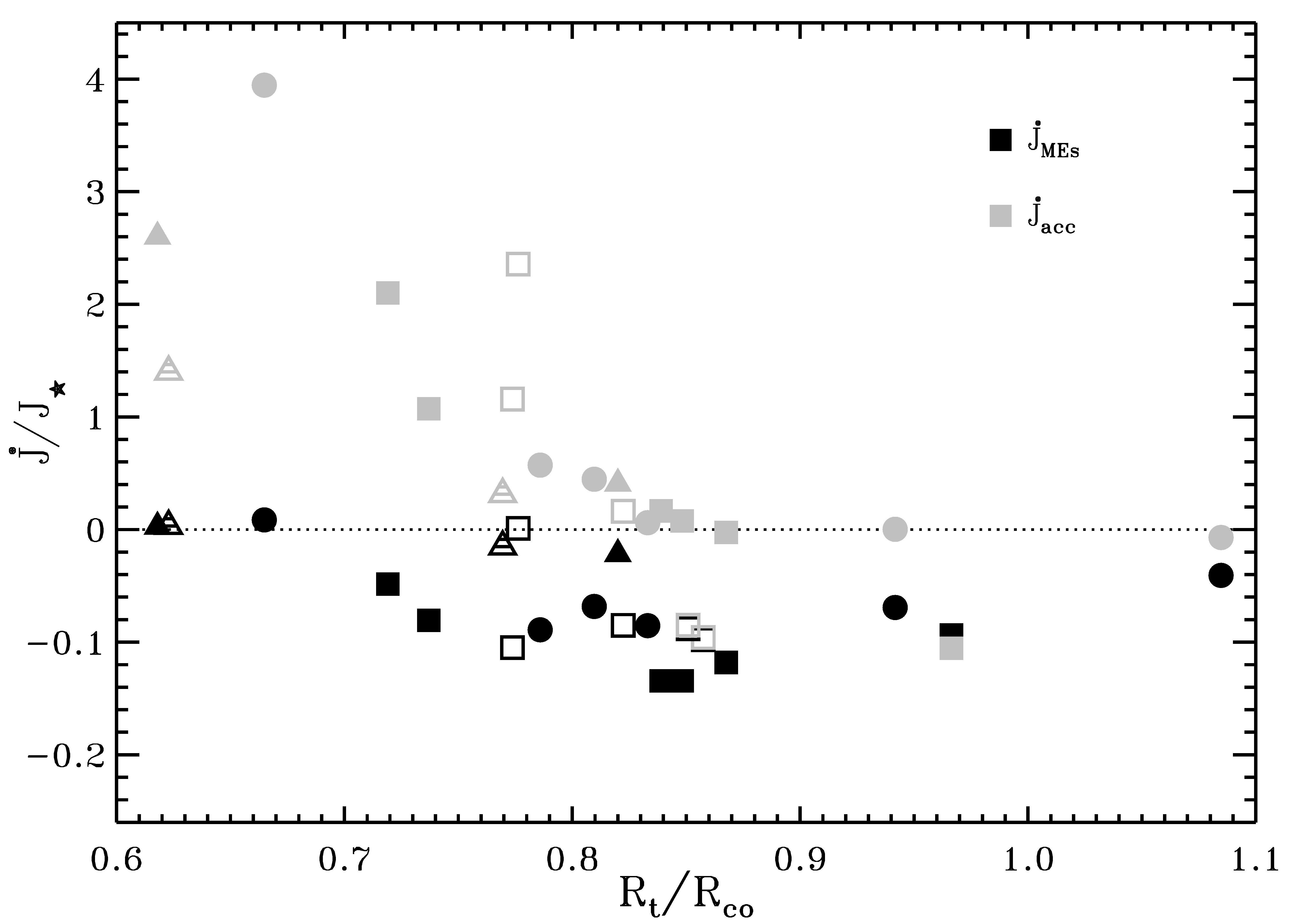}
        \caption{Star-disk interaction torques, normalized on the
          stellar angular momentum, as a function of the
          $R_\mathrm{t}/R_\mathrm{co}$ ratio. The normalized torque
          values are given in units of Eq. (\ref{eq_jnorm}). The
          torque exerted onto the star by accretion funnels is plotted
          with gray points while the magnetospheric ejections stellar
          torque corresponds to black points. We note the different
          linear scale used for spin-up (positive values) and
          spin-down (negative values) torques. Symbol shape and
          filling have the same meaning as in Fig. \ref{fig_p_rtrco}.} 
        \label{fig_torque_me_acc}
\end{figure}

\begin{figure*}[!t]
        \centering
        \includegraphics[width=\linewidth]{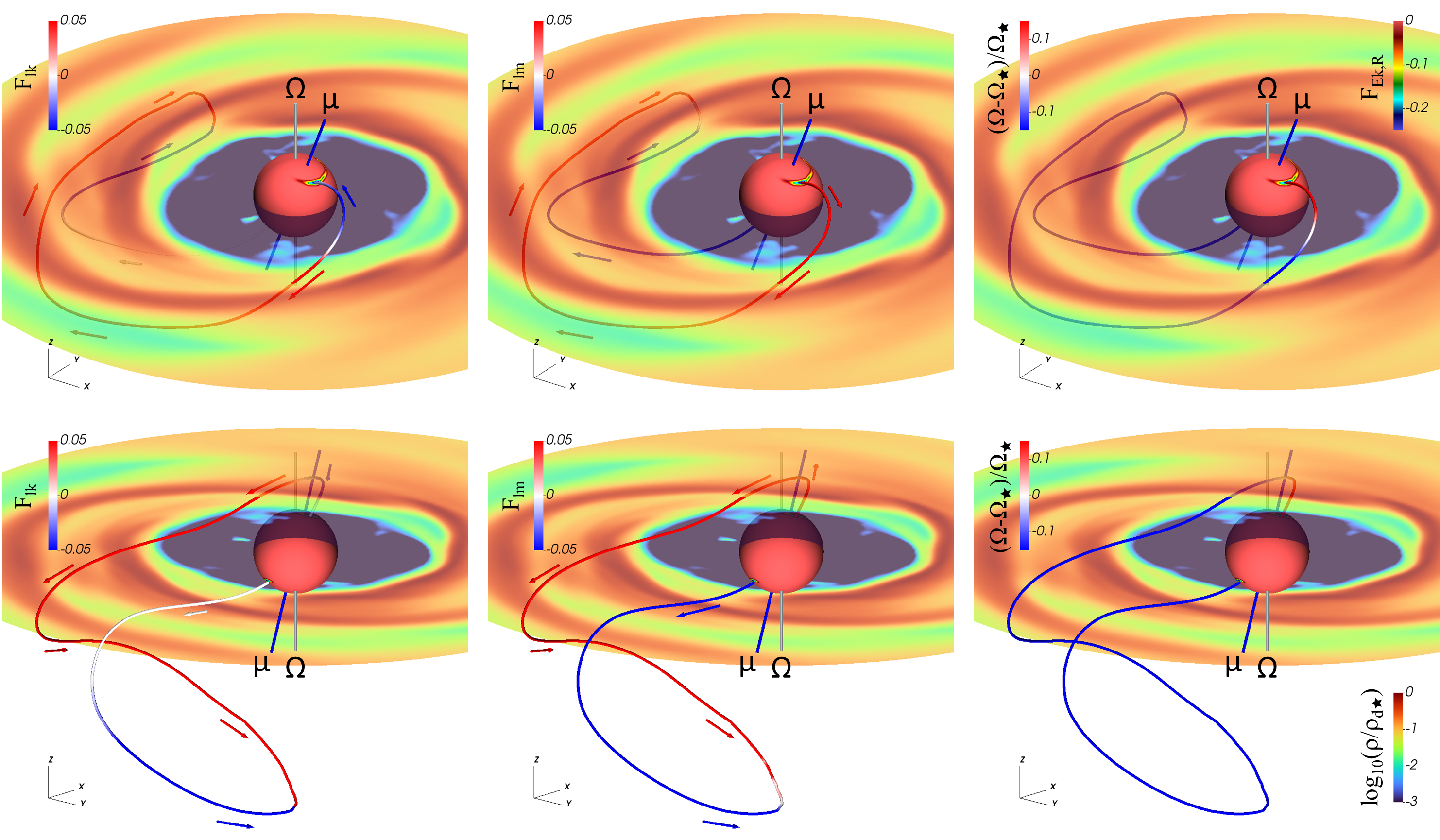}
        \caption{Projection of the kinetic (\textit{left};
                Eq. \ref{eq_flk}) and magnetic (\textit{center};
                Eq. \ref{eq_flm}) angular momentum fluxes along the magnetic
                field line anchored in the accretion spot in the northern
                hemisphere of the star. The arrows indicate the direction of
                the fluxes. Value of
                the stellar differential rotation
                $(\Omega-\Omega_\star)/\Omega_\star$ shown on the same field
                line (\textit{right}). The top (bottom) panels correspond to a view from above
                (below) the equatorial plane. An equatorial slice of the
                logarithmic density is shown in every panel. The accretion
                spot shape and position is highlighted on the stellar
                surface using the value of the radial flux of the specific
                kinetic energy: $F_\mathrm{Ek,R} = -0.5\rho v^2 v_R$. The
                direction of the stellar rotation and the dipolar magnetic
                moment is shown in each panel. The snapshots are taken from
                the stable case 6 after 20 stellar periods.} 
        \label{fig_amflux}
\end{figure*}

\subsubsection{Accretion torque}
\label{sec_accretion}
In Fig. \ref{fig_torque_me_acc}, we plot with gray symbols the
accretion torque divided by the stellar angular momentum as a
function of the $R_\mathrm{t}/R_\mathrm{co}$ ratio, to detect any
dependence of the torque on the accretion regime. Customarily, the
spin-up accretion torque has often been parametrized as 
\begin{equation}
\dot{J}_\mathrm{acc} = K_\mathrm{acc} \dot{M}_\mathrm{acc} \sqrt{G M_\star R_\mathrm{t}} \; ,
\label{eq_jacc}
\end{equation}
where the adimensional factor $K_\mathrm{acc}$ mainly takes into
account the deviation of the disk specific angular momentum from a
Keplerian profile. Different numerical works have estimated a value in
the range $0.5 < K_\mathrm{acc} < 1$
\citep[e.g.,][]{Pantolmos:2020aa,Ireland:2021aa,Ireland:2022aa}
indicating that the disk rotation in the truncation region is
sub-Keplerian but still faster 
than the stellar rotation. We ascribe this behavior to the action of
the MEs or conical winds, efficiently extracting disk angular momentum
before it is accreted, and to the rotating magnetosphere trying to
force stellar corotation. This qualitative picture could adequately
describe our strong accretion cases, corresponding to an unstable
regime with $R_\mathrm{t}/R_\mathrm{co} < 0.8$, but we noticed that in
the stable regime with $R_\mathrm{t}/R_\mathrm{co} > 0.85$ some cases
are characterized by a null or even negative spin-down accretion
torque corresponding to a $K_\mathrm{acc} \leq 0$ value. 

This apparently counter-intuitive result (accretion should transfer to
the star mass and therefore angular momentum at the same time)
requires a more in-depth discussion. Neglecting the flux along the
azimuthal direction, and assuming that the accretion funnels flow
parallel to the magnetic field lines, which is an appropriate
approximation for accretion in a stable regime, we can rewrite the
accretion torque from Eq. (\ref{eq_jdot}) as  
\begin{equation}
\dot{J}_\mathrm{acc} = \dot{M}_\mathrm{acc}\Lambda \qquad \mathrm{with} \qquad 
\Lambda = r\left(v_\phi-\frac{B_\phi B}{4\pi \rho u}\right) \; ,
\label{eq_Lambda}
\end{equation}
where the specific angular momentum $\Lambda$ is assumed to be
approximately constant for the whole accretion flow. We note that
$\Lambda$ is an invariant along magnetic field lines in a stationary
and axisymmetric solution, but this is not the case for a
3D model; see the discussion in Appendix
\ref{sec_alfven}. Even if $\Lambda$ is not strictly invariant along
the accretion flow, we estimate its value at the disk surface. Using
the continuity of the radial electric field, the toroidal magnetic
field at the disk surface inside a star-disk magnetospheric interaction
region has often been parametrized as a function of the star-disk
differential rotation \citep[see, e.g.,][]{Matt:2005ab} via 
\begin{equation}
B_\phi = r \frac{\Omega_\star-\Omega_\mathrm{d}}{\nu_\mathrm{m}}B H \; ,
\label{eq_bphi}
\end{equation}
where $\Omega_\mathrm{d}$ is the (possibly non-Keplerian) disk angular
velocity, $\eta_\mathrm{m}$ is the disk resistivity, controlling the
magnetic coupling between the stellar magnetosphere and the disk, $B$
the local vertical poloidal field in the disk and $H$ is the vertical
height scale of the disk. Combining Eqs. (\ref{eq_Lambda}) and
(\ref{eq_bphi}) we obtain the expression 
\begin{equation}
\dot{J}_\mathrm{acc} = \dot{M}_\mathrm{acc} \sqrt{G M_\star R_\mathrm{t}} \left. \left(\frac{\Omega_\mathrm{d}}{\Omega_\mathrm{K}} + C \frac{\Omega_\mathrm{d}-\Omega_\star}{\Omega_\mathrm{K}}\right) \right|_{R_\mathrm{t}} \; ,
\label{eq_jacc_exp}
\end{equation}
where $\Omega_\mathrm{K} = \sqrt{GM_\star/r^3}$ is the Keplerian
angular velocity. The first term in round parenthesis corresponds to
the kinetic angular momentum flux, while the term proportional to $C$
corresponds to the magnetic angular momentum flux. The positive
quantity $C$ can be approximated, for example, using our
parametrization for the magnetic resistivity, $\eta_\mathrm{m} =
\alpha_\mathrm{m} c_\mathrm{sd} H$, providing the expression $C \approx
1/\alpha_\mathrm{m} \beta M_\mathrm{A}$, where $\beta = 8\pi P/B^2$ is
the plasma beta and $M_\mathrm{A} = u/\sqrt{B/4\pi\rho}$ is the
injection Alfv\'{e}nic Mach number at the base of the funnel
flow. Since $\alpha_\mathrm{m} < 1$, the injection speed is
sub-Alfv\'{e}nic $M_\mathrm{A} < 1$ and in the truncation region
$\beta < 1$, the quantity $C$ is likely greater than one. For example,
from Figs. 5 and 6 in \citet{Zanni:2009aa} it is possible to estimate
a $C \approx 3$ value, but this is obviously just a single case and
$C$ is likely not a constant. The same Fig. 5 shows that the ratio
between the kinetic and the magnetic angular momentum flux can change
as the matter falls towards the star, with the magnetic flux typically
becoming dominant.  

The whole expression given in rounded parentheses in Eq. (\ref{eq_jacc_exp})
should correspond to the parameter $K_\mathrm{acc}$ in
Eq. (\ref{eq_jacc}). Even if (as expected) this quantity strongly
depends on the Keplerianity of the disk, it is far from being a
constant, particularly when the disk rotation period in the truncation
region becomes comparable to the stellar one. We note that while the kinetic angular momentum flux
always provides a spin-up torque, the magnetic part can become
negative (i.e., spin-down) when the disk rotation becomes slower than
the stellar one. In other words, since the magnetosphere tries to
force stellar corotation, the star magnetically transfers and loses
angular momentum to the infalling disk material when it rotates more slowly
than the star. Our cases that exhibit a spin-down accretion torque
suggest that the spin-down magnetic torque can even become more
important than the spin-up kinetic part. We tried to illustrate this
situation in Fig. \ref{fig_amflux} where in the left and center panels
we plot, for one of the cases that displays a spin-down accretion
torque, the projection of the angular momentum flux along the magnetic
field line anchored in the center of the main accretion spot in the
northern hemisphere of the star, as seen from above (upper panels) and
below (lower panels) the equatorial plane. The projection of the
kinetic angular momentum flux, defined as 
\begin{equation}
F_\mathrm{lk} = r\left(\rho v_\phi\vec{u} + P\hat{\phi}\right)\cdot\frac{\vec{B}}{B^2} \; ,
\label{eq_flk}
\end{equation}
is shown in the left panels while the projection of the magnetic angular momentum flux,
\begin{equation}
F_\mathrm{lm} = r\left(-\frac{B_\phi\vec{B}}{4\pi}+\frac{B^2}{8\pi}\hat{\phi}\right)\cdot\frac{\vec{B}}{B^2} \; ,
\label{eq_flm}
\end{equation}
is shown in the central panels\footnote{Analogously to the torque
  integral Eq. (\ref{eq_jdot}), in Eq. (\ref{eq_flm}) we subtracted
  the contribution of the stellar potential field; see the discussion
  in Appendix \ref{sec_potential}}. Positive (in red) and negative (in
blue) values correspond respectively to an angular momentum flux
parallel or anti-parallel to the magnetic field. We also plotted some
arrows to make the direction of the flux clearer. In the right column
we plot along the same field line the stellar differential rotation
$(\Omega-\Omega_\star)/\Omega_\star$ as seen from above (upper panels)
and below (lower panels) the midplane. In the upper-left panel we see
that the kinetic flux is actually accreting angular momentum towards
the accretion spot of star, while in the upper-central panel the
magnetic flux is extracting angular momentum from the star. Since this
case displays a spin-down accretion torque, the magnetic extraction
must be larger than the angular momentum kinetic accretion. In
agreement with our previous discussion, the upper-right panel shows
that the disk rotates more slowly than the star, thus triggering a magnetic
extraction of stellar angular momentum. We notice that since this
configuration has the infalling material flowing parallel along a trailing trajectory, 
it must rotate faster than the star as it approaches its surface.  

However, we must consider where the stellar angular momentum magnetically extracted along
the accretion funnel ends up. The lower panels in
Fig. \ref{fig_amflux} show that the extra angular momentum acquired by
the disk from the star is actually ejected along the same magnetic
field line on the other side of the disk. On the lower side of the
equator, the field line that we are considering is highly inflated and
twisted but still connected to the southern side of the star. This
configuration is analogous to the sketch in Fig. 7 in
\citet{Takasao:2022aa}. Angular momentum is extracted both from the
disk, as in a (conical) disk-wind configuration, and from the star
converging towards the tip of the closed field line and ejected as in
a magnetic slingshot. We notice that the whole magnetic field line in the
lower side is rotating more slowly than the star (lower-right panel),
dragging and slowing down the stellar rotation. This is the phenomenon
that was identified by \citet{Zanni:2013aa} as magnetospheric
ejections, whose contribution to the stellar torque is discussed
in greater detail in the next section.  

We also note that a similar solution, whereby a funnel flow accretes mass and
angular momentum but magnetically extracts it from the star, was also
found in \citet{Das:2022aa}. Coherently with our discussion, this
behavior was observed in cases in which the disk rotates more slowly than
the star at the base of the accretion funnel. Contrary to our
findings, the angular momentum transferred from the star to the disk
is not ejected but it is radially extracted along the disk by a
vigorous (albeit 2D) turbulence. 

\subsubsection{Magnetospheric ejections torque}
\label{sec_mes}

As already discussed earlier in this paper, we use the abbreviation MEs to refer to the
magnetospheric outflows that exploit magnetic surfaces that are still connecting the star with the disk, which
are twisted and inflated by the star-disk differential rotation. As
the magnetic twist becomes too strong and the expansion too large, these magnetic
field lines can undergo a reconnection event, launching magnetized
plasmoids and deflating the magnetic structure. Therefore, this process 
can be quite episodic and repeat quasi-periodically, possibly explaining
variable blue-shifted absorption components that are not rotationally 
modulated, observed in the atomic line profiles of stars such as GM Aur
\citep{Bouvier:2023ab}. From the point of view of the
disk, these ejections behave essentially as a disk wind and, as a
matter of fact, they have been associated with conical disk winds
\citep{Romanova:2009ab,Takasao:2022aa}, extracting an important
fraction of the disk angular momentum. Indeed, the sub-stellar disk
rotation shown in the example towards the end of Sect.
\ref{sec_accretion}, is likely due to the angular momentum extracted
from the disk by the MEs. Since the matter launched from the disk
is also connected to the star, MEs can also exchange angular momentum
with it. \citet{Zanni:2013aa} ascribed this angular momentum exchange
to the differential rotation between the MEs launched from the disk
and the star: an ME rotating slower (faster) than the star can provide
a spin-down (spin-up) torque. A typical MEs spin-down configuration
was shown in Sect. \ref{sec_accretion}. The rotation of the MEs
depends essentially on the position of the launching point in the
disk: if the disk rotation around the launching point is faster than
stellar, typically inside the corotation radius, the MEs will provide
a spin-up torque and vice versa. A qualitative parametrization of the
stellar torque exerted by the MEs  \citep[see, e.g.,][]{Gallet:2019aa} can be expressed as
\begin{equation}
\dot{J}_\mathrm{MEs} \propto \frac{B_\star^2 R_\star^6}{R_\mathrm{MEs}^3}\left(\frac{\Omega_\mathrm{MEs}-\Omega_\star}{\Omega_\mathrm{MEs}} \right) \; ,
\label{eq_jmes}
\end{equation} 
where $R_\mathrm{MEs}$ is the radial position of the MEs, $B_\star
R_\star^3/R_\mathrm{MEs}^3$ is the local dipolar field and
$\Omega_\mathrm{MEs}$ is the MEs angular speed.  

In Fig. \ref{fig_torque_me_acc} we plot using black symbols the
time-averaged MEs stellar torque divided by the stellar angular
momentum as a function of the $R_\mathrm{t}/R_\mathrm{co}$ ratio for
all the simulated cases. Clearly in most of the simulations the MEs
exert a stellar spin-down torque while this is negligible in
some cases. This is consistent with the fact that in the parameter
range that we consider here, the magnetic cavity can expand up to the
corotation radius, even in unstable cases. 
As a result, the MEs are likely launched from a region close to or beyond corotation and so, 
they rotate more slowly than the star, thereby providing a spin-down torque. 
In most cases, MEs are likely rotating more slowly than the star; 
furthermore, it is difficult to estimate 
the differential rotation factor in Eq. (\ref{eq_jmes}). 
Thus, we can simplify this expression to get 
\begin{equation}
\dot{J}_\mathrm{MEs} = K_\mathrm{MEs} \frac{B_\star^2
  R_\star^6}{R_\mathrm{MEs}^3} \qquad \mathrm{whith} \qquad
R_\mathrm{MEs} = \mathrm{max}\left[R_\mathrm{t},R_\mathrm{co}\right]
\; , 
\label{eq_jmes_simple}
\end{equation} 
analogously to the MEs torque parametrization in
\citet{Ireland:2022aa}, where the factor $K_\mathrm{MEs}$ should take
into account the uncertainties about the differential rotation and the
magnetic coupling between the star and the MEs. 

\begin{figure}[!t]
        \centering
         \includegraphics[width=\linewidth]{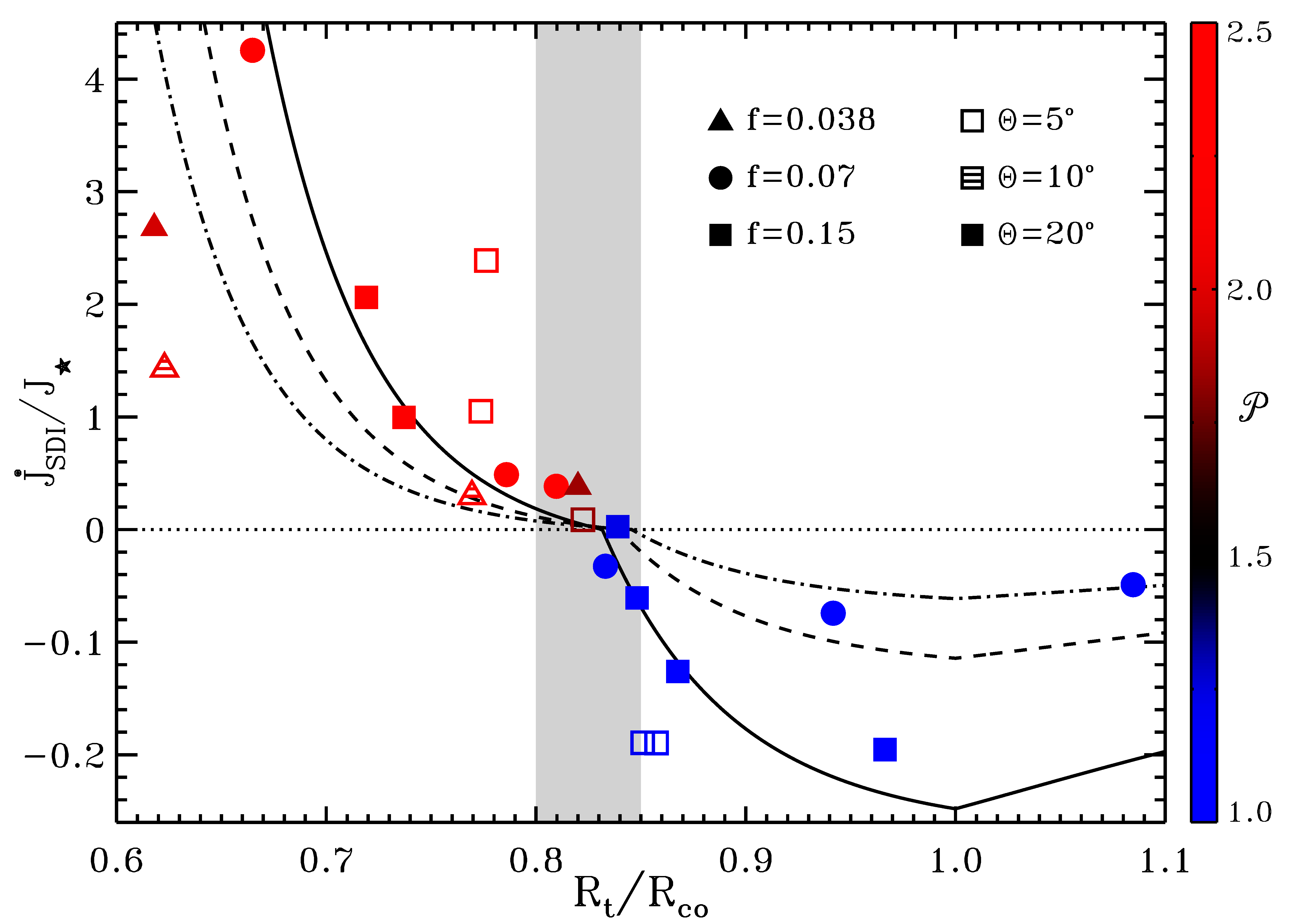}
        \caption{Normalized total SDI torque, $\dot{J}_\mathrm{SDI} =
          \dot{J}_\mathrm{acc}+\dot{J}_\mathrm{MEs}$, as a function of
          the $R_\mathrm{t}/R_\mathrm{co}$ ratio. The normalized
          torque values are given in units of
          Eq. (\ref{eq_jnorm}). We notice the different linear scale used
          for spin-up (positive values) and spin-down (negative
          values) torques. Symbol shape, filling and color have the
          same meaning as in Fig. \ref{fig_rt_y}. Black lines
          correspond to the parametrization Eq. (\ref{eq_tsdi}) for
          different values of the stellar rotation rate $f=0.15$
          (solid line), $f=0.07$ (dashed line) and $f=0.038$
          (dot-dashed line). The gray area marks the transition
          between unstable and stable regimes as in
          Fig. \ref{fig_p_rtrco}.} 
        \label{fig_torque_sdi}
\end{figure}

\subsubsection{SDI torque}
\label{sec_sdi}

We now consider the total star-disk interaction torque,
$\dot{J}_\mathrm{SDI}$, exerted onto the star along closed magnetic
field lines that steadily or intermittently connect the star with the
disk and the material launched from it. This torque is therefore
defined by the sum $\dot{J}_\mathrm{SDI} =
\dot{J}_\mathrm{acc}+\dot{J}_\mathrm{MEs}$ of the accretion and MEs
torques discussed in the previous sections. In
Fig. \ref{fig_torque_sdi}, we plot (as a function of
$R_\mathrm{t}/R_\mathrm{co}$) the time-averaged values of the
$\dot{J}_\mathrm{SDI}$ divided by the stellar angular
momentum (also listed in Table \ref{tab_data_sim}). This quantity provides
the inverse of the spin-up and spin-down 
timescale. We recall that to express this normalized torque in
physical units it must be multiplied by Eq. (\ref{eq_jnorm}).   
We used different colors to identify the accretion regimes, going from
red for unstable cases, $\mathcal{P}>1.5$, to blue for stable cases,
$\mathcal{P} < 1.5$.  
It is clearly possible to notice how stable cases correspond to an SDI
spin-down torque, while unstable regimes tend to provide a spin-up
stellar torque. As a consequence, the instability threshold
$\left(R_\mathrm{t}/R_\mathrm{co}\right)_\mathrm{stab}\approx
0.8-0.85$ and the mass accretion rate in Eqs. (\ref{eq_stab_f}) and
(\ref{eq_stab_p}) can also provide an estimate of the boundary between
star-disk interaction spin-up and spin-down torques.  

To keep the rotation period approximately constant, the
torque exerted onto the star should balance the spin-up due to the
stellar contraction. Assuming a star with the structure of a $n=3/2$
polytrope, the Kelvin-Helmholtz contraction timescale
$\tau_\mathrm{KH}=-R_\star/\dot{R}_\star$ can be estimated as
\citep{Collier-Cameron:1993aa} 
\begin{equation}
\begin{aligned}
\tau_\mathrm{KH} & = \frac{3GM_\star^2}{28\pi R_\star^3\sigma T_\mathrm{eff}^4} = \\
& = 7.3\times10^{6} \, \left(\frac{M_\star}{M_\sun}\right)^2
\left(\frac{R_\star}{2R_\sun}\right)^{-3}
\left(\frac{T_\mathrm{eff}}{4000 \, \mathrm{K}}\right)^{-4} \mathrm{yr} \; .
\end{aligned}
\label{eq_kh}
\end{equation}
Fig. \ref{fig_torque_sdi} shows that in a stable regime the SDI
spin-down timescale can be comparable to the stellar contraction time
scale, thus favoring spin equilibrium, while in an unstable regime the
accretion spin-up timescale quickly becomes shorter than the
contraction one to produce a rapid stellar spin-up. 

We also tried to parametrize the total $\dot{J}_\mathrm{SDI}$ torque
by adding up Eq. (\ref{eq_jacc}) for the accretion torque and
Eq. (\ref{eq_jmes_simple}) for the MEs torque, 
\begin{equation}
\label{eq_tsdi}
\dot{J}_\mathrm{SDI} = \dot{J}_\mathrm{acc}+\dot{J}_\mathrm{MEs} =
K_\mathrm{acc}\dot{M}_\mathrm{acc} \sqrt{GM_{\star}R_\mathrm{t}} +
K_\mathrm{MEs} \frac{B_\star^2 R_\star^6}{R_\mathrm{MEs}^3} \; , 
\end{equation}
analogously to \citet{Ireland:2022aa}, who showed that this simplified
expression can reasonably fit the total SDI torque spanning spin-up
and spin-down configurations. The accretion torque takes into account
the spin-up, while the MEs torque the spin-down, as confirmed by the
value of the fitting parameters $K_\mathrm{acc}$ and $K_\mathrm{MEs}$
in Table \ref{tab_fit}.  
We note that even in cases that displayed a spin-down accretion torque,
in ultimate analysis this effect was due to the presence of
magnetospheric ejections (see the discussion at the end of Sect.
\ref{sec_accretion}), thus justifying our simplified parametrization
with $K_\mathrm{acc} > 0$. 

Combining Eqs. (\ref{eq_tsdi}) and (\ref{eq_rt_yaccf}) it is possible
to express the parametrization for $\dot{J}_\mathrm{SDI}/J_\star$ as a
function of $R_\mathrm{t}/R_\mathrm{co}$ and $f$ only. In
Fig. \ref{fig_torque_sdi} we plot the analytical approximation
Eq. (\ref{eq_tsdi}) for values $f=0.15$ (solid line), $f=0.07$ (dashed
line) and $f=0.038$ (dot-dashed line), showing a good agreement
between the parametrization and the numerical results\footnote{We
  fitted Eq. (\ref{eq_tsdi}) taking $M_\mathrm{acc}$ and
  $R_\mathrm{t}$ as independent quantities. To plot this
  parametrization in Fig. \ref{fig_torque_sdi} we expressed
  $\dot{M}_\mathrm{acc}$ as a function of $R_\mathrm{t}/R_\mathrm{co}$
  using Eq. (\ref{eq_rt_yaccf}). The steep dependence of
  $\dot{M}_\mathrm{acc}$ on the value of $R_\mathrm{t}/R_\mathrm{co}$ (see, e.g., Eq. \ref{eq_stab_f}) amplifies the scatter of the data
  points around the best-fit models.}.  
 
\subsection{Stellar wind torque}
\label{sec_sw}

In this section, we evaluate the contribution of a stellar wind (SW) torque
to the stellar angular momentum evolution; in particular, in unstable
accretion regimes, where we found that the star-disk
interaction torque determines a short spin-up timescale due to the
combination of accretion and contraction (see Sect. \ref{sec_sdi}). 
We contextualize our SW solutions
in the framework of the Accretion Powered Stellar Wind model 
\citep{Matt:2005aa,Matt:2008aa,Matt:2008ab,Matt:2012ab,Matt:2012aa}, 
according to which the wind mass-loss rate and the spin-down
torque can be enhanced by extracting a fraction of the energy
deposited onto the star by the accretion funnels via, for example,
the excitation and dissipation of Alfv\'{e}n waves 
\citep{Decampli:1981aa, Cranmer:2008ab,Cranmer:2009aa}. 
Our setup cannot self-consistently handle this kind of accretion/ejection coupling.
In our simulations, the trans-sonic SW is thermally driven and accelerated, 
so its mass-loss rate is mainly determined by the density and
temperature at the base of the stellar wind fixed as initial conditions (see Sect. \ref{sec_params}) and by the size of the area taken by the open stellar magnetic 
flux, which, on the other hand, we can relate to accretion and stellar parameters 
and the position of the truncation radius in particular, as we discuss below.

The torque of a trans-Alfv\'{e}nic stellar wind is customarily
expressed as \citep[e.g.,][]{Weber:1967aa,Mestel:1984aa,Kawaler:1988aa,Pantolmos:2017aa} 
\begin{equation}
\dot{J}_\mathrm{SW} = \dot{M}_\mathrm{SW} \langle r_\mathrm{A}^2 \rangle \Omega_\star \; ,
\label{eq_jsw}
\end{equation} 
where $\dot{M}_\mathrm{SW}$ is the wind mass-loss rate, $\langle
\Lambda \rangle = \dot{J}_\mathrm{SW}/\dot{M}_\mathrm{SW} = \langle
r_\mathrm{A}^2 \rangle\Omega_\star$ is its average specific angular
momentum and $\langle r_\mathrm{A} \rangle = \langle r_\mathrm{A}^2
\rangle^{1/2}$ is the average cylindrical radius of the Alfv\'{e}n
surface. We note that the equivalence $\langle \Lambda \rangle =
\langle r_\mathrm{A}^2 \rangle\Omega_\star$ is strictly valid only for
axisymmetric winds (see the discussion in Appendix \ref{sec_alfven}),
but we use this parametrization for consistency with previous
works, even if in our 3D nonaxisymmetric simulations the quantity
$\langle r_\mathrm{A} \rangle$ does not exactly correspond to the
average Alfv\'{e}n radius.     
It was shown in \citet{Reville:2015aa} that the average
Alfv\'{e}n radius can be expressed as a function of the stellar open
magnetic flux exploited by the stellar wind $\Phi_\mathrm{SW}$ as 
\begin{equation}
\frac{\langle r_\mathrm{A} \rangle}{R_\star} = K_\mathrm{A} \left\{
\frac{\Upsilon_\mathrm{SW}}{[1 + (f/f_0)^2]^{1/2}} \right\}^{m_\mathrm{A}} \; ,
\label{eq_ra}
\end{equation} 
where $\Upsilon_\mathrm{SW}$ is the magnetization parameter
\begin{equation}
\Upsilon_\mathrm{SW} = \frac{\Phi_\mathrm{SW}^2}{4 \pi R_\star^2 \dot{M}_\mathrm{SW} v_\mathrm{esc}}
\label{eq_ysw}
\end{equation}
and $\Phi_\mathrm{SW}$ is the wind unsigned open flux
\begin{equation}
\Phi_\mathrm{SW} = \int_{S_\mathrm{SW}} \vert \vec{B} \cdot d\vec{S} \vert \; ,
\label{eq_phisw}
\end{equation}
integrated over the surface $S_\mathrm{SW}$ into which open magnetic field lines 
extending to infinity are anchored.
We note that Eq. (\ref{eq_phisw}) can be defined independently of the stellar magnetic topology. 
\begin{figure}[!t]
        \centering
        \includegraphics[width=\linewidth]{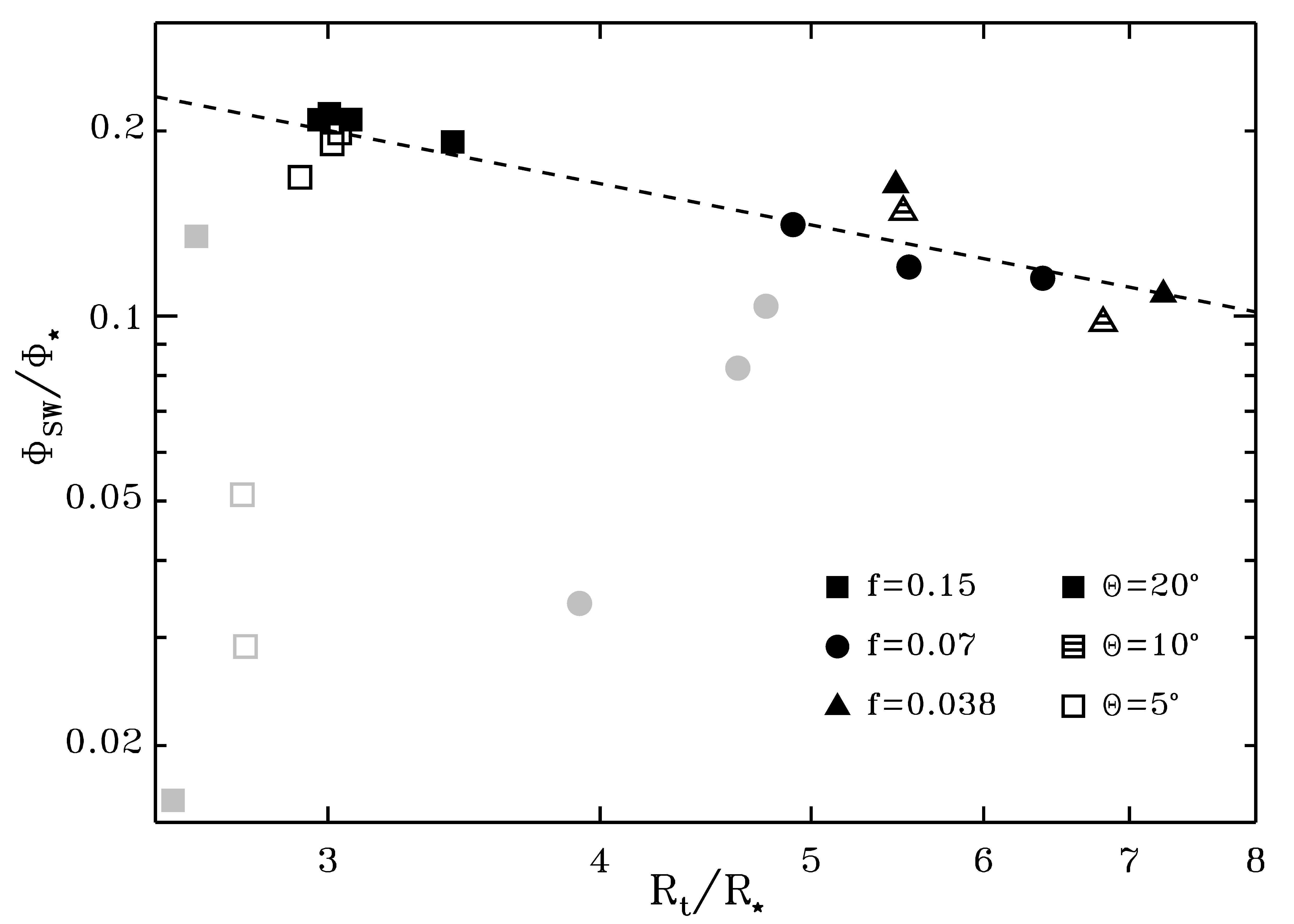}
        \caption{Stellar wind fractional open flux
                $\Phi_\mathrm{SW}/\Phi_\star$ versus truncation radius
                $R_\mathrm{t}/R_\star$. Symbols are the same as in
                Fig. \ref{fig_p_rtrco}. Gray points correspond to
                simulations in which the stellar wind launching region is
                significantly quenched by high-density accretion, with a
                disk-star initial density contrast of
                $\rho_{\mathrm{d}\star}/\rho_\star = 800,1600$. These points
                are excluded from the fit of the function
                Eq. (\ref{eq_swflux}) plotted with a dashed line.} 
        \label{fig_wflux_rt}
\end{figure}
While for nonaccreting stars the amount of open flux is 
determined by the properties of the stellar wind itself, such as its
mass-loss rate, it was shown that in accreting systems it is
mainly determined by the star-disk interaction
\citep{Pantolmos:2020aa,Ireland:2021aa,Ireland:2022aa}. In fact, the
differential rotation between the star and disk tends to open the
magnetic surfaces at latitudes higher than the location where the
field connects to $R_\mathrm{t}$. We therefore expect the 
fractional open flux to scale with the truncation radius as 
\begin{equation}
  \frac{\Phi_\mathrm{SW}}{\Phi_\star} = K_\Phi \left(
    \frac{R_\mathrm{t}}{R_\star} \right)^{m_{\Phi}} \; ,
  \label{eq_swflux}
\end{equation}
where $\Phi_\star$ is the total unsigned stellar flux
($\Phi_\star=2\pi B_{\star}R_{\star}^2$ for a dipolar topology) and
$m_\Phi < 0$, so that the closer the disk is truncated to the star
and the smaller the closed magnetosphere connected to the disk
becomes, the more fractional open stellar flux is available for the wind.
In Fig. \ref{fig_wflux_rt} we plot
the stellar wind fractional open flux $\Phi_\mathrm{SW}/\Phi_\star$
(given in Table \ref{tab_data_sim}) as
a function of the truncation radius $R_\mathrm{t}/R_\star$. First we
noticed that for high values of the disk-star initial density contrast
$\rho_{\mathrm{d}\star}/\rho_\star = 800,1600$ the accretion flow
tended to quench the stellar wind, leaving almost no open flux to be
exploited (gray points). We ascribed this behavior to a diffusion of
the accretion material into the wind launching region due to the high
density contrast and the poorer grid resolution around the stellar
poles. On the other hand the other cases showed a convincing
correlation between the fractional open flux and the truncation radius
with an expected $m_\Phi < 0$ (see Table \ref{tab_fit}).  

\begin{figure}[!t]
        \centering
        \includegraphics[width=\linewidth]{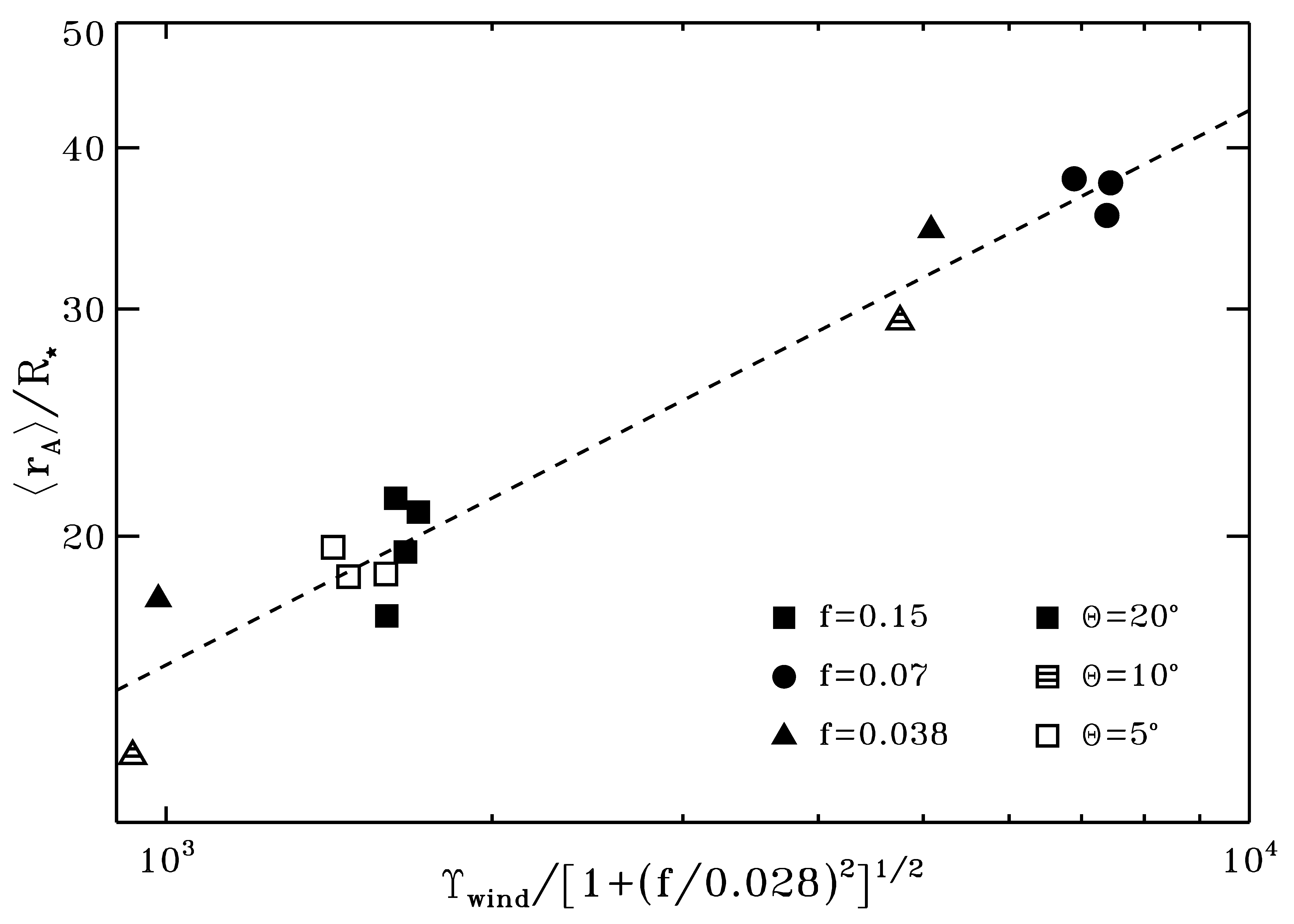}
        \caption{Average Alfv\'{e}n radius of the stellar wind,
          $\langle r_\mathrm{A} \rangle/R_\star$, versus the magnetization
          parameter,
          $\Upsilon_\mathrm{SW}/[1+(f/0.028)^2]^{1/2}$. Symbols are
          the same as in Fig. \ref{fig_p_rtrco}. The dashed line shows
          the best fit function Eq. (\ref{eq_ra}).} 
        \label{fig_ra_ywind}
\end{figure}

In Fig. \ref{fig_ra_ywind}, we plot the average Alfv\'{e}n radius of
the stellar wind, $\langle r_\mathrm{A} \rangle/R_\star$, as a function
of the magnetization parameter,
$\Upsilon_\mathrm{SW}/[1+(f/f_0)^2]^{1/2}$, and the best fit of
Eq. (\ref{eq_ra}) (excluding the quenched cases of
Fig. \ref{fig_wflux_rt}). Table \ref{tab_data_sim} provides the $\langle
r_\mathrm{A} \rangle/R_\star$ and $\Upsilon_\mathrm{SW}$ values 
for the cases considered in our analysis. Taking the value $f_0 = 0.028$ from
\citet{Ireland:2021aa,Ireland:2022aa}, our best fit parameters
$K_\mathrm{A}$ and $m_\mathrm{A}$ (see Table \ref{tab_fit}) are in
excellent agreement with torque models of stellar winds in isolated
\citep{Reville:2015aa} and accreting stars
\citep{Ireland:2021aa,Ireland:2022aa}.  

\begin{figure}[!t]
        \centering
        \includegraphics[width=\linewidth]{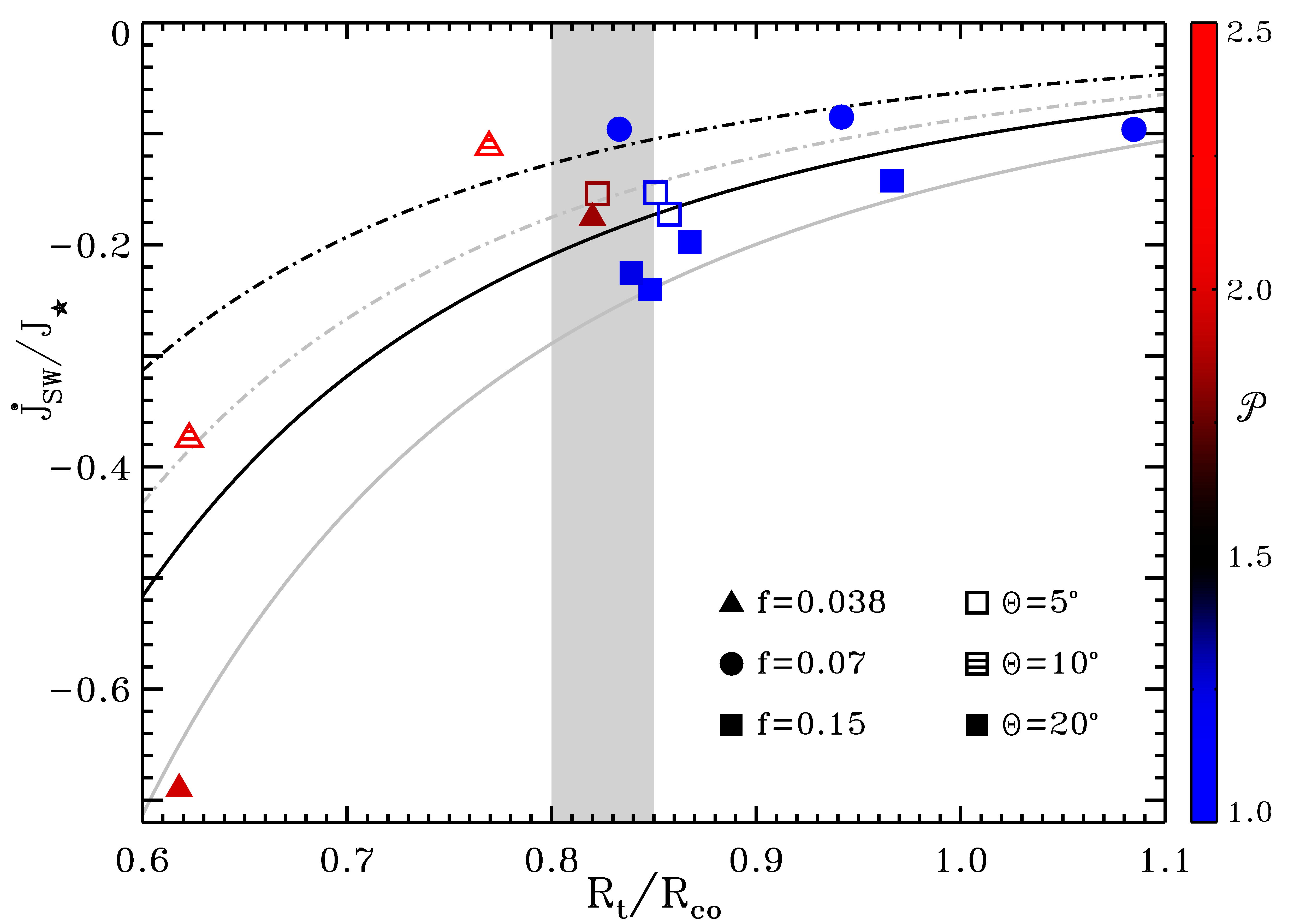}
        \caption{Normalized stellar wind torque,
                $\dot{J}_\mathrm{SW}/J_\star$, as a function of the
                $R_\mathrm{t}/R_\mathrm{co}$ ratio. The normalized torque
                values are given in units of Eq. (\ref{eq_jnorm}). Symbol
                shape, filling and color are the same as in
                Fig. \ref{fig_rt_y}. The gray area marks the transition
                between unstable and stable regimes as in
                Figs. \ref{fig_p_rtrco} and \ref{fig_torque_sdi}. The lines
                correspond to the torque parametrization Eq. (\ref{eq_jsw})
                for different values of the stellar rotation rate $f=0.15$
                (solid lines) and $f=0.038$ (dot-dashed lines) and wind mass
                ejection efficiency
                $\dot{M}_\mathrm{SW}/\dot{M}_\mathrm{acc}=0.01$ (black
                lines) and $\dot{M}_\mathrm{SW}/\dot{M}_\mathrm{acc}=0.1$
                (gray lines).} 
        \label{fig_torque_sw}
\end{figure}

To provide an estimate of the efficiency of the stellar wind
torque, we plot in Fig. \ref{fig_torque_sw} the inverse of the wind
spin-down timescale $\dot{J}_\mathrm{SW}/J_\star$ computed from our
simulations as a function of the $R_\mathrm{t}/R_\mathrm{co}$ ratio so
as to be directly compared with the SDI torques in
Figs. \ref{fig_torque_me_acc} and \ref{fig_torque_sdi}. By combining
Eq. (\ref{eq_ra}) with Eqs. (\ref{eq_rt_yaccf}, \ref{eq_ysw},
\ref{eq_swflux}) it is possible to express the parametrization for
$\dot{J}_\mathrm{SW}/J_\star$ as a function of
$R_\mathrm{t}/R_\mathrm{co}$, $f$ and the wind mass ejection
efficiency $\dot{M}_\mathrm{SW}/\dot{M}_\mathrm{acc}$. We plot the
analytical parametrization for values $f=0.038,0.15$ (dot-dashed and
solid lines, respectively) and
$\dot{M}_\mathrm{SW}/\dot{M}_\mathrm{acc}=0.01,0.1$ (black and gray
lines respectively), roughly encompassing the parameter space covered
by our simulations. Despite the fact that these curves do not represent a best fit
to the data points (as each simulation has a different
$\dot{M}_\mathrm{SW}/\dot{M}_\mathrm{acc}$ value; see Table
\ref{tab_data_sim}), they show that there is a good agreement between
our results and the torque parametrization. The increase in the wind
torque for smaller $R_\mathrm{t}/R_\mathrm{co}$ values corresponds,
for all other stellar parameters fixed, to an increase in the
accretion rate that determines a growth of the fractional open flux,
as discussed before, and the wind mass-loss (for a given ejection rate
$\dot{M}_\mathrm{SW}/\dot{M}_\mathrm{acc}$). 
Both the data points and the torque parametrization in
Fig. \ref{fig_torque_sw} show that in a stable regime (blue points) a
stellar wind with an ejection efficiency in the range
$\dot{M}_\mathrm{SW}/\dot{M}_\mathrm{acc}=0.01 - 0.1$ can provide a
spin-down torque comparable or even higher than the SDI one (see
Fig. \ref{fig_torque_sdi}) to oppose the stellar contraction. On the
other hand, despite the increase in the wind spin-down torque
efficiency in unstable regimes (red points), the stellar wind is not
able to balance the spin-up due to accretion and contraction (see
Fig. \ref{fig_torque_sdi}). We found that with our torque
parametrization, a wind with an ejection efficiency
$\dot{M}_\mathrm{SW}/\dot{M}_\mathrm{acc} > 1$ would be
needed. We do notice, however, that the unstable region
$R_\mathrm{t}/R_\mathrm{co} < 0.8$ is poorly sampled, mainly due to
the exclusion of the ``quenched'' cases of Fig. \ref{fig_wflux_rt}. A
more thorough investigation of stellar winds in unstable regimes would
be needed to better assess their torque efficiency.  

\section{Discussion}
\label{sec_disc}

In this section, we briefly summarize and discuss our main results to
compare them with the most relevant and recent results about the
numerical modeling of the magnetospheric star-disk
interaction. Subsequently, we apply our findings to different
observational data to test the possibility of determining the
accretion regime and spin-up/down configuration of CTTs.  

\subsection{Accretion regimes and disk truncation}

Our results confirm the existence of two distinct accretion
regimes: 1) a stable regime, where the disk is truncated to form two main
accretion funnels and spots; and 2) an unstable regime where the
truncation region of the disk is fragmented into many accretion
channels impacting the star at different latitudes and azimuths. We
found that the main parameter controlling the transition between
stable and unstable regimes is the ratio between the position of the
truncation and the corotation radius, where the interchange
instability starts to develop for $R_\mathrm{t}/R_\mathrm{co} < 0.8 -
0.85$. We therefore qualitatively confirmed the seminal results
obtained by \citet{Blinova:2016aa} even if they found an instability
threshold $R_\mathrm{t}/R_\mathrm{co} \approx 0.7$. Despite using a
similar criterion to determine the position of the truncation radius
based on the $\beta_{T} = 1$ condition, this small difference over the
$R_\mathrm{t}/R_\mathrm{co}$ boundary between stable and unstable
regimes could be ascribed to a different method to extract the
time-averaged disk truncation radius and to identify the accretion
regimes. As proposed by the same authors, we also confirmed
  the presence of an ordered unstable regime characterized by fewer, but azimuthally wider, intruding accretion fingers;
  however, our threshold for ordered unstable accretion is 
$R_\mathrm{t}/R_\mathrm{co} < 0.7$, which is slightly larger than their estimate
$R_\mathrm{t}/R_\mathrm{co} < 0.6$. More recently \citet{Zhu:2025aa},
relying on global numerical models of an MRI-turbulent disk
interacting with a stellar magnetosphere, found a similar behavior,
with the instability amplitude growing for smaller
$R_\mathrm{t}/R_\mathrm{co}$ values. The same instability has also
been observed in the MRI-turbulent simulations of
\citet{Takasao:2022aa}, but they concluded that even in cases that
should be interchange stable (their Model A) according to the
stability criterion Eq. (\ref{eq_spruit}), the truncation becomes
unstable due to another mechanism, the magneto-gradient-driven
instability proposed by \citet{Hirabayashi:2016aa}. We suppose that
this is due to the fact that the authors identify the truncation
region using the $\beta = 1$ criterion. Actually, we show in
Fig. \ref{fig_rho_2d} that in cases that we classify as stable by
looking at the almost circular $\beta_\mathrm{T} = 1$ line, the $\beta
= 1$ isocontour can be highly perturbed and asymmetric.

Despite qualitatively agreeing on the dependence of the instability threshold
  on the  $R_\mathrm{t}/R_\mathrm{co}$ ratio, our study and the cited works
  use different definitions of the truncation radius, $R_\mathrm{t}$,
  and find different scalings with the stellar and disk parameters. We recall that our definition of the truncation radius
Eq. (\ref{eq_rt_avg}) shows a strong correlation with the position of
the corotation radius and the rotation rate $R_\mathrm{t}/R_\star
\propto f^{-0.51}$ and relatively weak dependence on the accretion
parameter $R_\mathrm{t}/R_\star \propto
\Upsilon_\mathrm{acc}^{0.072}$, contrary to the classical
\citet{Ghosh:1979aa} parametrization. 
Defining the position of the truncation radius as the maximum radial
size of the magnetospheric cavity delimited by the $\beta_\mathrm{T} = 1$
curve,
\citet{Blinova:2016aa} also found a flatter scaling
of $R_\mathrm{t}/R_\star$ with $\Upsilon_\mathrm{acc}$, with an
exponent $m_\Upsilon \approx 0.2$ considerably smaller than the
\citeauthor{Ghosh:1979aa} $m_\Upsilon = 2/7$ value, but they did not
take into account any $f$ dependence. They ascribed this $m_\Upsilon$
discrepancy to a stellar magnetic field radial profile different from
perfectly dipolar $B\propto R^{-3}$, while our explanation is based on
the development of the interchange instability and a consequent
important dependence of $R_\mathrm{t}/R_\star$ on $f$. If our
interpretation is correct, the \citet{Blinova:2016aa} $m_\Upsilon$
difference with our best-fit and the scatter of their data
points could be due to the fact that they neglected the $f$
dependence possibly suggested by their Fig. 1, in particular
if the ordered unstable cases are not taken into account, for which
our proposed scaling likely provides only 
an upper limit to $R_\mathrm{t}/R_\star$; see the discussion in Sect. \ref{sec_rtrunc}.
As a matter of fact, if we neglect the scaling with $f$ and perform
the same analysis as \citet{Blinova:2016aa}, who fitted separately 
stable and unstable cases with $\Upsilon_\mathrm{acc}$ power-laws (see 
their Fig. A.2), we obtain the exponents $m_\Upsilon = 0.20$ and $0.22$ for
stable and unstable cases respectively, almost identical to the 
\citet{Blinova:2016aa} result, but with a larger scatter with respect to
our best-fit expression Eq. (\ref{eq_rt_yaccf}).
\citet{Takasao:2022aa}, derived a scaling
$R_\mathrm{t}/R_{\star} \propto \Upsilon_\mathrm{acc}^{m_\Upsilon}$
with a ``classical'' $m_\Upsilon \approx 2/7$ value. As already said,
this study computed the magnetospheric radius based on the $\beta = 1$
location and it relied on a small number of simulations. Besides, they
considered only strong accretors ($\Upsilon_\mathrm{acc} \approx 8$)
and included simulations that are outside our parameter space
($R_\mathrm{t}/R_\mathrm{co} \approx 0.4$, Model C),
for which we expect a stronger dependence on $\Upsilon_\mathrm{acc}$.
Recently, \citet{Zhu:2025aa} used another criterion to define the truncation
radius. Since in the magnetospheric interaction region the disk
becomes sub-Keplerian, $R_\mathrm{t}$ was defined as the position
where the average azimuthal velocity is equal to half the local
Keplerian velocity \citep{Zhu:2024aa}. Despite the difference in definition,
  their estimate of $R_\mathrm{t}$ is not in good agreement with a \citeauthor{Ghosh:1979aa} 
scaling and it shows a clear dependence on the position of the
corotation radius (see their Table 1).

Our parametrization of the truncation radius Eq. (\ref{eq_rt_yaccf})
can be used to express the threshold between stable and unstable
regimes in terms of more observable quantities. For example we can
define a mass accretion rate limit as a function of the stellar
parameters and the instability threshold of
$\left(R_\mathrm{t}/R_\mathrm{co}\right)_\mathrm{stab} \approx
0.8-0.85$ in terms of the rotation rate, $f$, via 
\begin{equation}
\begin{split}
\dot{M}_\mathrm{acc,stab} = 3.1 \times & 10^{-9}
\left(\frac{B_\star}{\mathrm{kG}}\right)^2 
\left(\frac{M_\star}{M_\sun}\right)^{-1/2}
\left(\frac{R_\star}{2 R_\sun}\right)^{5/2} \times \\
& \times \left(\frac{f}{0.1}\right)^{2.17} 
\left(\frac{R_\mathrm{t}/R_\mathrm{co}}{0.825}\right)_\mathrm{stab}^{-13.9} M_\sun\ \mathrm{yr^{-1}} \; ,
\label{eq_stab_f}
\end{split}
\end{equation}
or the stellar rotation period, $P_\star$, via 
\begin{equation}
\begin{split}
\dot{M}_\mathrm{acc,stab} = 1.2 \times & 10^{-9}
\left(\frac{B_\star}{\mathrm{kG}}\right)^2 
\left(\frac{M_\star}{M_\sun}\right)^{-1.58}
\left(\frac{R_\star}{2 R_\sun}\right)^{5.75} \times \\
& \times \left(\frac{P_\star}{5 \, \mathrm{days}}\right)^{-2.17} 
\left(\frac{R_\mathrm{t}/R_\mathrm{co}}{0.825}\right)_\mathrm{stab}^{-13.9} M_\sun\ \mathrm{yr^{-1}} \; .
\end{split}
\label{eq_stab_p}
\end{equation}
For $\dot{M}_\mathrm{acc} > \dot{M}_\mathrm{acc,stab}$ the system
should start to develop an interchange instability. We note, however,
how this expression strongly depends on the
$R_\mathrm{t}/R_\mathrm{co}$ instability threshold due to the small
value of the $m_\Upsilon$ exponent. 

\subsection{SDI and stellar wind torques}

We performed a thorough analysis of the torques exerted on the star by the star–disk interaction,
  transmitted along magnetic field lines that steadily or intermittently connect the star and the disk.
We have found that the transition from a spin-up to a spin-down SDI torque is mainly
determined, once again, by the $R_\mathrm{t}/R_\mathrm{co}$
ratio. Actually, this is a common feature of many analytical
\citep[e.g.,][]{Ghosh:1979ab,Collier-Cameron:1993aa,Matt:2005ab} and
numerical models \citep[e.g., ][]{Long:2005aa,Zanni:2013aa, Zhu:2025aa} 
and it reflects the simple idea that in order to balance the accretion torque (or even to be spun down), a star must be
magnetically connected mainly to material (in accretion or ejection)
that rotates more slowly than the star itself, a condition which is favored
if the disk is truncated close to the corotation radius; see for
example the cartoon in Fig. 7 of \citet{Zhu:2025aa}. An important
result of this work is that we clearly show that SDI spin-up and
spin-down configurations approximately coincide, respectively, with
unstable and stable accretion regimes, with the spin-up/spin-down,
unstable/stable transition happening around
$R_\mathrm{t}/R_\mathrm{co} = 0.8-0.85$. This is clearly shown in
Fig. \ref{fig_torque_sdi} while, using our truncation radius scaling
Eq. (\ref{eq_rt_yaccf}), the condition $\dot{J}_\mathrm{SDI} = 0$ of
Eq. (\ref{eq_tsdi}) can be written as 
\begin{equation}
\left. \frac{R_\mathrm{t}}{R_\mathrm{co}}\right\vert_{\dot{J}_\mathrm{SDI}=0} = 0.836 \, \left(\frac{f}{0.1}\right)^{-0.01} \; ,
\label{eq_null_torque}
\end{equation}
which provides approximately the same transition radius and a very
weak dependence on the stellar rotation. Using high resolution
simulations of MRI active disks, \citet{Zhu:2025aa} obtained a  
``zero torque'' configuration for $R_\mathrm{t}/R_\mathrm{co} = 0.82$,
close to our estimate.  

In our picture, the phenomenon that we call magnetospheric ejections
plays an important role. We recall that these are variable outflows
that exploit magnetic field lines that connect the star with the disk
undergoing cycles of inflation, reconnection and deflation, due to the
buildup of toroidal field caused by the star-disk differential
rotation. The same phenomenon was recently identified as a
``load-fire-reload'' mechanism in the 3D simulations of
\citet{Tu:2026aa}. The side of the MEs anchored in the disk behaves as
a disk wind efficiently extracting angular momentum from the disk so
as to dampen the accretion torque $\dot{J}_\mathrm{acc} =
K_\mathrm{acc} \dot{M}_\mathrm{acc}\sqrt{GM_\star R_\mathrm{t}}$. The
same effect was ascribed by \citet{Takasao:2022aa} to conical
disk winds, which we tend to identify as the outflow exploiting the
outermost magnetic field lines of the MEs and the innermost disk open
field lines. Based on the simulations of \citet{Takasao:2022aa},
\citet{Takasao:2025aa} estimated that conical winds can extract around
$80\%$ of the accretion torque (i.e., $K_\mathrm{acc} \approx
0.2$). We note that their parametrization for the conical wind torque
is analogous to the expression used for the X-Wind \citep{Shu:1994aa},
where the torque efficiency is assumed to be $100\%$, or
$K_\mathrm{acc} = 0$, to exert no torque onto the star. 
For example, applying this idea to the specific case of the star RU Lup,
\citet{Armeni:2025aa} have estimated that a conical/X-wind type outflow
should extract around $40\%$ of the mass accretion rate to cancel the 
accretion torque. On the other hand we have shown in Sect. \ref{sec_accretion} 
how the proportionality factor $K_\mathrm{acc}$ is not constant (see
Fig. \ref{fig_wflux_rt}) and in particular conditions can even become
negative (i.e., spin-down).  

Additionally, we have evaluated the torque exerted by MEs directly
onto the star along the magnetic field side connected to it, finding
that in most of our simulations the MEs exert a stellar spin-down
torque, coherently with the fact that their launching region in the
disk is located close/beyond corotation, which in most cases
corresponds the maximum extent of the magnetic cavity, so that MEs
tend to rotate more slowly than the star. Evaluating the total SDI torque,
given by the sum of the accretion and MEs torques, we found that in
stable regimes with $R_\mathrm{t} > 0.85 R_\mathrm{co}$, corresponding
to a weak propeller configuration, the spin-down is comparable to the
Kelvin-Helmholtz contraction timescale, while the spin-up timescale in
unstable regimes with $R_\mathrm{t} < 0.8 R_\mathrm{co}$ can be much
shorter, raising questions about the possibility of preventing these
stars from spinning up. 
 
We therefore evaluated the spin-down torque exerted by the stellar
winds modeled in our simulations. Our results confirmed the scaling of
the wind Alfv\'{e}n radius with the stellar open magnetic flux
proposed by \citet{Reville:2015aa} (Eq. \ref{eq_ra}) and showed that
in accreting systems this depends mostly on the position of the
truncation radius (Eq. \ref{eq_swflux}), the closer the disk is
truncated to the star, the more magnetic flux is opened. The scaling
Eq. (\ref{eq_swflux}) establishes an interesting correlation between
the accretion spin-up torque and the stellar wind spin-down: in fact,
an increase in the mass accretion rate reduces the truncation radius
to open more magnetic flux and trigger a stronger stellar wind
torque to oppose the increase in the accretion spin-up. This property
has been exploited, for example, by the stellar evolution models of
\citet{Amard:2023aa}, who showed, using the axisymmetric torque models
from \citet{Ireland:2022aa}, that stellar spin-up can be prevented
without relying on extremely massive stellar winds. These same
axisymmetric models were employed by \citet{Gehrig:2025aa} to
show that even relatively light stellar winds can sensibly change the
$R_\mathrm{t}/R_\mathrm{co}$ ratios in synthetic stellar populations
of accreting Class II stars.  
It must be said, however, that this flux opening effect is likely
stronger in axisymmetric models, since the truncation radius is more
sensitive to accretion rate variations than in our 3D simulations; see
the discussion in Sect. \ref{sec_rtrunc}. Besides, we found a
flatter exponent $m_\Phi=-0.69$ than \citet[][$m_\Phi\approx
-1.3$]{Ireland:2021aa,Ireland:2022aa}. This is possibly due to the
fact that in axisymmetry the truncation radius has a well defined
location, while in 3D solutions the magnetic cavity can span a range
of radii around the average $R_\mathrm{t}$.  

In conclusion, we found that while stellar winds can significantly
contribute to the spin-down torque in stable accretion regimes even
for relatively low mass ejection rates, $\dot{M}_\mathrm{SW} = 0.01
\dot{M}_\mathrm{acc}$, they seem to be unable to balance the accretion
torque in unstable configurations.  

\subsection{Comparison with spectropolarimetric observations}
\label{sec_spec}

The next step is testing our findings on accretion regimes, the position of the
truncation radius, and stellar torques by applying our models to a
sample of CTTs - Class II objects observed by the spectropolarimeters
ESPaDOnS \citep{Donati:2003aa} and SPIRou \citep{Donati:2020ab}
installed at the Canada–France–Hawaii Telescope (CFHT). Using
Zeeman–Doppler-Imaging (ZDI) techniques
\citep{Semel:1989aa,Donati:1997aa}, these observations can provide a
reconstruction of the large-scale stellar magnetic topology to
supply an estimate of the dipolar stellar field. Beside being a key
parameter of our models, it is widely accepted that this is the
magnetic field component mainly responsible for the large-scale
interaction with the stellar surroundings, while multipolar harmonics
are likely to have a more local influence, for example on the position
and size of the accretion spots
\citep[e.g.,][]{Mohanty:2008aa,Romanova:2011aa}. 
Our data sample, consisting of 20 different CTTs is listed in Table
\ref{tab_spec_data}. Nine stars that have been observed at multiple epochs
are listed twice, for a total of 29 entries, each observation corresponding 
to the minimum and maximum dipolar field intensity encompassed by 
the available data. We report both the
stellar parameters derived from observations and the outcome of our
models: position of the truncation radius given by Eq. (\ref{eq_rt_yaccf}); 
SDI torques determined by  Eq. (\ref{eq_tsdi}); stellar wind 
torques, based on Eqs. (\ref{eq_jsw}-\ref{eq_ysw}). 
The stellar wind torque was computed by assuming a mass ejection efficiency
$\dot{M}_\mathrm{SW} = 1\% \dot{M}_\mathrm{acc}$, which roughly
corresponds to the upper limit found by
\citet{Cranmer:2008ab,Cranmer:2009aa}. An ejection efficiency of
$10\%$ should determine a $\approx 38\%$ increase in the stellar wind
torque. In Table \ref{tab_spec_data} we also list the net stellar spin 
evolution timescale $\tau_\Omega$, taking into account both stellar torques 
and contraction, whose definition is provided in Appendix \ref{sec_spol}.
Stars are listed according to their decreasing $R_\mathrm{t}/R_\mathrm{co}$ ratios.

\begin{figure}[!t]
        \centering
        \includegraphics[width=\linewidth]{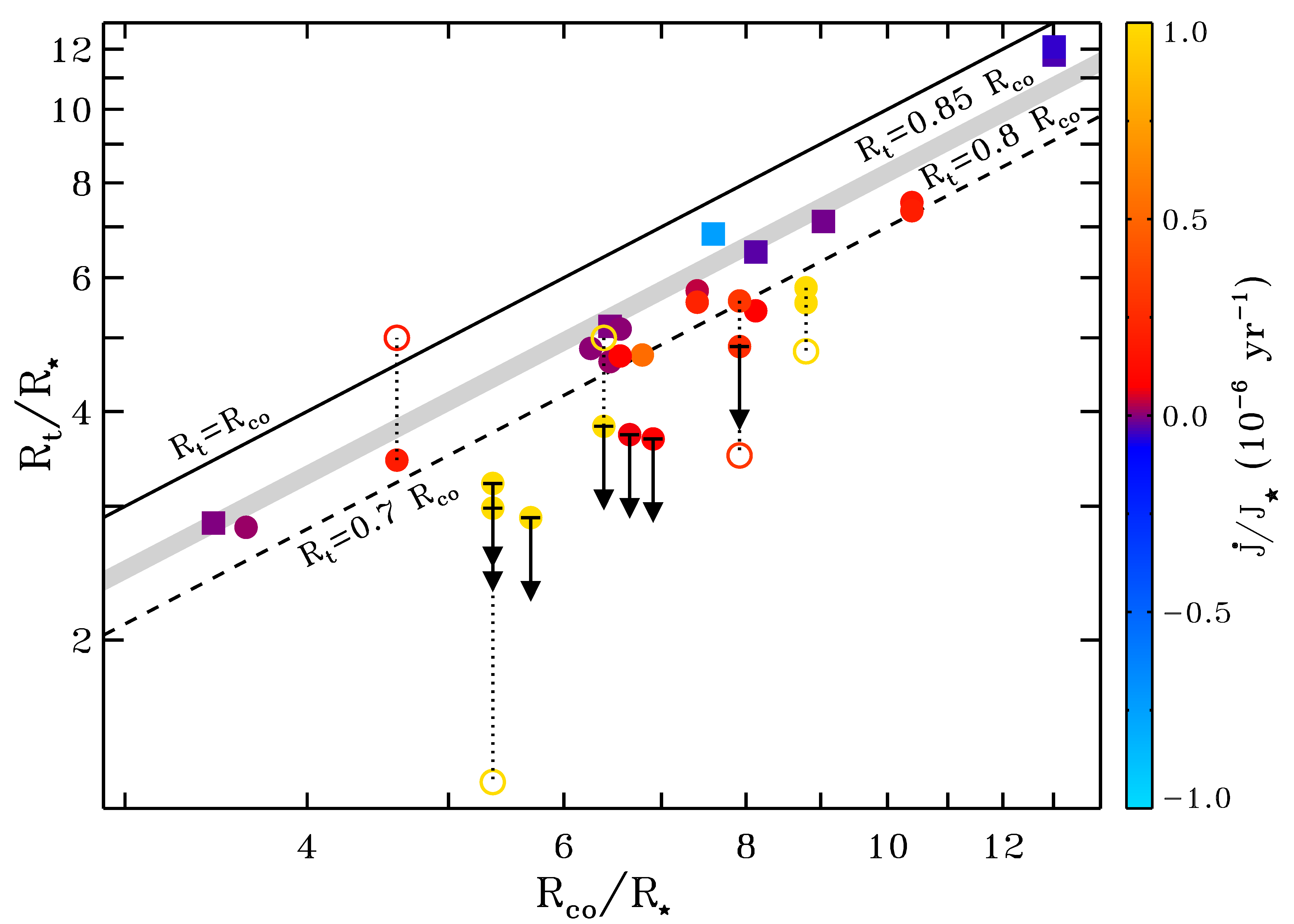}
        \caption{Truncation radius, $R_\mathrm{t}$,
                (Eq. \ref{eq_rt_yaccf}) vs. corotation radius
                $R_\mathrm{co}$ for the stellar sample observed in
                spectropolarimetry listed in Table \ref{tab_spec_data}
                (filled points). The colors correspond to the total
                normalized stellar torque, given by the sum of the SDI
                torque Eq. (\ref{eq_tsdi}) and a stellar wind torque
                Eqs. (\ref{eq_jsw}-\ref{eq_ysw}) assuming a
                $\dot{M}_\mathrm{SW} = 1\% \dot{M}_\mathrm{acc}$ ejection
                efficiency. Squares correspond to a total null/spin-down
                torque ($\dot{J} \leq 0$), while circles to a total spin-up
                torque. The gray band shows the $R_\mathrm{t}/R_\mathrm{co}
                = 0.8-0.85$ transition region between stable and unstable
                accretion regimes. The solid line corresponds to the
                $R_\mathrm{t}=R_\mathrm{co}$ propeller regime, while the
                $R_\mathrm{t}=0.7 \, R_\mathrm{co}$ dashed line marks the
                transition to an unstable ordered accretion
                regime. The arrows highlight CTTs with $R_\mathrm{t} \lesssim 
                0.6 \, R_\mathrm{co}$ for which our scaling Eq. (\ref{eq_rt_yaccf})
                likely provides an upper limit.
                Empty points correspond to interferometric estimates
                of the size of the Br$\gamma$ emitting region, connected
                with the corresponding truncation radius estimates.} 
        \label{fig_spol}
\end{figure}

We summarize the data given in Table \ref{tab_spec_data} in Fig. \ref{fig_spol}, where we plot our
estimate of the truncation radius, $R_\mathrm{t}$, according to
Eq. (\ref{eq_rt_yaccf}) versus the stellar corotation radius
$R_\mathrm{co}$. The color of the points corresponds to the value of
the normalized stellar torque, given by the sum of the SDI torque
Eq. (\ref{eq_tsdi}) and a stellar wind torque with
$\dot{M}_\mathrm{SW} = 1\% \dot{M}_\mathrm{acc}$, Eq.(\ref{eq_jsw}),
with the blue-light blue values corresponding to a total spin-down
torque, purple points to a zero-torque condition and the red-yellow
points indicating a spin-up situation. Both Fig. \ref{fig_spol} and Table 
\ref{tab_spec_data} show that  only 5 stars in our sample (6 entries, $\sim 20\%$) 
are expected to be in a stable accretion regime 
(i.e., $R_\mathrm{t}/R_\mathrm{co} \gtrsim 0.8-0.85$), while 
the remaining $80\%$ is supposed to be in an unstable regime, among 
which 9 stars (11 entries, $\sim 38\%$ of the total) should be in an ordered 
unstable accretion regime with $R_\mathrm{t}/R_\mathrm{co} \lesssim
0.7$. Nevertheless, our conclusions are based on numerical models with typical 
magnetic obliquities $\Theta \leq 20^{\circ}$, whereas different stars in 
our sample  display larger misalignments. As pointed out by \citet{Kulkarni:2008aa}, 
a larger $\Theta \geq 30^{\circ}$ rotation axis/magnetic moment 
misalignment can have a stabilizing effect.

Looking at the stellar torques, only two stars
in our sample, IRAS04125+2902 (with two entries) and AA Tau, are characterized
by a spin-down SDI torque (i.e., their $R_\mathrm{t}/R_\mathrm{co}$ ratio is
larger than $\approx 0.84$), coherently with
Eq. (\ref{eq_null_torque}). Even including the effect of a stellar wind, 
only 7 stars ($\sim 25\%$) are subject to a total spin-down torque; see Table 
\ref{tab_spec_data} and the histogram in Fig. \ref{fig_hists}a.
As already stated in Sect \ref{sec_sdi}, we must not forget that in order to 
prevent the star from spinning up, the stellar spin-down torque must balance the rotational
acceleration due to contraction. Using our estimates, only in one case of 
our sample (AA Tau) the total spin-down torque can overcome the 
Kelvin-Helmholtz contraction to effectively prevent the star from 
spinning up; see Fig. \ref{fig_hists}b. The same histogram also shows
that around 10 stars are characterized by a spin-up timescale (taking into account 
both the stellar torques and contraction) longer than $3 \times 10^6$ years, a typical average 
value for the disk lifetime \citep[see, e.g.,][]{Galli:2015aa}, suggesting that for these stars the spin-up process could be relatively slow. On the other hand these timescale estimates provide
only a snapshot of the system, while a time-dependent stellar model must be employed to properly 
assess the stellar spin evolution \citep{Gallet:2019aa,Amard:2023aa}.

This overall picture is consistent with recent results by
\citet{Pittman:2025aa}. Using a completely different method to
determine the position of the truncation radius, based on measurements
of the H$\alpha$ line flux \citep{Pittman:2025ab}, and a torque
parametrization taken from  \citet{Zhu:2025aa}, they also find that most of 
the stars in their sample, around $80\%$, are in an unstable/spin-up regime.  
Their Fig. 4 is clearly analogous to our Fig. \ref{fig_spol} but,
while our $R_\mathrm{t}$ versus $R_\mathrm{co}$ distribution reflects the
weak dependence of the truncation radius with the accretion parameter
$\Upsilon_\mathrm{acc}$ and the strong one with the stellar rotation
found in Sect. \ref{sec_rtrunc}, which brings our $R_\mathrm{t}$
estimates close to $R_\mathrm{co}$, their points are more
scattered, especially for small $R_\mathrm{t}/R_\mathrm{co}$
ratios. This difference could be ascribed to two factors. First, since
we pointed out in Sect. \ref{sec_rtrunc} that the dependence of the
truncation radius with the accretion parameter $\Upsilon_\mathrm{acc}$
tends to steepen for small $R_\mathrm{t}/R_\mathrm{co}$ values, our
best-fit parametrization likely provides only an upper limit to
$R_\mathrm{t}/R_\star$ for small $R_\mathrm{t}/R_\mathrm{co} \lesssim 0.6$ 
ratios outside the parameter space investigated in this work; see the arrows in Fig. \ref{fig_spol}. 
Second, our definition of $R_\mathrm{t}$ via Eq. (\ref{eq_rt_avg})
provides an average between the maximum and minimum radial extent of
the magnetospheric cavity while the estimates of $R_\mathrm{t}$ given
in \citet{Pittman:2025aa} are obtained from synthetic
H$\alpha$ line profiles based on an axisymmetric accretion model
assuming that the complex accretion flows arising from an unstable
regime primarily affects the width of the H$\alpha$ lines. If most of the line flux is
emitted by the inner accretion columns formed by the unstable tongues truncated
around the minimum radial extent of the stellar cavity 
(see our Fig. \ref{fig_rho_2d}), the method used by \citet{Pittman:2025ab},
which relies on the H$\alpha$ flux produced by an axisymmetric funnel flow
emerging from a single $R_\mathrm{t}$, likely yields a
smaller value for the disk truncation radius compared to our
estimates. Indeed, while the \citet{Pittman:2025ab} stellar sample is
characterized by a median value for the truncation radius of 2.8 stellar 
radii, the truncation radii that we obtain by applying our $R_\mathrm{t}$ 
parametrization to our observational sample range from 3 to 12 stellar radii, 
with a median of $5.13 \, R_\star$.   

In Fig. \ref{fig_spol} we also show, for a subsample of stars observed
with the VLTI/GRAVITY interferometer \citep{Gravity:2017aa}, the
estimated size of the Br$\gamma$ emitting region, shown as empty
circles connected to our corresponding estimate of the truncation
radius. Although we are limited to a small number of observations, no
clear correlation appears between these two quantities, other than the
fact that they differ by less than a factor of two. For the targets
exhibiting a smaller Br$\gamma$ emitting region than our estimated
$R_\mathrm{t}$, this could be due to the fact that most of the
Br$\gamma$ flux from the stellar magnetosphere is expected to be
produced in a region extending from  0.5 to 0.9 $R_\mathrm{t}$
\citep{Tessore:2023aa}. In contrast, for the stars showing a larger
Br$\gamma$ region, winds and outflows are also expected to contribute
to the Br$\gamma$ flux, thus extending this region beyond
$R_\mathrm{t}$ \citep{Gravity:2023ab}.
   
\section{Conclusions}
\label{sec_concl}

This study was based on 3D MHD time-dependent numerical
simulations of a rotating stellar dipolar magnetosphere interacting with an accretion disk. 
The disk was modeled as viscous and resistive using an $\alpha$ prescription \citep{Shakura:1973aa}.
Our study includes 21 cases that span a broad parameter space varying the stellar rotation rate, the dipole strength and obliquity, and the disk density.
First, we characterized the stable and unstable accretion regimes
  (distinguished by whether an interchange instability develops at the magnetospheric boundary)
  to investigate how disk truncation occurs in the two cases.
  We also provided a parameterization for the disk truncation radius that encompasses both stable and unstable 
configurations. Then we examined the torques exerted onto the star by the flow components 
simulated in our models in different accretion regimes, providing handy parametrizations depending on 
observable stellar parameters. Finally, we tested our results by applying our findings 
to a sample of CTTs observed in spectropolarimetry and interferometry. The following points summarize the results of this work:

\begin{enumerate}

\item By estimating the amplitude of the interchange instability, based on a
      characterization of the geometry of the magnetospheric cavity, we found that stable
      accretion occurs for $R_\mathrm{t} > 0.80 - 0.85 R_\mathrm{co}$ and becomes unstable
      otherwise, with the accretion flow fragmenting into different accretion streams 
      forming multiple columns impacting the stellar photosphere. For $R_\mathrm{t}\lesssim 0.7
      R_\mathrm{co}$ the unstable accretion tongues become fewer and more azimuthally 
      extended, suggesting the transition to an ordered unstable regime, in line
      with the \citet{Blinova:2016aa} results.

\item Our definition of the truncation radius, based on the geometrical shape of the magnetospheric
      cavity, exhibits a weak dependence on the accretion parameter, $\Upsilon_\mathrm{acc} \propto
      B_\star^2/\dot{M}_\mathrm{acc}$, and a strong correlation with the stellar rotation rate and the position of the corotation radius. 
      Our scaling is very different from the ``classical'' \citet{Ghosh:1979aa} parametrization. 
      In stable cases, this behavior could be due to a weak propeller regime that tends
      to keep the truncation radius oscillating around corotation. 
      In the unstable regime, the size of the magnetospheric cavity is mainly
      determined by the onset of the interchange instability. Furthermore, in the cases
      that we simulated in this study, the instability starts to fragment the disk into multiple accretion streams 
      close to corotation before reaching the \citeauthor{Ghosh:1979aa} radius.     
      Our best-fit expression is limited to the parameter space that we explored; however, it does cover the majority of the CTTs sample observed in spectropolarimetry.
      For small ratios of $R_\mathrm{t}/R_\mathrm{co} \lesssim 0.6$, which is below the parameter space investigated in this paper, our scaling 
      Eq. (\ref{eq_rt_yaccf}) likely provides an upper limit to the $R_\mathrm{t}/R_\star$ value, particularly in the ordered unstable regime.
        
\item The combination of our truncation radius parametrization and interchange instability 
      threshold predicts that even for a relatively strong kG stellar dipolar field, 
      only weak accretors with $\dot{M}_\mathrm{acc} \lesssim 10^{-9}$ can be in a stable 
      regime, as indicated by Eqs. (\ref{eq_stab_f}) and (\ref{eq_stab_p}).
        
\item We analyzed the torques associated with the circumstellar flows, which in our picture 
      directly rely on the star-disk interaction, magnetospheric accretion, and
      magnetospheric ejections. In unstable regimes, the stellar torque is mainly determined 
      by the spin-up due to accretion, which can become negligible or even negative (spin-down)
      in stable regimes. We show that this counterintuitive behavior could be due to the fact
      that in a stable regime, the magnetic field lines channeling the inflowing gas are
      connected to the disk near the corotation radius, where the disk rotation
      can become slower than stellar; as a result, Laplace forces can transfer angular momentum 
      from the star to the disk. In our picture, this excess angular momentum acquired by the disk
      can be ejected by magnetospheric ejections and conical winds in the hemisphere that is opposite with respect to the
      accretion funnel. In addition, since the magnetospheric cavity tends to extend up to the
      corotation radius, both in the stable and unstable cases (see Fig. \ref{fig_rho_2d}), 
      magnetospheric ejections can be typically launched from the disk beyond $R_\mathrm{co}$ to extract angular momentum both from the star and the disk. This becomes particularly
      relevant in stable and propeller regimes.
     
\item By inspecting the total star-disk interaction torque, given by the sum of the accretion 
      and the MEs torques, and expressing it as a function of the relevant stellar parameters, 
      we found that stable (unstable) regimes correspond quite precisely to spin-down (spin-up) 
      SDI torque configurations. On the other hand, in stable regimes the SDI torque timescale can become comparable to the stellar contraction timescale, thereby favoring stellar 
      spin-equilibrium; whereas in an unstable regime, the spin-up timescale due to accretion is significantly
      shorter, leading to a rapid increase in the stellar rotation rate. 
      
\item We therefore investigated and parametrized the stellar torque exerted by stellar winds.
      We confirmed the scaling of the SW specific angular momentum with the open magnetic stellar
      flux \citep{Reville:2015aa}. In accreting systems, this is mainly determined by the position
      of the truncation radius, confirming the results from 
      \citet{Pantolmos:2020aa,Ireland:2021aa,Ireland:2022aa}. We found that a stellar wind with
      a mass ejection efficiency of $\dot{M}_\mathrm{SW}/\dot{M}_\mathrm{acc} = 0.01-0.1$ in stable
      regimes can exert a spin-down torque comparable or even stronger than the SDI torque, further
      counteracting the stellar spin-up caused by contraction. In contrast, in unstable regimes the
      SW torque does not seem to be efficient enough to hinder the combined effects
      of strong accretion torques and contraction. 

\item To test our findings, we applied our truncation radius and stellar torque expressions
      to a sample of CTTs-Class II objects observed in spectropolarimetry to obtain
      an estimate of the stellar dipolar field strength. We found that about $20\%$ of 
      stars in our sample are expected to be in a stable accretion regime.
      Around $25\%$ of the sample is subject to a total spin-down torque, taking into account 
      the sum of SDI and SW torques, showing a relatively good correspondence between stable regimes 
      and spin-down torques, the slightly higher percentage with respect to the stable cases 
      due to the inclusion of the SW torque (whose spin-down effect is not correlated with the
      accretion regime). On the other hand, only one star in the sample, AA Tau, is predicted to
      experience a spin-down torque that can balance the spin-up due to stellar contraction. 
      All the other stars are either subject to a spin-up torque or the spin-down torque is
      insufficient to counteract contraction. As a consequence, the angular momentum evolution of
      young accreting stars still remains an open issue.       
        
\end{enumerate}

Certainly, our conclusions come with a range of limitations. As already stated, our estimate of the  
spin-up and spin-down timescales is based on a snapshot of the system, assuming that the star
rotates as a solid body, while the long-term stellar angular momentum evolution can be  
properly assessed only employing state-of-the-art stellar evolution models \citep{Amard:2023aa}.
Just as an example, the radiative core-convective envelope decoupling \citep{Gallet:2013aa} can
considerably change the moment of inertia of the part of the star that is directly affected
by the external torques, thereby modifying the associated timescales. 
In addition, our study is based on computationally cheaper laminar ``alpha'' disk models,
which allowed us to improve the statistics of our results, but we had to make specific choices, 
albeit physically motivated, about the value of the $\alpha_\mathrm{v}$ and $\alpha_\mathrm{m}$ coefficients parameterizing the turbulence-driven anomalous viscosity and magnetic resistivity. 
Numerical experiments based on MRI-active disks \citep{Takasao:2022aa,Zhu:2025aa} self-consistently
take care of these effects, but we did not notice any major qualitative difference
with the outcome of our models, at least concerning the aspects we focus on here.
On the other hand, as of today, both kinds of numerical models do not factor in the role played by
an intrinsic large-scale disk magnetic field interacting with the stellar magnetosphere. Besides
possibly modifying the global magnetic topology of the star-disk interaction
\citep{Ferreira:2000aa,Ferreira:2006aa}, a disk magnetic field could mitigate the steep radial decay 
of the magnetospheric field and strengthen the magnetic coupling of the star with the disk
and its environment.

\begin{acknowledgements}
  The authors thank G. Lesur for assistance with the use of the HPC
  centers utilized in this work. GP further thanks N. Vlahakis for the
  valuable discussions during the development of the script. 
  This project has received funding from the European
  Research Council (ERC) under the European Union’s Horizon 2020
  research and innovation programme (grant agreement No 742095; {\it
    SPIDI}: Star-Planets-Inner Disk-Interactions);
  https://www.spidi-eu.org/. GP acknowledges support by the action
  ``Visiting Professors/Visiting Researchers'' through the National
  Recovery and Resilience Plan, ``Greece 2.0'', funded by the European
  Union. CZ acknowledges financial support from the Large Grant INAF-2024 ``Spectral 
  Key features of Young stellar objects: Wind-Accretion LinKs Explored in the 
  infraRed (SKYWALKER)''. This work was performed using HPC resources 
  from GENCI–TGCC (Grants [2022-A0110412953] and [2023-A0140414183])
  and the Dahu supercomputer of the GRICAD infrastructure
  (https://gricad.univ-grenoble-alpes.fr) that is supported by
  Grenoble research communities.
\end{acknowledgements}

%-------------------------------------------------------------------

\bibliographystyle{aa}
\bibliography{3Dsdi_torques}

@article{Johnstone:2014ab,
  author =        {{Johnstone}, C.~P. and {Jardine}, M. and
                   {Gregory}, S.~G. and {Donati}, J.-F. and
                   {Hussain}, G.},
  journal =       {\mnras},
  month =         feb,
  pages =         {3202-3220},
  title =         {{Classical T Tauri stars: magnetic fields, coronae
                   and star-disc interactions}},
  volume =        {437},
  year =          {2014},
  doi =           {10.1093/mnras/stt2107},
}

@article{Hartmann:2016aa,
  author =        {{Hartmann}, L. and {Herczeg}, G. and {Calvet}, N.},
  journal =       {\araa},
  month =         sep,
  pages =         {135-180},
  title =         {{Accretion onto Pre-Main-Sequence Stars}},
  volume =        {54},
  year =          {2016},
  doi =           {10.1146/annurev-astro-081915-023347},
}

@article{Romanova:2015aa,
  author =        {{Romanova}, M.~M. and {Owocki}, S.~P.},
  journal =       {\ssr},
  month =         oct,
  pages =         {339-389},
  title =         {{Accretion, Outflows, and Winds of Magnetized Stars}},
  volume =        {191},
  year =          {2015},
  doi =           {10.1007/s11214-015-0200-9},
}

@article{Calvet:1998aa,
  author =        {{Calvet}, N. and {Gullbring}, E.},
  journal =       {\apj},
  month =         dec,
  pages =         {802-818},
  title =         {{The Structure and Emission of the Accretion Shock in
                   T Tauri Stars}},
  volume =        {509},
  year =          {1998},
  doi =           {10.1086/306527},
}

@article{Gullbring:2000aa,
  author =        {{Gullbring}, E. and {Calvet}, N. and {Muzerolle}, J. and
                   {Hartmann}, L.},
  journal =       {\apj},
  month =         dec,
  pages =         {927-932},
  title =         {{The Structure and Emission of the Accretion Shock in
                   T Tauri Stars. II. The Ultraviolet-Continuum
                   Emission}},
  volume =        {544},
  year =          {2000},
  doi =           {10.1086/317253},
}

@article{Hartmann:1994aa,
  author =        {{Hartmann}, Lee and {Hewett}, Robert and
                   {Calvet}, Nuria},
  journal =       {\apj},
  month =         may,
  pages =         {669},
  title =         {{Magnetospheric Accretion Models for T Tauri Stars.
                   I. Balmer Line Profiles without Rotation}},
  volume =        {426},
  year =          {1994},
  doi =           {10.1086/174104},
}

@article{Muzerolle:2001aa,
  author =        {{Muzerolle}, J. and {Calvet}, N. and {Hartmann}, L.},
  journal =       {\apj},
  month =         apr,
  pages =         {944-961},
  title =         {{Emission-Line Diagnostics of T Tauri Magnetospheric
                   Accretion. II. Improved Model Tests and Insights into
                   Accretion Physics}},
  volume =        {550},
  year =          {2001},
  doi =           {10.1086/319779},
}

@article{Edwards:2006aa,
  author =        {{Edwards}, S. and {Fischer}, W. and {Hillenbrand}, L. and
                   {Kwan}, J.},
  journal =       {\apj},
  month =         jul,
  pages =         {319-341},
  title =         {{Probing T Tauri Accretion and Outflow with 1 Micron
                   Spectroscopy}},
  volume =        {646},
  year =          {2006},
  doi =           {10.1086/504832},
}

@article{Edwards:2013aa,
  author =        {{Edwards}, Suzan and {Kwan}, John and
                   {Fischer}, William and {Hillenbrand}, Lynne and
                   {Finn}, Kimberly and {Fedorenko}, Kristina and
                   {Feng}, Wanda},
  journal =       {\apj},
  month =         dec,
  number =        {2},
  pages =         {148},
  title =         {{Interpreting Near-infrared Hydrogen Line Ratios in T
                   Tauri Stars}},
  volume =        {778},
  year =          {2013},
  doi =           {10.1088/0004-637X/778/2/148},
  eid =           {148},
}

@article{Sousa:2021aa,
  author =        {{Sousa}, A.~P. and {Bouvier}, J. and
                   {Alencar}, S.~H.~P. and {Donati}, J. -F. and
                   {Alecian}, E. and {Roquette}, J. and {Perraut}, K. and
                   {Dougados}, C. and {Carmona}, A. and {Covino}, S. and
                   {Fugazza}, D. and {Molinari}, E. and {Moutou}, C. and
                   {Santerne}, A. and {Grankin}, K. and
                   {Artigau}, {\'E}. and {Delfosse}, X. and
                   {Hebrard}, G. and {SPIRou Consortium}},
  journal =       {\aap},
  month =         may,
  pages =         {A68},
  title =         {{Star-disk interaction in the T Tauri star V2129
                   Ophiuchi: An evolving accretion-ejection structure}},
  volume =        {649},
  year =          {2021},
  doi =           {10.1051/0004-6361/202140346},
  eid =           {A68},
}

@article{Gravity:2023ab,
  author =        {{GRAVITY Collaboration} and {Wojtczak}, J.~A. and
                   {Labadie}, L. and {Perraut}, K. and {Tessore}, B. and
                   {Soulain}, A. and {Ganci}, V. and {Bouvier}, J. and
                   {Dougados}, C. and {Al{\'e}cian}, E. and
                   {Nowacki}, H. and {Cozzo}, G. and {Brandner}, W. and
                   {Caratti O Garatti}, A. and {Garcia}, P. and
                   {Garcia Lopez}, R. and {Sanchez-Bermudez}, J. and
                   {Amorim}, A. and {Benisty}, M. and {Berger}, J. -P. and
                   {Bourdarot}, G. and {Caselli}, P. and
                   {Cl{\'e}net}, Y. and {de Zeeuw}, P.~T. and
                   {Davies}, R. and {Drescher}, A. and {Duvert}, G. and
                   {Eckart}, A. and {Eisenhauer}, F. and {Eupen}, F. and
                   {F{\"o}rster-Schreiber}, N.~M. and {Gendron}, E. and
                   {Gillessen}, S. and {Grant}, S. and {Grellmann}, R. and
                   {Hei{\ss}el}, G. and {Henning}, Th. and {Hippler}, S. and
                   {Horrobin}, M. and {Hubert}, Z. and {Jocou}, L. and
                   {Kervella}, P. and {Lacour}, S. and
                   {Lapeyr{\`e}re}, V. and {Le Bouquin}, J. -B. and
                   {L{\'e}na}, P. and {Lutz}, D. and {Mang}, F. and
                   {Ott}, T. and {Paumard}, T. and {Perrin}, G. and
                   {Scheithauer}, S. and {Shangguan}, J. and
                   {Shimizu}, T. and {Spezzano}, S. and {Straub}, O. and
                   {Straubmeier}, C. and {Sturm}, E. and
                   {van Dishoeck}, E. and {Vincent}, F. and
                   {Widmann}, F.},
  journal =       {\aap},
  month =         jan,
  pages =         {A59},
  title =         {{The GRAVITY young stellar object survey. IX.
                   Spatially resolved kinematics of hot hydrogen gas in
                   the star-disk interaction region of T Tauri stars}},
  volume =        {669},
  year =          {2023},
  doi =           {10.1051/0004-6361/202244675},
  eid =           {A59},
}

@article{Herbst:2007aa,
  author =        {{Herbst}, W. and {Eisl{\"o}ffel}, J. and {Mundt}, R. and
                   {Scholz}, A.},
  journal =       {Protostars and Planets V},
  pages =         {297-311},
  title =         {{The Rotation of Young Low-Mass Stars and Brown
                   Dwarfs}},
  year =          {2007},
}

@article{Bouvier:2014aa,
  author =        {{Bouvier}, J. and {Matt}, S.~P. and {Mohanty}, S. and
                   {Scholz}, A. and {Stassun}, K.~G. and {Zanni}, C.},
  journal =       {Protostars and Planets VI},
  month =         {September},
  pages =         {433-450},
  title =         {{Angular Momentum Evolution of Young Low-Mass Stars
                   and Brown Dwarfs: Observations and Theory}},
  year =          {2014},
  doi =           {10.2458/azu_uapress_9780816531240-ch019},
}

@article{Gallet:2013aa,
  author =        {{Gallet}, F. and {Bouvier}, J.},
  journal =       {\aap},
  month =         aug,
  pages =         {A36},
  title =         {{Improved angular momentum evolution model for
                   solar-like stars}},
  volume =        {556},
  year =          {2013},
  doi =           {10.1051/0004-6361/201321302},
  eid =           {A36},
}

@article{Smith:2023aa,
  author =        {{Smith}, Gareth D. and {Gillen}, Edward and
                   {Hodgkin}, Simon T. and {Alves}, Douglas R. and
                   {Anderson}, David R. and {Battley}, Matthew P. and
                   {Burleigh}, Matthew R. and {Casewell}, Sarah L. and
                   {Gill}, Samuel and {Goad}, Michael R. and
                   {Henderson}, Beth A. and {Jenkins}, James S. and
                   {Kendall}, Alicia and {Moyano}, Maximiliano and
                   {Ramsay}, Gavin and {Tilbrook}, Rosanna H. and
                   {Vines}, Jose I. and {West}, Richard G. and
                   {Wheatley}, Peter J.},
  journal =       {\mnras},
  month =         jul,
  number =        {1},
  pages =         {169-188},
  title =         {{NGTS clusters survey - V. Rotation in the Orion
                   star-forming complex}},
  volume =        {523},
  year =          {2023},
  doi =           {10.1093/mnras/stad1435},
}

@article{Koenigl:1991aa,
  author =        {{Koenigl}, A.},
  journal =       {\apjl},
  month =         mar,
  pages =         {L39-L43},
  title =         {{Disk accretion onto magnetic T Tauri stars}},
  volume =        {370},
  year =          {1991},
  doi =           {10.1086/185972},
}

@article{Collier-Cameron:1993aa,
  author =        {{Collier Cameron}, A. and {Campbell}, C.~G.},
  journal =       {\aap},
  month =         jul,
  pages =         {309},
  title =         {{Rotational evolution of magnetic T Tauri stars with
                   accretion discs}},
  volume =        {274},
  year =          {1993},
}

@article{Armitage:1996aa,
  author =        {{Armitage}, P.~J. and {Clarke}, C.~J.},
  journal =       {\mnras},
  month =         may,
  pages =         {458-468},
  title =         {{Magnetic braking of T Tauri stars}},
  volume =        {280},
  year =          {1996},
  doi =           {10.1093/mnras/280.2.458},
}

@article{Agapitou:2000aa,
  author =        {{Agapitou}, Vasso and {Papaloizou}, John C.~B.},
  journal =       {\mnras},
  month =         {Sep},
  number =        {2},
  pages =         {273-288},
  title =         {{Accretion disc-stellar magnetosphere interaction:
                   field line inflation and the effect on the spin-down
                   torque}},
  volume =        {317},
  year =          {2000},
  doi =           {10.1046/j.1365-8711.2000.03541.x},
}

@article{Uzdensky:2002ab,
  author =        {{Uzdensky}, D.~A. and {K{\"o}nigl}, A. and
                   {Litwin}, C.},
  journal =       {\apj},
  month =         feb,
  pages =         {1205-1215},
  title =         {{Magnetically Linked Star-Disk Systems. II. Effects
                   of Plasma Inertia and Reconnection in the
                   Magnetosphere}},
  volume =        {565},
  year =          {2002},
  doi =           {10.1086/324724},
}

@article{Matt:2005ab,
  author =        {{Matt}, S. and {Pudritz}, R.~E.},
  journal =       {\mnras},
  month =         jan,
  pages =         {167-182},
  title =         {{The spin of accreting stars: dependence on magnetic
                   coupling to the disc}},
  volume =        {356},
  year =          {2005},
  doi =           {10.1111/j.1365-2966.2004.08431.x},
}

@article{Zanni:2009aa,
  author =        {{Zanni}, C. and {Ferreira}, J.},
  journal =       {\aap},
  month =         dec,
  pages =         {1117-1133},
  title =         {{MHD simulations of accretion onto a dipolar
                   magnetosphere. I. Accretion curtains and the
                   disk-locking paradigm}},
  volume =        {508},
  year =          {2009},
  doi =           {10.1051/0004-6361/200912879},
}

@article{Romanova:2009ab,
  author =        {{Romanova}, M.~M. and {Ustyugova}, G.~V. and
                   {Koldoba}, A.~V. and {Lovelace}, R.~V.~E.},
  journal =       {\mnras},
  month =         nov,
  number =        {4},
  pages =         {1802-1828},
  title =         {{Launching of conical winds and axial jets from the
                   disc-magnetosphere boundary: axisymmetric and 3D
                   simulations}},
  volume =        {399},
  year =          {2009},
  doi =           {10.1111/j.1365-2966.2009.15413.x},
}

@article{Takasao:2022aa,
  author =        {{Takasao}, Shinsuke and {Tomida}, Kengo and
                   {Iwasaki}, Kazunari and {Suzuki}, Takeru K.},
  journal =       {\apj},
  month =         dec,
  number =        {1},
  pages =         {73},
  title =         {{Three-dimensional Simulations of Magnetospheric
                   Accretion in a T Tauri Star: Accretion and Wind
                   Structures Just Around the Star}},
  volume =        {941},
  year =          {2022},
  doi =           {10.3847/1538-4357/ac9eb1},
  eid =           {73},
}

@article{Shu:1994aa,
  author =        {{Shu}, F. and {Najita}, J. and {Ostriker}, E. and
                   {Wilkin}, F. and {Ruden}, S. and {Lizano}, S.},
  journal =       {\apj},
  month =         jul,
  pages =         {781-796},
  title =         {{Magnetocentrifugally driven flows from young stars
                   and disks. 1: A generalized model}},
  volume =        {429},
  year =          {1994},
  doi =           {10.1086/174363},
}

@article{Matt:2005aa,
  author =        {{Matt}, S. and {Pudritz}, R.~E.},
  journal =       {\apjl},
  month =         oct,
  pages =         {L135-L138},
  title =         {{Accretion-powered Stellar Winds as a Solution to the
                   Stellar Angular Momentum Problem}},
  volume =        {632},
  year =          {2005},
  doi =           {10.1086/498066},
}

@article{Zanni:2013aa,
  author =        {{Zanni}, C. and {Ferreira}, J.},
  journal =       {\aap},
  month =         feb,
  pages =         {A99},
  title =         {{MHD simulations of accretion onto a dipolar
                   magnetosphere. II. Magnetospheric ejections and
                   stellar spin-down}},
  volume =        {550},
  year =          {2013},
  doi =           {10.1051/0004-6361/201220168},
  eid =           {A99},
}

@article{Ferreira:2000aa,
  author =        {{Ferreira}, J. and {Pelletier}, G. and {Appl}, S.},
  journal =       {\mnras},
  month =         feb,
  pages =         {387-397},
  title =         {{Reconnection X-winds: spin-down of low-mass
                   protostars}},
  volume =        {312},
  year =          {2000},
  doi =           {10.1046/j.1365-8711.2000.03215.x},
}

@article{Robinson:2021aa,
  author =        {{Robinson}, Connor E. and {Espaillat}, Catherine C. and
                   {Owen}, James E.},
  journal =       {\apj},
  month =         feb,
  number =        {1},
  pages =         {16},
  title =         {{Synthetic Light Curves of Accretion Variability in T
                   Tauri Stars}},
  volume =        {908},
  year =          {2021},
  doi =           {10.3847/1538-4357/abd410},
  eid =           {16},
}

@article{Tessore:2023aa,
  author =        {{Tessore}, B. and {Soulain}, A. and {Pantolmos}, G. and
                   {Bouvier}, J. and {Pinte}, C. and {Perraut}, K.},
  journal =       {\aap},
  month =         mar,
  pages =         {A129},
  title =         {{Spectroscopic and interferometric signatures of
                   magnetospheric accretion in young stars}},
  volume =        {671},
  year =          {2023},
  doi =           {10.1051/0004-6361/202245039},
  eid =           {A129},
}

@article{Shakura:1973aa,
  author =        {{Shakura}, N.~I. and {Sunyaev}, R.~A.},
  journal =       {\aap},
  month =         {Jun},
  pages =         {33-51},
  title =         {{Reprint of 1973A\&amp;A....24..337S. Black holes in
                   binary systems. Observational appearance.}},
  volume =        {500},
  year =          {1973},
}

@article{Romanova:2008aa,
  author =        {{Romanova}, Marina M. and {Kulkarni}, Akshay K. and
                   {Lovelace}, Richard V.~E.},
  journal =       {\apjl},
  month =         feb,
  number =        {2},
  pages =         {L171},
  title =         {{Unstable Disk Accretion onto Magnetized Stars: First
                   Global Three-dimensional Magnetohydrodynamic
                   Simulations}},
  volume =        {673},
  year =          {2008},
  doi =           {10.1086/527298},
}

@article{Kulkarni:2008aa,
  author =        {{Kulkarni}, A.~K. and {Romanova}, M.~M.},
  journal =       {\mnras},
  month =         may,
  pages =         {673-687},
  title =         {{Accretion to magnetized stars through the
                   Rayleigh-Taylor instability: global 3D simulations}},
  volume =        {386},
  year =          {2008},
  doi =           {10.1111/j.1365-2966.2008.13094.x},
}

@article{Blinova:2016aa,
  author =        {{Blinova}, A.~A. and {Romanova}, M.~M. and
                   {Lovelace}, R.~V.~E.},
  journal =       {\mnras},
  month =         jul,
  pages =         {2354-2369},
  title =         {{Boundary between stable and unstable regimes of
                   accretion. Ordered and chaotic unstable regimes}},
  volume =        {459},
  year =          {2016},
  doi =           {10.1093/mnras/stw786},
}

@article{Romanova:2012vx,
  author =        {{Romanova}, M.~M. and {Ustyugova}, G.~V. and
                   {Koldoba}, A.~V. and {Lovelace}, R.~V.~E.},
  journal =       {\mnras},
  month =         mar,
  number =        {1},
  pages =         {63-77},
  title =         {{MRI-driven accretion on to magnetized stars: global
                   3D MHD simulations of magnetospheric and boundary
                   layer regimes}},
  volume =        {421},
  year =          {2012},
  doi =           {10.1111/j.1365-2966.2011.20055.x},
}

@article{Zhu:2024aa,
  author =        {{Zhu}, Zhaohuan and {Stone}, James M. and
                   {Calvet}, Nuria},
  journal =       {\mnras},
  month =         feb,
  number =        {2},
  pages =         {2883-2911},
  title =         {{A global 3D simulation of magnetospheric accretion -
                   I. Magnetically disrupted discs and surface
                   accretion}},
  volume =        {528},
  year =          {2024},
  doi =           {10.1093/mnras/stad3712},
}

@article{Zhu:2025aa,
  author =        {{Zhu}, Zhaohuan},
  journal =       {\mnras},
  month =         mar,
  number =        {4},
  pages =         {3701-3729},
  title =         {{Global 3D simulations of magnetospheric accretion -
                   II. Hotspots, equilibrium torque, episodic wind, and
                   mid-plane outflow}},
  volume =        {537},
  year =          {2025},
  doi =           {10.1093/mnras/staf250},
}

@article{Kulkarni:2009aa,
  author =        {{Kulkarni}, A.~K. and {Romanova}, M.~M.},
  journal =       {\mnras},
  month =         sep,
  number =        {2},
  pages =         {701-714},
  title =         {{Possible quasi-periodic oscillations from unstable
                   accretion: 3D magnetohydrodynamic simulations}},
  volume =        {398},
  year =          {2009},
  doi =           {10.1111/j.1365-2966.2009.15186.x},
}

@article{Kurosawa:2013aa,
  author =        {{Kurosawa}, Ryuichi and {Romanova}, M.~M.},
  journal =       {\mnras},
  month =         {May},
  number =        {3},
  pages =         {2673-2689},
  title =         {{Spectral variability of classical T Tauri stars
                   accreting in an unstable regime}},
  volume =        {431},
  year =          {2013},
  doi =           {10.1093/mnras/stt365},
}

@article{Armeni:2023aa,
  author =        {{Armeni}, A. and {Stelzer}, B. and {Claes}, R.~A.~B. and
                   {Manara}, C.~F. and {Frasca}, A. and
                   {Alcal{\'a}}, J.~M. and {Walter}, F.~M. and
                   {K{\'o}sp{\'a}l}, {\'A}. and {Campbell-White}, J. and
                   {Gangi}, M. and {Mauco}, K. and {Tychoniec}, L.},
  journal =       {\aap},
  month =         nov,
  pages =         {A14},
  title =         {{PENELLOPE. V. The magnetospheric structure and the
                   accretion variability of the classical T Tauri star
                   HM Lup}},
  volume =        {679},
  year =          {2023},
  doi =           {10.1051/0004-6361/202347051},
  eid =           {A14},
}

@article{Armeni:2024aa,
  author =        {{Armeni}, A. and {Stelzer}, B. and {Frasca}, A. and
                   {Manara}, C.~F. and {Walter}, F.~M. and
                   {Alcal{\'a}}, J.~M. and {Schneider}, P.~C. and
                   {Sicilia-Aguilar}, A. and {Campbell-White}, J. and
                   {Fiorellino}, E. and {Gameiro}, J.~F. and
                   {Gangi}, M.},
  journal =       {\aap},
  month =         oct,
  pages =         {A225},
  title =         {{Evidence for magnetic boundary layer accretion in RU
                   Lup: A spectrophotometric analysis}},
  volume =        {690},
  year =          {2024},
  doi =           {10.1051/0004-6361/202451065},
  eid =           {A225},
}

@article{Romanova:2025aa,
  author =        {{Romanova}, M.~M. and {Espaillat}, C.~C. and
                   {Wendeborn}, J. and {Donati}, J.-F. and
                   {Petrov}, P.~P. and {Lovelace}, R.~V.~E.},
  journal =       {\mnras},
  month =         mar,
  number =        {1},
  pages =         {480-502},
  title =         {{Unstable accretion in TW Hya: 3D simulations and
                   comparisons with observations}},
  volume =        {538},
  year =          {2025},
  doi =           {10.1093/mnras/staf148},
}

@article{Ghosh:1979ab,
  author =        {{Ghosh}, P. and {Lamb}, F.~K.},
  journal =       {\apj},
  month =         nov,
  pages =         {296-316},
  title =         {{Accretion by rotating magnetic neutron stars. III.
                   Accretion torques and period changes in pulsating
                   X-ray sources.}},
  volume =        {234},
  year =          {1979},
  doi =           {10.1086/157498},
}

@article{Ostriker:1995aa,
  author =        {{Ostriker}, Eve C. and {Shu}, Frank H.},
  journal =       {\apj},
  month =         jul,
  pages =         {813},
  title =         {{Magnetocentrifugally Driven Flows from Young Stars
                   and Disks. IV. The Accretion Funnel and Dead Zone}},
  volume =        {447},
  year =          {1995},
  doi =           {10.1086/175920},
}

@article{Long:2005aa,
  author =        {{Long}, M. and {Romanova}, M.~M. and
                   {Lovelace}, R.~V.~E.},
  journal =       {\apj},
  month =         dec,
  pages =         {1214-1222},
  title =         {{Locking of the Rotation of Disk-Accreting Magnetized
                   Stars}},
  volume =        {634},
  year =          {2005},
  doi =           {10.1086/497000},
}

@article{Pantolmos:2020aa,
  author =        {{Pantolmos}, G. and {Zanni}, C. and {Bouvier}, J.},
  journal =       {\aap},
  month =         nov,
  pages =         {A129},
  title =         {{Magnetic torques on T Tauri stars: Accreting versus
                   non-accreting systems}},
  volume =        {643},
  year =          {2020},
  doi =           {10.1051/0004-6361/202038569},
  eid =           {A129},
}

@article{Mignone:2007aa,
  author =        {{Mignone}, A. and {Bodo}, G. and {Massaglia}, S. and
                   {Matsakos}, T. and {Tesileanu}, O. and {Zanni}, C. and
                   {Ferrari}, A.},
  journal =       {\apjs},
  month =         may,
  pages =         {228-242},
  title =         {{PLUTO: A Numerical Code for Computational
                   Astrophysics}},
  volume =        {170},
  year =          {2007},
  doi =           {10.1086/513316},
}

@article{Miyoshi:2010aa,
  author =        {{Miyoshi}, Takahiro and {Terada}, Naoki and
                   {Matsumoto}, Yosuke and {Fukazawa}, Keiichiro and
                   {Umeda}, Takayuki and {Kusano}, Kanya},
  journal =       {IEEE Transactions on Plasma Science},
  month =         sep,
  number =        {9},
  pages =         {2236-2242},
  title =         {{The HLLD Approximate Riemann Solver for
                   Magnetospheric Simulation}},
  volume =        {38},
  year =          {2010},
  doi =           {10.1109/TPS.2010.2057451},
}

@article{Dedner:2002aa,
  author =        {{Dedner}, A. and {Kemm}, F. and {Kr{\"o}ner}, D. and
                   {Munz}, C. -D. and {Schnitzer}, T. and
                   {Wesenberg}, M.},
  journal =       {Journal of Computational Physics},
  month =         jan,
  number =        {2},
  pages =         {645-673},
  title =         {{Hyperbolic Divergence Cleaning for the MHD
                   Equations}},
  volume =        {175},
  year =          {2002},
  doi =           {10.1006/jcph.2001.6961},
}

@article{Mignone:2012ac,
  author =        {{Mignone}, A. and {Zanni}, C. and {Tzeferacos}, P. and
                   {van Straalen}, B. and {Colella}, P. and {Bodo}, G.},
  journal =       {\apjs},
  month =         jan,
  number =        {1},
  pages =         {7},
  title =         {{The PLUTO Code for Adaptive Mesh Computations in
                   Astrophysical Fluid Dynamics}},
  volume =        {198},
  year =          {2012},
  doi =           {10.1088/0067-0049/198/1/7},
  eid =           {7},
}

@article{Ferreira:1993aa,
  author =        {{Ferreira}, J. and {Pelletier}, G.},
  journal =       {\aap},
  month =         sep,
  pages =         {625},
  title =         {{Magnetized accretion-ejection structures. 1. General
                   statements}},
  volume =        {276},
  year =          {1993},
}

@article{Reville:2015aa,
  author =        {{R{\'e}ville}, V. and {Brun}, A.~S. and
                   {Strugarek}, A. and {Matt}, S.~P. and {Bouvier}, J. and
                   {Folsom}, C.~P. and {Petit}, P.},
  journal =       {\apj},
  month =         dec,
  pages =         {99},
  title =         {{From Solar to Stellar Corona: The Role of Wind,
                   Rotation, and Magnetism}},
  volume =        {814},
  year =          {2015},
  doi =           {10.1088/0004-637X/814/2/99},
  eid =           {99},
}

@article{Rucinski:1988aa,
  author =        {{Rucinski}, S.~M.},
  journal =       {\aj},
  month =         jun,
  pages =         {1895},
  title =         {{Rotational Properties of Composite Polytrope
                   Models}},
  volume =        {95},
  year =          {1988},
  doi =           {10.1086/114784},
}

@article{Salvesen:2016aa,
  author =        {{Salvesen}, Greg and {Simon}, Jacob B. and
                   {Armitage}, Philip J. and {Begelman}, Mitchell C.},
  journal =       {\mnras},
  month =         mar,
  number =        {1},
  pages =         {857-874},
  title =         {{Accretion disc dynamo activity in local simulations
                   spanning weak-to-strong net vertical magnetic flux
                   regimes}},
  volume =        {457},
  year =          {2016},
  doi =           {10.1093/mnras/stw029},
}

@article{Stone:2007aa,
  author =        {{Stone}, James M. and {Gardiner}, Thomas},
  journal =       {\apj},
  month =         dec,
  number =        {2},
  pages =         {1726-1735},
  title =         {{The Magnetic Rayleigh-Taylor Instability in Three
                   Dimensions}},
  volume =        {671},
  year =          {2007},
  doi =           {10.1086/523099},
}

@article{Carlyle:2017aa,
  author =        {{Carlyle}, Jack and {Hillier}, Andrew},
  journal =       {\aap},
  month =         sep,
  pages =         {A101},
  title =         {{The non-linear growth of the magnetic
                   Rayleigh-Taylor instability}},
  volume =        {605},
  year =          {2017},
  doi =           {10.1051/0004-6361/201730802},
  eid =           {A101},
}

@article{Arons:1976aa,
  author =        {{Arons}, J. and {Lea}, S.~M.},
  journal =       {\apj},
  month =         aug,
  pages =         {914-936},
  title =         {{Accretion onto magnetized neutron stars: structure
                   and interchange instability of a model
                   magnetosphere.}},
  volume =        {207},
  year =          {1976},
  doi =           {10.1086/154562},
}

@article{Wang:1985aa,
  author =        {{Wang}, Y. -M. and {Robertson}, J.~A.},
  journal =       {\apj},
  month =         dec,
  pages =         {85-108},
  title =         {{Late stages of the Rayleigh-Taylor instability - A
                   numerical study in the context of accreting neutron
                   stars}},
  volume =        {299},
  year =          {1985},
  doi =           {10.1086/163684},
}

@article{Spruit:1995aa,
  author =        {{Spruit}, H.~C. and {Stehle}, R. and
                   {Papaloizou}, J.~C.~B.},
  journal =       {\mnras},
  month =         aug,
  pages =         {1223-1231},
  title =         {{Interchange instability in and accretion disc with a
                   poloidal magnetic field}},
  volume =        {275},
  year =          {1995},
  doi =           {10.1093/mnras/275.4.1223},
}

@article{Kluzniak:2007aa,
  author =        {{Klu{\'z}niak}, W. and {Rappaport}, S.},
  journal =       {\apj},
  month =         {Dec},
  number =        {2},
  pages =         {1990-2005},
  title =         {{Magnetically Torqued Thin Accretion Disks}},
  volume =        {671},
  year =          {2007},
  doi =           {10.1086/522954},
}

@article{Pringle:1972aa,
  author =        {{Pringle}, J.~E. and {Rees}, M.~J.},
  journal =       {\aap},
  month =         oct,
  pages =         {1},
  title =         {{Accretion Disc Models for Compact X-Ray Sources}},
  volume =        {21},
  year =          {1972},
}

@article{Bessolaz:2008aa,
  author =        {{Bessolaz}, N. and {Zanni}, C. and {Ferreira}, J. and
                   {Keppens}, R. and {Bouvier}, J.},
  journal =       {\aap},
  month =         jan,
  pages =         {155-162},
  title =         {{Accretion funnels onto weakly magnetized young
                   stars}},
  volume =        {478},
  year =          {2008},
  doi =           {10.1051/0004-6361:20078328},
}

@article{Romanova:2002aa,
  author =        {{Romanova}, M.~M. and {Ustyugova}, G.~V. and
                   {Koldoba}, A.~V. and {Lovelace}, R.~V.~E.},
  journal =       {\apj},
  month =         oct,
  pages =         {420-438},
  title =         {{Magnetohydrodynamic Simulations of Disk-Magnetized
                   Star Interactions in the Quiescent Regime: Funnel
                   Flows and Angular Momentum Transport}},
  volume =        {578},
  year =          {2002},
  doi =           {10.1086/342464},
}

@article{Ireland:2021aa,
  author =        {{Ireland}, Lewis G. and {Zanni}, Claudio and
                   {Matt}, Sean P. and {Pantolmos}, George},
  journal =       {\apj},
  month =         jan,
  number =        {1},
  pages =         {4},
  title =         {{Magnetic Braking of Accreting T Tauri Stars: Effects
                   of Mass Accretion Rate, Rotation, and Dipolar Field
                   Strength}},
  volume =        {906},
  year =          {2021},
  doi =           {10.3847/1538-4357/abc828},
  eid =           {4},
}

@article{Ghosh:1979aa,
  author =        {{Ghosh}, P. and {Lamb}, F.~K.},
  journal =       {\apj},
  month =         aug,
  pages =         {259-276},
  title =         {{Accretion by rotating magnetic neutron stars. II -
                   Radial and vertical structure of the transition zone
                   in disk accretion}},
  volume =        {232},
  year =          {1979},
  doi =           {10.1086/157285},
}

@article{Kulkarni:2013aa,
  author =        {{Kulkarni}, A.~K. and {Romanova}, M.~M.},
  journal =       {\mnras},
  month =         aug,
  pages =         {3048-3061},
  title =         {{Analytical hotspot shapes and magnetospheric radius
                   from 3D simulations of magnetospheric accretion}},
  volume =        {433},
  year =          {2013},
  doi =           {10.1093/mnras/stt945},
}

@article{Wang:1987aa,
  author =        {{Wang}, Y. -M.},
  journal =       {\aap},
  month =         sep,
  pages =         {257-264},
  title =         {{Disc accretion by magnetized neutron stars : a
                   reassessment of the torque.}},
  volume =        {183},
  year =          {1987},
}

@article{Bozzo:2009aa,
  author =        {{Bozzo}, E. and {Stella}, L. and {Vietri}, M. and
                   {Ghosh}, P.},
  journal =       {\aap},
  month =         jan,
  number =        {3},
  pages =         {809-818},
  title =         {{Can disk-magnetosphere interaction models and beat
                   frequency models for quasi-periodic oscillation in
                   accreting X-ray pulsars be reconciled?}},
  volume =        {493},
  year =          {2009},
  doi =           {10.1051/0004-6361:200810658},
}

@article{Ireland:2022aa,
  author =        {{Ireland}, Lewis G. and {Matt}, Sean P. and
                   {Zanni}, Claudio},
  journal =       {\apj},
  month =         apr,
  number =        {1},
  pages =         {65},
  title =         {{Magnetic Braking of Accreting T Tauri Stars II:
                   Torque Formulation Spanning Spin-up and Spin-down
                   Regimes}},
  volume =        {929},
  year =          {2022},
  doi =           {10.3847/1538-4357/ac59b2},
  eid =           {65},
}

@article{Jacq:2021aa,
  author =        {{Jacquemin-Ide}, J. and {Lesur}, G. and
                   {Ferreira}, J.},
  journal =       {\aap},
  month =         mar,
  pages =         {A192},
  title =         {{Magnetic outflows from turbulent accretion disks. I.
                   Vertical structure and secular evolution}},
  volume =        {647},
  year =          {2021},
  doi =           {10.1051/0004-6361/202039322},
  eid =           {A192},
}

@article{Das:2022aa,
  author =        {{Das}, Pushpita and {Porth}, Oliver and
                   {Watts}, Anna L.},
  journal =       {\mnras},
  month =         sep,
  number =        {3},
  pages =         {3144-3161},
  title =         {{GRMHD simulations of accreting neutron stars with
                   non-dipole fields}},
  volume =        {515},
  year =          {2022},
  doi =           {10.1093/mnras/stac1817},
}

@article{Bouvier:2023ab,
  author =        {{Bouvier}, J. and {Sousa}, A. and {Pouilly}, K. and
                   {Almenara}, J.~M. and {Donati}, J. -F. and
                   {Alencar}, S.~H.~P. and {Frasca}, A. and
                   {Grankin}, K. and {Carmona}, A. and {Pantolmos}, G. and
                   {Zaire}, B. and {Bonfils}, X. and {Bayo}, A. and
                   {Rebull}, L.~M. and {Alonso-Santiago}, J. and
                   {Gameiro}, J.~F. and {Cook}, N.~J. and {Artigau}, E.},
  journal =       {\aap},
  month =         apr,
  pages =         {A5},
  title =         {{Stable accretion and episodic outflows in the young
                   transition disk system GM Aurigae. A semester-long
                   optical and near-infrared spectrophotometric
                   monitoring campaign}},
  volume =        {672},
  year =          {2023},
  doi =           {10.1051/0004-6361/202245342},
  eid =           {A5},
}

@article{Gallet:2019aa,
  author =        {{Gallet}, F. and {Zanni}, C. and {Amard}, L.},
  journal =       {\aap},
  month =         {Dec},
  pages =         {A6},
  title =         {{Rotational evolution of solar-type protostars during
                   the star-disk interaction phase}},
  volume =        {632},
  year =          {2019},
  doi =           {10.1051/0004-6361/201935432},
  eid =           {A6},
}

@article{Matt:2008aa,
  author =        {{Matt}, S. and {Pudritz}, R.~E.},
  journal =       {\apj},
  month =         may,
  pages =         {1109-1118},
  title =         {{Accretion-powered Stellar Winds. II. Numerical
                   Solutions for Stellar Wind Torques}},
  volume =        {678},
  year =          {2008},
  doi =           {10.1086/533428},
}

@article{Matt:2008ab,
  author =        {{Matt}, Sean and {Pudritz}, Ralph E.},
  journal =       {\apj},
  month =         jul,
  number =        {1},
  pages =         {391-399},
  title =         {{Accretion-powered Stellar Winds. III.
                   Spin-Equilibrium Solutions}},
  volume =        {681},
  year =          {2008},
  doi =           {10.1086/587453},
}

@article{Matt:2012ab,
  author =        {{Matt}, S.~P. and {MacGregor}, K.~B. and
                   {Pinsonneault}, M.~H. and {Greene}, T.~P.},
  journal =       {\apjl},
  month =         aug,
  pages =         {L26},
  title =         {{Magnetic Braking Formulation for Sun-like Stars:
                   Dependence on Dipole Field Strength and Rotation
                   Rate}},
  volume =        {754},
  year =          {2012},
  doi =           {10.1088/2041-8205/754/2/L26},
  eid =           {L26},
}

@article{Matt:2012aa,
  author =        {{Matt}, S.~P. and {Pinz{\'o}n}, G. and
                   {Greene}, T.~P. and {Pudritz}, R.~E.},
  journal =       {\apj},
  month =         jan,
  pages =         {101},
  title =         {{Spin Evolution of Accreting Young Stars. II. Effect
                   of Accretion-powered Stellar Winds}},
  volume =        {745},
  year =          {2012},
  doi =           {10.1088/0004-637X/745/1/101},
  eid =           {101},
}

@article{Decampli:1981aa,
  author =        {{Decampli}, W.~M.},
  journal =       {\apj},
  month =         feb,
  pages =         {124-146},
  title =         {{T Tauri winds}},
  volume =        {244},
  year =          {1981},
  doi =           {10.1086/158691},
}

@article{Cranmer:2008ab,
  author =        {{Cranmer}, S.~R.},
  journal =       {\apj},
  month =         dec,
  pages =         {316-334},
  title =         {{Turbulence-driven Polar Winds from T Tauri Stars
                   Energized by Magnetospheric Accretion}},
  volume =        {689},
  year =          {2008},
  doi =           {10.1086/592566},
}

@article{Cranmer:2009aa,
  author =        {{Cranmer}, S.~R.},
  journal =       {\apj},
  month =         nov,
  pages =         {824-843},
  title =         {{Testing Models of Accretion-Driven Coronal Heating
                   and Stellar Wind Acceleration for T Tauri Stars}},
  volume =        {706},
  year =          {2009},
  doi =           {10.1088/0004-637X/706/1/824},
}

@article{Weber:1967aa,
  author =        {{Weber}, E.~J. and {Davis}, Jr., L.},
  journal =       {\apj},
  month =         apr,
  pages =         {217-227},
  title =         {{The Angular Momentum of the Solar Wind}},
  volume =        {148},
  year =          {1967},
  doi =           {10.1086/149138},
}

@inproceedings{Mestel:1984aa,
  author =        {{Mestel}, L.},
  booktitle =     {Cool Stars, Stellar Systems, and the Sun},
  editor =        {{Baliunas}, S.~L. and {Hartmann}, L.},
  pages =         {49},
  series =        {Lecture Notes in Physics, Berlin Springer Verlag},
  title =         {{Angular Momentum Loss During Pre-Main Sequence
                   Contraction}},
  volume =        {193},
  year =          {1984},
  doi =           {10.1007/3-540-12907-3_179},
}

@article{Kawaler:1988aa,
  author =        {{Kawaler}, S.~D.},
  journal =       {\apj},
  month =         oct,
  pages =         {236-247},
  title =         {{Angular momentum loss in low-mass stars}},
  volume =        {333},
  year =          {1988},
  doi =           {10.1086/166740},
}

@article{Pantolmos:2017aa,
  author =        {{Pantolmos}, G. and {Matt}, S.~P.},
  journal =       {\apj},
  month =         nov,
  pages =         {83},
  title =         {{Magnetic Braking of Sun-like and Low-mass Stars:
                   Dependence on Coronal Temperature}},
  volume =        {849},
  year =          {2017},
  doi =           {10.3847/1538-4357/aa9061},
  eid =           {83},
}

@article{Hirabayashi:2016aa,
  author =        {{Hirabayashi}, Kota and {Hoshino}, Masahiro},
  journal =       {\apj},
  month =         may,
  number =        {2},
  pages =         {87},
  title =         {{Instability of Non-uniform Toroidal Magnetic Fields
                   in Accretion Disks}},
  volume =        {822},
  year =          {2016},
  doi =           {10.3847/0004-637X/822/2/87},
  eid =           {87},
}

@article{Tu:2026aa,
  author =        {{Tu}, Yisheng and {Li}, Zhi-Yun and {Zhu}, Zhaohuan and
                   {Hu}, Xiao and {Hsu}, Chun-Yen},
  journal =       {\apj},
  month =         apr,
  number =        {2},
  pages =         {187},
  title =         {{Modeling YSO Jets in 3D. II. Accretion-fed,
                   Star-anchored Poynting Jets in the Low-density Polar
                   Cavity Powered by Disk─Magnetosphere Interaction}},
  volume =        {1000},
  year =          {2026},
  doi =           {10.3847/1538-4357/ae4598},
  eid =           {187},
}

@article{Takasao:2025aa,
  author =        {{Takasao}, Shinsuke and {Kunitomo}, Masanobu and
                   {Suzuki}, Takeru K. and {Iwasaki}, Kazunari and
                   {Tomida}, Kengo},
  journal =       {\apj},
  month =         feb,
  number =        {1},
  pages =         {111},
  title =         {{Spin-down of Solar-mass Protostars in Magnetospheric
                   Accretion Paradigm}},
  volume =        {980},
  year =          {2025},
  doi =           {10.3847/1538-4357/ada364},
  eid =           {111},
}

@article{Armeni:2025aa,
  author =        {{Armeni}, A. and {Stelzer}, B. and {Frasca}, A. and
                   {Manara}, C.~F. and {Campbell-White}, J. and
                   {Gameiro}, J.~F. and {Gangi}, M.},
  journal =       {\aap},
  month =         dec,
  pages =         {A263},
  title =         {{Spinning-down RU Lup: Constraints on the physics of
                   the outflow from high-resolution spectroscopy}},
  volume =        {704},
  year =          {2025},
  doi =           {10.1051/0004-6361/202554624},
  eid =           {A263},
}

@article{Amard:2023aa,
  author =        {{Amard}, L. and {Matt}, S.~P.},
  journal =       {\aap},
  month =         oct,
  pages =         {A7},
  title =         {{Effects of accretion on the structure and rotation
                   of forming stars}},
  volume =        {678},
  year =          {2023},
  doi =           {10.1051/0004-6361/202346148},
  eid =           {A7},
}

@article{Gehrig:2025aa,
  author =        {{Gehrig}, Lukas and {Gaidos}, Eric and
                   {Venuti}, Laura and {Cody}, Ann Marie and
                   {Turner}, Neal J.},
  journal =       {\aap},
  month =         apr,
  pages =         {L18},
  title =         {{Do accretion-powered stellar winds help spin down T
                   Tauri stars?}},
  volume =        {696},
  year =          {2025},
  doi =           {10.1051/0004-6361/202553730},
  eid =           {L18},
}

@inproceedings{Donati:2003aa,
  author =        {{Donati}, J.-F.},
  booktitle =     {Solar Polarization},
  editor =        {{Trujillo-Bueno}, Javier and
                   {Sanchez Almeida}, Jorge},
  month =         jan,
  pages =         {41},
  series =        {Astronomical Society of the Pacific Conference
                   Series},
  title =         {{ESPaDOnS: An Echelle SpectroPolarimetric Device for
                   the Observation of Stars at CFHT}},
  volume =        {307},
  year =          {2003},
}

@article{Donati:2020ab,
  author =        {{Donati}, J.-F. and {Kouach}, D. and {Moutou}, C. and
                   {Doyon}, R. and {Delfosse}, X. and {Artigau}, E. and
                   {Baratchart}, S. and {Lacombe}, M. and {Barrick}, G. and
                   {H{\'e}brard}, G. and {Bouchy}, F. and
                   {Saddlemyer}, L. and {Par{\`e}s}, L. and {Rabou}, P. and
                   {Micheau}, Y. and {Dolon}, F. and {Reshetov}, V. and
                   {Challita}, Z. and {Carmona}, A. and {Striebig}, N. and
                   {Thibault}, S. and {Martioli}, E. and {Cook}, N. and
                   {Fouqu{\'e}}, P. and {Vermeulen}, T. and
                   {Wang}, S.~Y. and {Arnold}, L. and {Pepe}, F. and
                   {Boisse}, I. and {Figueira}, P. and {Bouvier}, J. and
                   {Ray}, T.~P. and {Feugeade}, C. and {Morin}, J. and
                   {Alencar}, S. and {Hobson}, M. and {Castilho}, B. and
                   {Udry}, S. and {Santos}, N.~C. and {Hernandez}, O. and
                   {Benedict}, T. and {Vall{\'e}e}, P. and {Gallou}, G. and
                   {Dupieux}, M. and {Larrieu}, M. and {Perruchot}, S. and
                   {Sottile}, R. and {Moreau}, F. and {Usher}, C. and
                   {Baril}, M. and {Wildi}, F. and {Chazelas}, B. and
                   {Malo}, L. and {Bonfils}, X. and {Loop}, D. and
                   {Kerley}, D. and {Wevers}, I. and {Dunn}, J. and
                   {Pazder}, J. and {Macdonald}, S. and {Dubois}, B. and
                   {Carri{\'e}}, E. and {Valentin}, H. and {Henault}, F. and
                   {Yan}, C.~H. and {Steinmetz}, T.},
  journal =       {\mnras},
  month =         nov,
  number =        {4},
  pages =         {5684-5703},
  title =         {{SPIRou: NIR velocimetry and spectropolarimetry at
                   the CFHT}},
  volume =        {498},
  year =          {2020},
  doi =           {10.1093/mnras/staa2569},
}

@article{Semel:1989aa,
  author =        {{Semel}, M.},
  journal =       {\aap},
  month =         nov,
  pages =         {456-466},
  title =         {{Zeeman-Doppler imaging of active stars. I - Basic
                   principles.}},
  volume =        {225},
  year =          {1989},
}

@article{Donati:1997aa,
  author =        {{Donati}, J.-F. and {Brown}, S.~F.},
  journal =       {\aap},
  month =         oct,
  pages =         {1135-1142},
  title =         {{Zeeman-Doppler imaging of active stars. V.
                   Sensitivity of maximum entropy magnetic maps to field
                   orientation.}},
  volume =        {326},
  year =          {1997},
}

@article{Mohanty:2008aa,
  author =        {{Mohanty}, Subhanjoy and {Shu}, Frank H.},
  journal =       {\apj},
  month =         nov,
  number =        {2},
  pages =         {1323-1338},
  title =         {{Magnetocentrifugally Driven Flows from Young Stars
                   and Disks. VI. Accretion with a Multipole Stellar
                   Field}},
  volume =        {687},
  year =          {2008},
  doi =           {10.1086/591924},
}

@article{Romanova:2011aa,
  author =        {{Romanova}, M.~M. and {Long}, M. and {Lamb}, F.~K. and
                   {Kulkarni}, A.~K. and {Donati}, J. -F.},
  journal =       {\mnras},
  month =         feb,
  number =        {2},
  pages =         {915-928},
  title =         {{Global 3D simulations of disc accretion on to the
                   classical T Tauri star V2129 Oph}},
  volume =        {411},
  year =          {2011},
  doi =           {10.1111/j.1365-2966.2010.17724.x},
}

@article{Galli:2015aa,
  author =        {{Galli}, P.~A.~B. and {Bertout}, C. and
                   {Teixeira}, R. and {Ducourant}, C.},
  journal =       {\aap},
  month =         aug,
  pages =         {A26},
  title =         {{Evolution of the T Tauri star population in the
                   Lupus association}},
  volume =        {580},
  year =          {2015},
  doi =           {10.1051/0004-6361/201525804},
  eid =           {A26},
}

@article{Pittman:2025aa,
  author =        {{Pittman}, Caeley V. and {Espaillat}, Catherine C. and
                   {Zhu}, Zhaohuan and {Thanathibodee}, Thanawuth and
                   {Robinson}, Connor E. and {Calvet}, Nuria and
                   {K{\'o}sp{\'a}l}, {\'A}gnes},
  journal =       {\apj},
  month =         nov,
  number =        {2},
  pages =         {181},
  title =         {{The ODYSSEUS Survey. Using Accretion and Stellar
                   Rotation to Reveal the Star─Disk Connection in T
                   Tauri Stars}},
  volume =        {993},
  year =          {2025},
  doi =           {10.3847/1538-4357/ae03b3},
  eid =           {181},
}

@article{Pittman:2025ab,
  author =        {{Pittman}, Caeley V. and {Espaillat}, Catherine C. and
                   {Robinson}, Connor E. and {Thanathibodee}, Thanawuth and
                   {Lopez}, Sophia and {Calvet}, Nuria and
                   {Zhu}, Zhaohuan and {Walter}, Frederick M. and
                   {Wendeborn}, John and {Manara}, Carlo F. and
                   {Campbell-White}, Justyn and {Claes}, Rik and
                   {Fang}, Min and {Frasca}, Antonio and
                   {Gameiro}, Jorge F. and {Gangi}, Manuele and
                   {Hern{\'a}ndez}, Jesus and
                   {K{\'o}sp{\'a}l}, {\'A}gnes and {Mauc{\'o}}, Karina and
                   {Muzerolle}, James and {Siwak}, Micha{\l} and
                   {Tychoniec}, {\L}ukasz and {Venuti}, Laura},
  journal =       {\apj},
  month =         oct,
  number =        {1},
  pages =         {134},
  title =         {{The ODYSSEUS Survey. Characterizing Magnetospheric
                   Geometries and Hotspot Structures in T Tauri Stars}},
  volume =        {992},
  year =          {2025},
  doi =           {10.3847/1538-4357/adef35},
  eid =           {134},
}

@article{Gravity:2017aa,
  author =        {{GRAVITY Collaboration} and {Abuter}, R. and
                   {Accardo}, M. and {Amorim}, A. and {Anugu}, N. and
                   {{\'A}vila}, G. and {Azouaoui}, N. and {Benisty}, M. and
                   {Berger}, J.~P. and {Blind}, N. and {Bonnet}, H. and
                   {Bourget}, P. and {Brandner}, W. and {Brast}, R. and
                   {Buron}, A. and {Burtscher}, L. and {Cassaing}, F. and
                   {Chapron}, F. and {Choquet}, {\'E}. and
                   {Cl{\'e}net}, Y. and {Collin}, C. and
                   {Coud{\'e} Du Foresto}, V. and {de Wit}, W. and
                   {de Zeeuw}, P.~T. and {Deen}, C. and
                   {Delplancke-Str{\"o}bele}, F. and {Dembet}, R. and
                   {Derie}, F. and {Dexter}, J. and {Duvert}, G. and
                   {Ebert}, M. and {Eckart}, A. and {Eisenhauer}, F. and
                   {Esselborn}, M. and {F{\'e}dou}, P. and {Finger}, G. and
                   {Garcia}, P. and {Garcia Dabo}, C.~E. and
                   {Garcia Lopez}, R. and {Gendron}, E. and {Genzel}, R. and
                   {Gillessen}, S. and {Gonte}, F. and {Gordo}, P. and
                   {Grould}, M. and {Gr{\"o}zinger}, U. and {Guieu}, S. and
                   {Haguenauer}, P. and {Hans}, O. and {Haubois}, X. and
                   {Haug}, M. and {Haussmann}, F. and {Henning}, Th. and
                   {Hippler}, S. and {Horrobin}, M. and {Huber}, A. and
                   {Hubert}, Z. and {Hubin}, N. and {Hummel}, C.~A. and
                   {Jakob}, G. and {Janssen}, A. and {Jochum}, L. and
                   {Jocou}, L. and {Kaufer}, A. and {Kellner}, S. and
                   {Kendrew}, S. and {Kern}, L. and {Kervella}, P. and
                   {Kiekebusch}, M. and {Klein}, R. and {Kok}, Y. and
                   {Kolb}, J. and {Kulas}, M. and {Lacour}, S. and
                   {Lapeyr{\`e}re}, V. and {Lazareff}, B. and
                   {Le Bouquin}, J.-B. and {L{\`e}na}, P. and
                   {Lenzen}, R. and {L{\'e}v{\^e}que}, S. and
                   {Lippa}, M. and {Magnard}, Y. and {Mehrgan}, L. and
                   {Mellein}, M. and {M{\'e}rand}, A. and
                   {Moreno-Ventas}, J. and {Moulin}, T. and
                   {M{\"u}ller}, E. and {M{\"u}ller}, F. and
                   {Neumann}, U. and {Oberti}, S. and {Ott}, T. and
                   {Pallanca}, L. and {Panduro}, J. and {Pasquini}, L. and
                   {Paumard}, T. and {Percheron}, I. and {Perraut}, K. and
                   {Perrin}, G. and {Pfl{\"u}ger}, A. and {Pfuhl}, O. and
                   {Phan Duc}, T. and {Plewa}, P.~M. and {Popovic}, D. and
                   {Rabien}, S. and {Ram{\'\i}rez}, A. and {Ramos}, J. and
                   {Rau}, C. and {Riquelme}, M. and {Rohloff}, R.-R. and
                   {Rousset}, G. and {Sanchez-Bermudez}, J. and
                   {Scheithauer}, S. and {Sch{\"o}ller}, M. and
                   {Schuhler}, N. and {Spyromilio}, J. and
                   {Straubmeier}, C. and {Sturm}, E. and {Suarez}, M. and
                   {Tristram}, K.~R.~W. and {Ventura}, N. and
                   {Vincent}, F. and {Waisberg}, I. and {Wank}, I. and
                   {Weber}, J. and {Wieprecht}, E. and {Wiest}, M. and
                   {Wiezorrek}, E. and {Wittkowski}, M. and
                   {Woillez}, J. and {Wolff}, B. and {Yazici}, S. and
                   {Ziegler}, D. and {Zins}, G.},
  journal =       {\aap},
  month =         jun,
  pages =         {A94},
  title =         {{First light for GRAVITY: Phase referencing optical
                   interferometry for the Very Large Telescope
                   Interferometer}},
  volume =        {602},
  year =          {2017},
  doi =           {10.1051/0004-6361/201730838},
  eid =           {A94},
}

@article{Ferreira:2006aa,
  author =        {{Ferreira}, J. and {Dougados}, C. and {Cabrit}, S.},
  journal =       {\aap},
  month =         jul,
  number =        {3},
  pages =         {785-796},
  title =         {{Which jet launching mechanism(s) in T Tauri stars?}},
  volume =        {453},
  year =          {2006},
  doi =           {10.1051/0004-6361:20054231},
}

@article{Costigan:2014aa,
  author =        {{Costigan}, G. and {Vink}, Jorick S. and {Scholz}, A. and
                   {Ray}, T. and {Testi}, L.},
  journal =       {\mnras},
  month =         jun,
  number =        {4},
  pages =         {3444-3461},
  title =         {{Temperaments of young stars: rapid mass accretion
                   rate changes in T Tauri and Herbig Ae stars}},
  volume =        {440},
  year =          {2014},
  doi =           {10.1093/mnras/stu529},
}

@article{Matt:2010aa,
  author =        {{Matt}, S.~P. and {Pinz{\'o}n}, G. and
                   {de la Reza}, R. and {Greene}, T.~P.},
  journal =       {\apj},
  month =         may,
  pages =         {989-1000},
  title =         {{Spin Evolution of Accreting Young Stars. I. Effect
                   of Magnetic Star-Disk Coupling}},
  volume =        {714},
  year =          {2010},
  doi =           {10.1088/0004-637X/714/2/989},
}

@article{Donati:2025aa,
  author =        {{Donati}, J.-F. and {Gaidos}, E. and {Moutou}, C. and
                   {Cristofari}, P.~I. and {Arnold}, L. and
                   {Barber}, M.~G. and {Mann}, A.~W.},
  journal =       {\aap},
  month =         jun,
  pages =         {L14},
  title =         {{Mass, gas, and Gauss around a T Tauri Star with
                   SPIRou}},
  volume =        {698},
  year =          {2025},
  doi =           {10.1051/0004-6361/202554628},
  eid =           {L14},
}

@article{Donati:2010ab,
  author =        {{Donati}, J.-F. and {Skelly}, M.~B. and {Bouvier}, J. and
                   {Gregory}, S.~G. and {Grankin}, K.~N. and
                   {Jardine}, M.~M. and {Hussain}, G.~A.~J. and
                   {M{\'e}nard}, F. and {Dougados}, C. and {Unruh}, Y. and
                   {Mohanty}, S. and {Auri{\`e}re}, M. and {Morin}, J. and
                   {Far{\`e}s}, R. and {MAPP Collaboration}},
  journal =       {\mnras},
  month =         dec,
  pages =         {1347-1361},
  title =         {{Magnetospheric accretion and spin-down of the
                   prototypical classical T Tauri star AA Tau}},
  volume =        {409},
  year =          {2010},
  doi =           {10.1111/j.1365-2966.2010.17409.x},
}

@article{Donati:2010aa,
  author =        {{Donati}, J.-F. and {Skelly}, M.~B. and {Bouvier}, J. and
                   {Jardine}, M.~M. and {Gregory}, S.~G. and {Morin}, J. and
                   {Hussain}, G.~A.~J. and {Dougados}, C. and
                   {M{\'e}nard}, F. and {Unruh}, Y.},
  journal =       {\mnras},
  month =         mar,
  pages =         {1426-1436},
  title =         {{Complex magnetic topology and strong differential
                   rotation on the low-mass T Tauri star V2247 Oph}},
  volume =        {402},
  year =          {2010},
  doi =           {10.1111/j.1365-2966.2009.15998.x},
}

@article{Donati:2024ad,
  author =        {{Donati}, J.-F. and {Cristofari}, P.~I. and
                   {Alencar}, S.~H.~P. and {K{\'o}sp{\'a}l}, {\'A}. and
                   {Bouvier}, J. and {Moutou}, C. and {Carmona}, A. and
                   {Gregorio-Hetem}, J. and {M{\'e}nard}, F. and
                   {Artigau}, E. and {Doyon}, R. and {Takami}, M. and
                   {Shang}, H. and {Dias do Nascimento}, J. and
                   {M{\'e}nard}, F. and {Gaidos}, E. and
                   {SPIRou Science Team}},
  journal =       {\mnras},
  month =         dec,
  number =        {4},
  pages =         {3363-3382},
  title =         {{SPIRou observations of the young planet-hosting star
                   PDS 70}},
  volume =        {535},
  year =          {2024},
  doi =           {10.1093/mnras/stae2506},
}

@article{Donati:2011ab,
  author =        {{Donati}, J.-F. and {Bouvier}, J. and {Walter}, F.~M. and
                   {Gregory}, S.~G. and {Skelly}, M.~B. and
                   {Hussain}, G.~A.~J. and {Flaccomio}, E. and
                   {Argiroffi}, C. and {Grankin}, K.~N. and
                   {Jardine}, M.~M. and {M{\'e}nard}, F. and
                   {Dougados}, C. and {Romanova}, M.~M.},
  journal =       {\mnras},
  month =         apr,
  number =        {4},
  pages =         {2454-2468},
  title =         {{Non-stationary dynamo and magnetospheric accretion
                   processes of the classical T Tauri star V2129 Oph}},
  volume =        {412},
  year =          {2011},
  doi =           {10.1111/j.1365-2966.2010.18069.x},
}

@article{Donati:2019aa,
  author =        {{Donati}, J.-F. and {Bouvier}, J. and
                   {Alencar}, S.~H. and {Hill}, C. and {Carmona}, A. and
                   {Folsom}, C.~P. and {M{\'e}nard}, F. and
                   {Gregory}, S.~G. and {Hussain}, G.~A. and
                   {Grankin}, K. and {Moutou}, C. and {Malo}, L. and
                   {Takami}, M. and {Herczeg}, G.~J.},
  journal =       {\mnras},
  month =         feb,
  pages =         {L1-L5},
  title =         {{The magnetic propeller accretion regime of LkCa 15}},
  volume =        {483},
  year =          {2019},
  doi =           {10.1093/mnrasl/sly207},
}

@article{Donati:2013aa,
  author =        {{Donati}, J.-F. and {Gregory}, S.~G. and
                   {Alencar}, S.~H.~P. and {Hussain}, G. and
                   {Bouvier}, J. and {Jardine}, M.~M. and
                   {M{\'e}nard}, F. and {Dougados}, C. and
                   {Romanova}, M.~M. and {MaPP Collaboration}},
  journal =       {\mnras},
  month =         nov,
  pages =         {881-897},
  title =         {{Magnetospheric accretion on the fully convective
                   classical T Tauri star DN Tau}},
  volume =        {436},
  year =          {2013},
  doi =           {10.1093/mnras/stt1622},
}

@article{Donati:2008ab,
  author =        {{Donati}, J.-F. and {Jardine}, M.~M. and
                   {Gregory}, S.~G. and {Petit}, P. and {Paletou}, F. and
                   {Bouvier}, J. and {Dougados}, C. and {M{\'e}nard}, F. and
                   {Collier Cameron}, A. and {Harries}, T.~J. and
                   {Hussain}, G.~A.~J. and {Unruh}, Y. and {Morin}, J. and
                   {Marsden}, S.~C. and {Manset}, N. and
                   {Auri{\`e}re}, M. and {Catala}, C. and {Alecian}, E.},
  journal =       {\mnras},
  month =         may,
  pages =         {1234-1251},
  title =         {{Magnetospheric accretion on the T Tauri star BP
                   Tauri}},
  volume =        {386},
  year =          {2008},
  doi =           {10.1111/j.1365-2966.2008.13111.x},
}

@article{Hussain:2009aa,
  author =        {{Hussain}, G.~A.~J. and {Collier Cameron}, A. and
                   {Jardine}, M.~M. and {Dunstone}, N. and
                   {Ramirez Velez}, J. and {Stempels}, H.~C. and
                   {Donati}, J.-F. and {Semel}, M. and {Aulanier}, G. and
                   {Harries}, T. and {Bouvier}, J. and {Dougados}, C. and
                   {Ferreira}, J. and {Carter}, B.~D. and
                   {Lawson}, W.~A.},
  journal =       {\mnras},
  month =         sep,
  number =        {1},
  pages =         {189-200},
  title =         {{Surface magnetic fields on two accreting TTauri
                   stars: CVCha and CRCha}},
  volume =        {398},
  year =          {2009},
  doi =           {10.1111/j.1365-2966.2009.14881.x},
}

@article{Freitas:2026aa,
  author =        {{Freitas}, T.~P. and {Bouvier}, J. and {Zaire}, B. and
                   {Alencar}, S.~H.~P. and {Sousa}, A.~P. and
                   {Rebull}, L. and {Bayo}, A. and {Frasca}, A. and
                   {Alonso-Santiago}, J. and {Grankin}, K. and
                   {Contreras Pe{\~n}a}, C. and {Cody}, A.~M. and
                   {Hillenbrand}, L.~A. and {Carmona}, A.},
  journal =       {\aap},
  month =         may,
  pages =         {A243},
  title =         {{The circumstellar environment of the young, low-mass
                   dipper star JH 223: Accretion and large-scale
                   magnetic field topology}},
  volume =        {709},
  year =          {2026},
  doi =           {10.1051/0004-6361/202558514},
  eid =           {A243},
}

@article{Bouvier:2020aa,
  author =        {{Bouvier}, J. and {Perraut}, K. and
                   {Le Bouquin}, J. -B. and {Duvert}, G. and
                   {Dougados}, C. and {Brandner}, W. and {Benisty}, M. and
                   {Berger}, J. -P. and {Al{\'e}cian}, E.},
  journal =       {\aap},
  month =         apr,
  pages =         {A108},
  title =         {{Probing the magnetospheric accretion region of the
                   young pre-transitional disk system DoAr 44 using
                   VLTI/GRAVITY}},
  volume =        {636},
  year =          {2020},
  doi =           {10.1051/0004-6361/202037611},
  eid =           {A108},
}

@article{Bouvier:2020ab,
  author =        {{Bouvier}, J. and {Alecian}, E. and
                   {Alencar}, S.~H.~P. and {Sousa}, A. and
                   {Donati}, J.-F. and {Perraut}, K. and {Bayo}, A. and
                   {Rebull}, L.~M. and {Dougados}, C. and {Duvert}, G. and
                   {Berger}, J.-P. and {Benisty}, M. and {Pouilly}, K. and
                   {Folsom}, C. and {Moutou}, C. and
                   {SPIRou Consortium}},
  journal =       {\aap},
  month =         nov,
  pages =         {A99},
  title =         {{Investigating the magnetospheric accretion process
                   in the young pre-transitional disk system DoAr 44
                   (V2062 Oph). A multiwavelength interferometric,
                   spectropolarimetric, and photometric observing
                   campaign}},
  volume =        {643},
  year =          {2020},
  doi =           {10.1051/0004-6361/202038892},
  eid =           {A99},
}

@article{Donati:2012aa,
  author =        {{Donati}, J.-F. and {Gregory}, S.~G. and
                   {Alencar}, S.~H.~P. and {Hussain}, G. and
                   {Bouvier}, J. and {Dougados}, C. and {Jardine}, M.~M. and
                   {M{\'e}nard}, F. and {Romanova}, M.~M.},
  journal =       {\mnras},
  month =         oct,
  number =        {4},
  pages =         {2948-2963},
  title =         {{Magnetometry of the classical T Tauri star GQ Lup:
                   non-stationary dynamos and spin evolution of young
                   Suns}},
  volume =        {425},
  year =          {2012},
  doi =           {10.1111/j.1365-2966.2012.21482.x},
}

@article{Donati:2024ab,
  author =        {{Donati}, J.-F. and {Cristofari}, P.~I. and
                   {Lehmann}, L.~T. and {Moutou}, C. and
                   {Alencar}, S.~H.~P. and {Bouvier}, J. and
                   {Arnold}, L. and {Delfosse}, X. and {Artigau}, E. and
                   {Cook}, N. and {K{\'o}sp{\'a}l}, {\'A}. and
                   {M{\'e}nard}, F. and {Baruteau}, C. and {Takami}, M. and
                   {Cabrit}, S. and {H{\'e}brard}, G. and {Doyon}, R. and
                   {SPIRou Science Team}},
  journal =       {\mnras},
  month =         jul,
  number =        {3},
  pages =         {3256-3278},
  title =         {{SPIRou spectropolarimetry of the T Tauri star TW
                   Hydrae: magnetic fields, accretion, and planets}},
  volume =        {531},
  year =          {2024},
  doi =           {10.1093/mnras/stae1227},
}

@article{Gravity:2020aa,
  author =        {{GRAVITY Collaboration} and {Garcia Lopez}, R. and
                   {Natta}, A. and {Caratti o Garatti}, A. and
                   {Ray}, T.~P. and {Fedriani}, R. and {Koutoulaki}, M. and
                   {Klarmann}, L. and {Perraut}, K. and
                   {Sanchez-Bermudez}, J. and {Benisty}, M. and
                   {Dougados}, C. and {Labadie}, L. and {Brandner}, W. and
                   {Garcia}, P.~J.~V. and {Henning}, Th. and
                   {Caselli}, P. and {Duvert}, G. and {de Zeeuw}, T. and
                   {Grellmann}, R. and {Abuter}, R. and {Amorim}, A. and
                   {Baub{\"o}ck}, M. and {Berger}, J.~P. and
                   {Bonnet}, H. and {Buron}, A. and {Cl{\'e}net}, Y. and
                   {Coud{\'e} Du Foresto}, V. and {de Wit}, W. and
                   {Eckart}, A. and {Eisenhauer}, F. and {Filho}, M. and
                   {Gao}, F. and {Garcia Dabo}, C.~E. and {Gendron}, E. and
                   {Genzel}, R. and {Gillessen}, S. and {Habibi}, M. and
                   {Haubois}, X. and {Haussmann}, F. and {Hippler}, S. and
                   {Hubert}, Z. and {Horrobin}, M. and
                   {Jimenez Rosales}, A. and {Jocou}, L. and
                   {Kervella}, P. and {Kolb}, J. and {Lacour}, S. and
                   {Le Bouquin}, J. -B. and {L{\'e}na}, P. and {Ott}, T. and
                   {Paumard}, T. and {Perrin}, G. and {Pfuhl}, O. and
                   {Ramirez}, A. and {Rau}, C. and {Rousset}, G. and
                   {Scheithauer}, S. and {Shangguan}, J. and
                   {Stadler}, J. and {Straub}, O. and {Straubmeier}, C. and
                   {Sturm}, E. and {van Dishoeck}, E. and {Vincent}, F. and
                   {von Fellenberg}, S. and {Widmann}, F. and
                   {Wieprecht}, E. and {Wiest}, M. and {Wiezorrek}, E. and
                   {Woillez}, J. and {Yazici}, S. and {Zins}, G.},
  journal =       {\nat},
  month =         aug,
  number =        {7822},
  pages =         {547-550},
  title =         {{A measure of the size of the magnetospheric
                   accretion region in TW Hydrae}},
  volume =        {584},
  year =          {2020},
  doi =           {10.1038/s41586-020-2613-1},
}

@article{Zaire:2024aa,
  author =        {{Zaire}, B. and {Donati}, J.-F. and {Alencar}, S.~P. and
                   {Bouvier}, J. and {Moutou}, C. and {Bellotti}, S. and
                   {Carmona}, A. and {Petit}, P. and
                   {K{\'o}sp{\'a}l}, {\'A}. and {Shang}, H. and
                   {Grankin}, K. and {Manara}, C. and {Alecian}, E. and
                   {Gregory}, S.~P. and {Fouqu{\'e}}, P. and
                   {the SLS consortium}},
  journal =       {\mnras},
  month =         sep,
  number =        {3},
  pages =         {2893-2915},
  title =         {{Magnetic field, magnetospheric accretion, and
                   candidate planet of the young star GM Aurigae
                   observed with SPIRou}},
  volume =        {533},
  year =          {2024},
  doi =           {10.1093/mnras/stae1955},
}

@article{Donati:2007aa,
  author =        {{Donati}, J.-F. and {Jardine}, M.~M. and
                   {Gregory}, S.~G. and {Petit}, P. and {Bouvier}, J. and
                   {Dougados}, C. and {M{\'e}nard}, F. and
                   {Collier Cameron}, A. and {Harries}, T.~J. and
                   {Jeffers}, S.~V. and {Paletou}, F.},
  journal =       {\mnras},
  month =         oct,
  number =        {4},
  pages =         {1297-1312},
  title =         {{Magnetic fields and accretion flows on the classical
                   T Tauri star V2129 Oph}},
  volume =        {380},
  year =          {2007},
  doi =           {10.1111/j.1365-2966.2007.12194.x},
}

@article{Donati:2020aa,
  author =        {{Donati}, J. -F. and {Bouvier}, J. and
                   {Alencar}, S.~H. and {Moutou}, C. and {Malo}, L. and
                   {Takami}, M. and {M{\'e}nard}, F. and {Dougados}, C. and
                   {Hussain}, G.~A. and {The Matysse Collaboration}},
  journal =       {\mnras},
  month =         feb,
  number =        {4},
  pages =         {5660-5670},
  title =         {{The magnetic field and accretion regime of CI Tau}},
  volume =        {491},
  year =          {2020},
  doi =           {10.1093/mnras/stz3368},
}

@article{Gravity:2023aa,
  author =        {{GRAVITY Collaboration} and {Soulain}, A. and
                   {Perraut}, K. and {Bouvier}, J. and {Pantolmos}, G. and
                   {Caratti O Garatti}, A. and {Caselli}, P. and
                   {Garcia}, P. and {Lopez}, R. Garcia and {Aimar}, N. and
                   {Amorin}, A. and {Benisty}, M. and {Berger}, J.-P. and
                   {Bourdarot}, G. and {Brandner}, W. and
                   {Cl{\'e}net}, Y. and {de Zeeuw}, T. and {Davies}, R. and
                   {Drescher}, A. and {Eckart}, A. and {Eisenhauer}, F. and
                   {Schreiber}, N.~M. F{\"o}rster and {Gendron}, E. and
                   {Genzuel}, R. and {Gillessen}, S. and
                   {Hei{\ss}el}, G. and {Henning}, Th. and {Hippler}, S. and
                   {Horrobin}, M. and {Jocou}, L. and {Kervella}, P. and
                   {Labadie}, L. and {Lacour}, S. and {Lapeyrere}, V. and
                   {Le Bouquin}, J.-B. and {L{\'e}na}, P. and {Lutz}, D. and
                   {Mang}, F. and {Ott}, T. and {Paumard}, T. and
                   {Perrin}, G. and {Sanchez}, J. and {Scheithauer}, S. and
                   {Shangguan}, J. and {Shimizu}, T. and {Straub}, O. and
                   {Straubmeier}, C. and {Sturm}, E. and
                   {Tacconi}, L.~J. and {Vincent}, F. and
                   {van Dishoeck}, E. and {Widmann}, F. and
                   {Wieprecht}, E. and {Wiezorrek}, E. and {Yazici}, S.},
  journal =       {\aap},
  month =         jun,
  pages =         {A203},
  title =         {{The GRAVITY young stellar object survey. X. Probing
                   the inner disk and magnetospheric accretion region of
                   CI Tau}},
  volume =        {674},
  year =          {2023},
  doi =           {10.1051/0004-6361/202346446},
  eid =           {A203},
}

@article{Donati:2024aa,
  author =        {{Donati}, J.-F. and {Finociety}, B. and
                   {Cristofari}, P.~I. and {Alencar}, S.~H.~P. and
                   {Moutou}, C. and {Delfosse}, X. and {Fouqu{\'e}}, P. and
                   {Arnold}, L. and {Baruteau}, C. and
                   {K{\'o}sp{\'a}l}, {\'A}. and {M{\'e}nard}, F. and
                   {Carmona}, A. and {Grankin}, K. and {Takami}, M. and
                   {Artigau}, E. and {Doyon}, R. and {H{\'e}brard}, G. and
                   {the SPIRou science team}},
  journal =       {\mnras},
  month =         may,
  number =        {1},
  pages =         {264-286},
  title =         {{The classical T Tauri star CI Tau observed with
                   SPIRou: magnetospheric accretion and planetary
                   formation}},
  volume =        {530},
  year =          {2024},
  doi =           {10.1093/mnras/stae675},
}

@article{Donati:2011aa,
  author =        {{Donati}, J.-F. and {Gregory}, S.~G. and
                   {Alencar}, S.~H.~P. and {Bouvier}, J. and
                   {Hussain}, G. and {Skelly}, M. and {Dougados}, C. and
                   {Jardine}, M.~M. and {M{\'e}nard}, F. and
                   {Romanova}, M.~M. and {Unruh}, Y.~C.},
  journal =       {\mnras},
  month =         oct,
  pages =         {472-487},
  title =         {{The large-scale magnetic field and poleward mass
                   accretion of the classical T Tauri star TW Hya}},
  volume =        {417},
  year =          {2011},
  doi =           {10.1111/j.1365-2966.2011.19288.x},
}

@article{Donati:2026aa,
  author =        {{Donati}, J.-F. and {Cristofari}, P.~I. and
                   {Carmona}, A. and {Lavail}, A. and {Moutou}, C. and
                   {Bouvier}, J. and {Perraut}, K. and
                   {Alencar}, S.~H.~P. and {M{\'e}nard}, F. and
                   {Audard}, M. and {Petit}, P. and {Alecian}, E. and
                   {Ray}, T.},
  journal =       {\aap},
  month =         apr,
  pages =         {A230},
  title =         {{Monitoring the magnetospheric accretion of the
                   classical T Tauri star DO Tau with SPIRou}},
  volume =        {708},
  year =          {2026},
  doi =           {10.1051/0004-6361/202558694},
  eid =           {A230},
}

@article{Gravity:2026aa,
  author =        {{GRAVITY Collaboration} and {Perraut}, K. and
                   {Bouvier}, J. and {Nowacki}, H. and {Sousa}, A. and
                   {Houll{\'e}}, M. and {Donati}, J.~F. and
                   {Alecian}, E. and {Alencar}, S. and {Audard}, M. and
                   {Berger}, J.-P. and {Bouarour}, Y.-I. and
                   {Bordier}, E. and {Bourdarot}, G. and {Carmona}, A. and
                   {Caratti O Garatti}, A. and {Dougados}, C. and
                   {Flock}, M. and {Garcia-Lopez}, R. and {Grankin}, K. and
                   {K{\'o}sp{\'a}l}, {\'A}. and {Labadie}, L. and
                   {Moutou}, C. and {Sanchez-Bermudez}, J. and
                   {Shang}, H. and {Takami}, M. and {Amorim}, A. and
                   {Brandner}, W. and {Cl{\'e}net}, Y. and {Davies}, R. and
                   {Dembet}, R. and {Drescher}, A. and {Eckart}, A. and
                   {Eisenhauer}, F. and {Fabricius}, M. and
                   {Feuchtgruber}, H. and {F{\"o}rster-Schreiber}, N.~M. and
                   {Garcia}, P. and {Gendron}, E. and {Genzel}, R. and
                   {Gillessen}, S. and {Henning}, T. and {Jocou}, L. and
                   {Joharle}, S. and {Kervella}, P. and {Kreidberg}, L. and
                   {Lacour}, S. and {Lapeyr{\`e}re}, V. and
                   {Le Bouquin}, J.-B. and {Lutz}, D. and {Mang}, F. and
                   {Ott}, T. and {Paumard}, T. and {Perrin}, G. and
                   {Rabien}, S. and {Ribeiro}, D.~C. and
                   {Sadun Bordoni}, M. and {Santos}, D. and
                   {Shangguan}, J. and {Shimizu}, T. and
                   {Straubmeier}, C. and {Sturm}, E. and {Tacconi}, L. and
                   {Vincent}, F.},
  journal =       {\aap},
  month =         apr,
  pages =         {A334},
  title =         {{The GRAVITY young stellar object survey: XV. The
                   star-disk interaction region of the T Tauri star DO
                   Tau}},
  volume =        {708},
  year =          {2026},
  doi =           {10.1051/0004-6361/202558438},
  eid =           {A334},
}

@article{Nowacki:2023aa,
  author =        {{Nowacki}, H. and {Alecian}, E. and {Perraut}, K. and
                   {Zaire}, B. and {Folsom}, C.~P. and {Pouilly}, K. and
                   {Bouvier}, J. and {Manick}, R. and {Pantolmos}, G. and
                   {Sousa}, A.~P. and {Dougados}, C. and
                   {Hussain}, G.~A.~J. and {Alencar}, S.~H.~P. and
                   {Le Bouquin}, J.~B.},
  journal =       {\aap},
  month =         oct,
  pages =         {A86},
  title =         {{Star-disk interactions in the strongly accreting T
                   Tauri star S CrA N}},
  volume =        {678},
  year =          {2023},
  doi =           {10.1051/0004-6361/202347145},
  eid =           {A86},
}

@article{Gravity:2024aa,
  author =        {{GRAVITY Collaboration} and {Nowacki}, H. and
                   {Perraut}, K. and {Labadie}, L. and {Bouvier}, J. and
                   {Dougados}, C. and {Benisty}, M. and
                   {Wojtczak}, J.~A. and {Soulain}, A. and {Alecian}, E. and
                   {Brandner}, W. and {Caratti o Garatti}, A. and
                   {Garcia Lopez}, R. and {Ganci}, V. and
                   {S{\'a}nchez-Berm{\'u}dez}, J. and {Berger}, J. -P. and
                   {Bourdarot}, G. and {Caselli}, P. and
                   {Cl{\'e}net}, Y. and {Davies}, R. and {Drescher}, A. and
                   {Eckart}, A. and {Eisenhauer}, F. and {Fabricius}, M. and
                   {Feuchtgruber}, H. and {F{\"o}rster-Schreiber}, N.~M. and
                   {Garcia}, P. and {Gendron}, E. and {Genzel}, R. and
                   {Gillessen}, S. and {Grant}, S. and {Henning}, T. and
                   {Jocou}, L. and {Kervella}, P. and {Kurtovic}, N. and
                   {Lacour}, S. and {Lapeyr{\`e}re}, V. and
                   {Le Bouquin}, J. -B. and {Lutz}, D. and {Mang}, F. and
                   {Ott}, T. and {Paumard}, T. and {Perrin}, G. and
                   {Rabien}, S. and {Ribeiro}, D. and
                   {Sadun Bordoni}, M. and {Scheithauer}, S. and
                   {Shangguan}, J. and {Shimizu}, T. and {Spezzano}, S. and
                   {Straubmeier}, C. and {Sturm}, E. and {Tacconi}, L. and
                   {van Dishoeck}, E. and {Vincent}, F. and
                   {Widmann}, F.},
  journal =       {\aap},
  month =         oct,
  pages =         {A123},
  title =         {{The GRAVITY young stellar object survey. XIV.
                   Investigating the magnetospheric accretion-ejection
                   processes in S CrA N}},
  volume =        {690},
  year =          {2024},
  doi =           {10.1051/0004-6361/202451254},
  eid =           {A123},
}

@article{Donati:2011ac,
  author =        {{Donati}, J.-F. and {Gregory}, S.~G. and
                   {Montmerle}, T. and {Maggio}, A. and {Argiroffi}, C. and
                   {Sacco}, G. and {Hussain}, G. and {Kastner}, J. and
                   {Alencar}, S.~H.~P. and {Audard}, M. and
                   {Bouvier}, J. and {Damiani}, F. and {G{\"u}del}, M. and
                   {Huenemoerder}, D. and {Wade}, G.~A.},
  journal =       {\mnras},
  month =         nov,
  number =        {3},
  pages =         {1747-1759},
  title =         {{The close classical T Tauri binary V4046 Sgr:
                   complex magnetic fields and distributed mass
                   accretion}},
  volume =        {417},
  year =          {2011},
  doi =           {10.1111/j.1365-2966.2011.19366.x},
}

\begin{appendix}
  \nolinenumbers

  \section{Simulation parameters and global properties}
  \label{sec_tab_params_props}

  Table \ref{tab_data_sim} lists the simulation parameters that are varied throughout the study (see also Sect. \ref{sec_params})
  and the global properties of each run used in the analysis presented in Sects. \ref{sec_bndry}, \ref{sec_rtrunc}, and \ref{sec_torques}.

   \begin{sidewaystable*}
        %[tbp]
        \caption{Input parameters and global properties of the numerical simulations.}
        \label{tab_data_sim}
        \centering
        \begin{tabular}{c c c c c c c c c c c c c c c c}
                \hline\hline\\[-2.ex]
                Case & $f$ & $P_\star/P_0$ & $R_\mathrm{co}/R_\star$ & $B_\star/B_0$ & $\Theta$  & $\rho_{\mathrm{d}\star}/\rho_\star$
                & $\dot{M}_\mathrm{acc}/\dot{M}_0$ & $\Upsilon_\mathrm{acc}$ & $R_\mathrm{t}/R_\star$ &$\dot{J}_\mathrm{SDI}/J_\star$ &$\dot{M}_\mathrm{SW}/\dot{M}\mathrm{acc}\tablefootmark{(a)}$
                &$\Phi_\mathrm{SW}/\Phi_\star$ \tablefootmark{(b)} & $\Upsilon_\mathrm{SW}$\tablefootmark{(a)}& $\langle r_\mathrm{A} \rangle/R_\star$\tablefootmark{(a)} & Accretion \\
                &&&&&&&&&&&&&[$10^4$]&& regime ($\mathcal{P}$) \\
                \hline\\[-2.ex]
                1  & \multirow{11}*{0.15} & \multirow{11}*{6.67} & \multirow{11}*{3.54} &12.5 &$5^{\circ}$& 25 & 0.294& 375.4 & 3.04&-0.189&0.068&0.198&0.870&18.7&Stable (1.00)  \\
                2  &&&&12.5&$5^{\circ}$&50  &0.427&258.6&3.01&-0189&0.041&0.191&0.803&18.6&Stable (1.22) \\
                3  &&&&12.5&$5^{\circ}$&200 &1.19 &92.82&2.91&0.087&0.013&0.168&0.777&19.6&Unstable  (1.83)\\
                4  &&&&12.5&$5^{\circ}$&800 &5.36 &20.62&2.74&1.05&---&\color{gray}0.051&---&---&Unstable (2.25) \\
                5  &&&&12.5&$5^{\circ}$&1600&12.4 &8.908&2.75&2.39&---&\color{gray}0.029&---&---&Unstable (1.41) \\
                6  &&&&12.5&$20^{\circ}$&25  &0.230&480.5&3.42&-0.195&0.080&0.192&0.871&17.3&Stable (1.04)  \\
                7  &&&&12.5&$20^{\circ}$&50  &0.566&195.3&3.07&-0.126&0.037&0.209&0.906&19.4&Stable (1.08)  \\
                8  &&&&12.5&$20^{\circ}$&100 &0.866&127.6&3.00&-0.061&0.025&0.213&0.931&20.9&Stable (1.11)  \\
                9  &&&&12.5&$20^{\circ}$&200 &1.30 &85.02&2.97&0.024&0.015&0.209&0.888&21.4&Stable (1.23)  \\
                10 &&&&12.5&$20^{\circ}$&800 &4.45 &24.80&2.61&0.996&---&\color{gray}0.135&---&---&Unstable (2.24)   \\
                11 &&&&12.5&$20^{\circ}$&1600&9.18 &12.03&2.55&2.06&---&\color{gray}0.016&---&---&Unstable (2.46)  \\
                \hline\\[-2.ex]
                12  & \multirow{6}*{0.07}& \multirow{6}*{14.3} & \multirow{6}*{5.89} & 30.4  & $20^{\circ}$ &25&0.0959& 6811 &6.39&-0.049&0.138&0.115&1.99&35.4&Stable (1.14)\\
                13 &&&&30.4&$20^{\circ}$&50  &0.266&2453 &5.54&-0.074&0.054&0.120&1.85&37.8&Stable (1.06) \\           
                14 &&&&30.4&$20^{\circ}$&200 &1.00 &647.7&4.90&-0.033&0.016&0.141&2.00&37.6&Stable (1.18)\\           
                15 &&&&30.4&$20^{\circ}$&800 &3.58 &182.5&4.77&0.384&---&\color{gray}0.104&---&---&Unstable (2.28)\\
                16 &&&&30.4&$20^{\circ}$&1600&5.02 &130.1&4.63&0.487&---&\color{gray}0.082&---&---&Unstable (2.33)\\
                17 &&&&12.5&$20^{\circ}$&1600&8.83 &12.51&3.91&4.25&---&\color{gray}0.034&---&---&Unstable (2.31)\\
                \hline\\[-2.ex]
                18 & \multirow{4}*{0.038}& \multirow{4}*{26.3} & \multirow{4}*{8.85} & 7 & $10^{\circ}$ & 100&0.335& 103.5 &5.51&1.43&0.075&0.148&0.157&13.5&Unstable (2.06) \\
                19 &&&&7 &$20^{\circ}$&100&0.562&61.69&5.47&2.68&0.048&0.163&0.166&17.7&Unstable (1.96)\\
                20 &&&&20&$10^{\circ}$&100&0.474&596.3&6.81&0.295&0.027&0.097&0.802&29.3&Unstable (2.24) \\
                21 &&&&20&$20^{\circ}$&100&0.583&485.0&7.25&0.380&0.026&0.108&0.857&34.5&Unstable (1.84) \\
                \hline\hline
        \end{tabular}  
        \begin{minipage}[]{0.99\textwidth}
                \tablefoot{\tablefoottext{a}{For the properties of the simulated
                                stellar winds listed in these columns we identified with a bar
                                the cases that have not been shown in Figs. \ref{fig_ra_ywind}
                                and \ref{fig_torque_sw} and have not been used to fit
                                Eqs. (\ref{eq_swflux}) and (\ref{eq_ra}).}\tablefoottext{b}{We
                                indicated in gray the values of the cases in which the stellar
                                wind appears to be quenched by the accretion flow, as in
                                Fig. \ref{fig_wflux_rt}.}} 
        \end{minipage}   
  \end{sidewaystable*}

\section{Temporal variability}
\label{sec_variability}

As discussed in Sect. \ref{sec_params}, after discarding the first 10 stellar periods to avoid
        initial transients, the analysis and the scaling relations presented in this work are based on
        quantities time-averaged over the final 20 stellar periods, from 10 to 30 $P_\star$. 
        Besides, we employed median values for the time-averaging procedure in order to minimize the 
        effect of short-lived transients. 
        
\begin{figure}
        \centering
        \includegraphics[width=\linewidth]{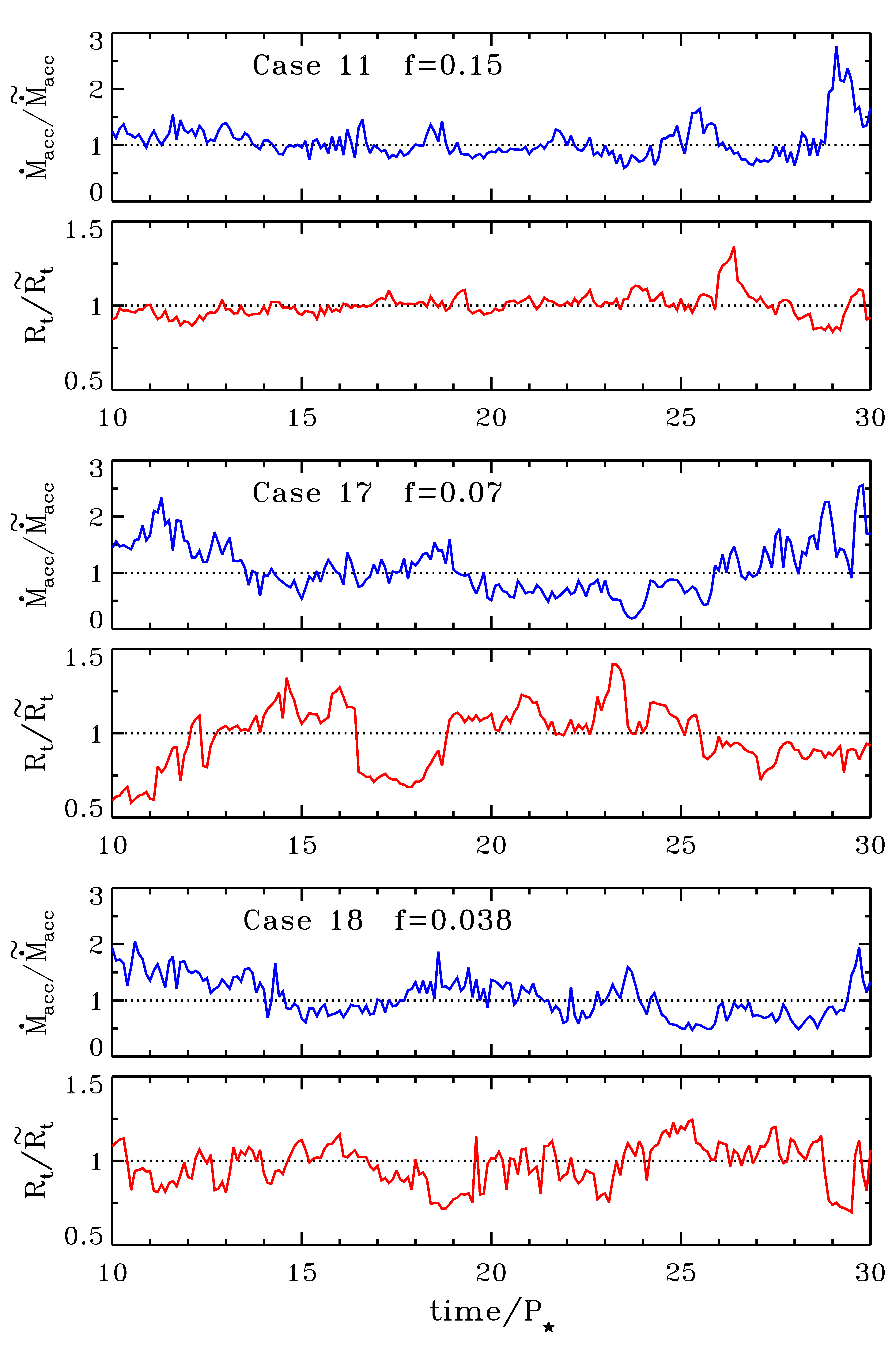}
        \caption{Temporal evolution of the mass accretion rate $\dot{M}_\mathrm{acc}/\tilde{\dot{M}}_{\mathrm{acc}}$ (blue curves)
        and the truncation radius $R_\mathrm{t}/\tilde{R}_\mathrm{t}$ (red curves) 
        normalized over the respective median values $\tilde{\dot{M}}_{\mathrm{acc}}$ 
        and $\tilde{R}_\mathrm{t}$, for unstable cases 11, 17 and 18, characterized by 
        rotation parameters $f=0.15,0.07$ and 0.038 respectively. 
        Time is given in units of the stellar rotation period $P_\star$.}
           
        \label{fig_mdot_rt_time}
\end{figure}

        In order to assess the impact of temporal variability on our analysis, we plot in Fig.
        \ref{fig_mdot_rt_time} the time evolution of the mass accretion rate
        $\dot{M}_\mathrm{acc}/\tilde{\dot{M}}_{\mathrm{acc}}$ (blue curves) and the truncation radius $R_\mathrm{t}/\tilde{R}_\mathrm{t}$ (red curves) normalized over the respective median values for 
        cases 11, 17 and 18, which are among the most unstable and time-variable ones (i.e., characterized by smaller $R_\mathrm{t}/R_\mathrm{co}$ values) for the three simulated stellar periods, 
        corresponding to $f=0.15,0.07$ and $0.038$ respectively. 
        After the initial 10 stellar periods, the mass accretion rates are characterized by a 
        time variability compatible with the observed one over timescales from days to weeks 
        \citep[$\leq 0.5$ dex,][]{Costigan:2014aa}. We note that since these three cases are characterized
        by different stellar periods and $f$ values, the timescale covered in the three panels is
        different, with the timespan of the $f=0.038$ simulation being around 1.8 and 3.9 times longer 
        than the $f=0.07$ and $f=0.15$ cases respectively. On the other hand, it seems that the different 
        duration of the numerical experiments has no significant impact on the simulated variability.
        
        In the cases presented in Fig. \ref{fig_mdot_rt_time}, which are among the most unstable and 
        variable ones, the variability of $R_\mathrm{t}$ corresponds to a standard deviation of
        $10-20\%$ around the mean value, with the largest variability observed for case 17, where the
        minimum and maximum values can differ from the median by up to $\approx 40\%$.
        The variability of the truncation radius can be affected by different factors, primarily 
        the variable shape of the magnetic cavity and the time variability of the accretion rate. 
        For example, as expected, the truncation radius shows a correlation with the mass accretion 
        rate, where an increase in $\dot{M}_\mathrm{acc}$ and a stronger push against the stellar 
        magnetosphere should determine a smaller value of $R_\mathrm{t}$. 
        In the plotted curves, mass accretion rates above the median value tend to correspond 
        to truncation radii below the median and vice versa.  
        
        In any case, the temporal evolution of $\dot{M}_\mathrm{acc}$ and $R_\mathrm{t}$ presented in Fig.
        \ref{fig_mdot_rt_time} strongly suggests that our time averages are clearly representative of 
        the long-term behavior of the simulations.

\section{Stellar torque integrals \\ with potential magnetic fields}
\label{sec_potential}

Integrating the angular momentum conservation in the system of Eqs. (\ref{eq_MHD}) on the stellar volume $V$ delimited by a closed surface ${\bf S}$ we can apply the Gauss theorem to get
\begin{equation}
\frac{\partial}{\partial t} \left( \int_V r\rho v_\phi \, \mathrm{d}V\right) = -\int_S r \left[ \rho v_{\phi} \vec{u} - \frac{B_{\phi} \vec{B}}{4 \pi}+\left(P+\frac{\vec{B}\cdot\vec{B}}{8\pi} \right)\hat{\phi}\right] \cdot \mathrm{d}\vec{S} \; ,
\label{eq_torque}
\end{equation} 
meaning that the temporal variation of the stellar angular momentum contained in the volume $V$ is equal to the integral of the angular momentum flux through the closed surface $\vec{S}$ delimiting the volume $V$.
We note that if the stellar magnetic field is potential (i.e., it is current-free), we have $\vec{J} = \nabla\times\vec{B}/4\pi = 0$, and therefore the magnetic field does not exert any force. Nevertheless, the magnetic part of the integrand in Eqs. (\ref{eq_jdot}) and (\ref{eq_torque}) is not necessarily zero, since a misaligned magnetosphere always has a toroidal component. However, when this is integrated over a closed surface, all the contributions to the integral cancel out, consistently yielding a null magnetic torque.

On the other hand, in order to evaluate the contribution to the torque of different flow components (the stellar wind, mass accretion and the magnetospheric ejections), we divided the stellar surface into open spherical sectors; see Sect. \ref{sec_torques}. We therefore must subtract the potential field contribution to the angular momentum flux to calculate the integrals on open surfaces. Our numerical method assumes that the magnetic field $\vec{B} = \vec{B}_0+\vec{B}_1$ can be decomposed in a potential component $\vec{B}_0$ with $\vec{J}_0 = \nabla\times\vec{B}_0/4\pi = 0$ and a deviation from it $\vec{B}_1$ with $\vec{J}_1 = \nabla\times\vec{B}_1/4\pi \neq 0$, so that the moment of the Laplace force can be rewritten as
\begin{equation}
\begin{aligned}
\nabla\cdot\left(-\frac{r B_\phi \vec{B}}{4\pi} \right)+ \frac{\partial}{\partial \phi} \left(\frac{\vec{B}\cdot\vec{B}}{8\pi} \right) & = \\
= -r \,\left.\vec{J}\times\vec{B}\right|_\phi & = \\
= -r \,\left.\vec{J}_1\times\vec{B}\right|_\phi & = \\
= \nabla\cdot\left(-\frac{r B_{1\phi} \vec{B}}{4\pi} -\frac{r B_{0\phi}\vec{B}_1}{4\pi}\right) & + 
\frac{\partial}{\partial \phi} \left(\frac{\vec{B}_1\cdot\vec{B}_1}{8\pi} +
\frac{\vec{B}_0\cdot\vec{B}_1}{4\pi} \right) \; .                          
\end{aligned}
\end{equation}
The expression of the torque $\dot{J}$ becomes
\begin{equation}
\begin{split}
\dot{J} = -\int_S
r & \Biggl[\rho v_\phi \vec{u} - \frac{B_{1\phi} \vec{B}}{4\pi} 
- \frac{B_{0\phi} \vec{B}_1}{4\pi} + \\
& +\left(P + \frac{\vec{B}_1\cdot\vec{B}_1}{8\pi} +
\frac{\vec{B}_0\cdot\vec{B}_1}{4\pi} \right)\hat{\phi}\Biggr] \cdot \mathrm{d}\vec{S} \; ,
\end{split}
\end{equation}
where clearly the magnetic part of the integrand can be different from zero only if a nonpotential
component $\vec{B}_1$ of the field is present, even when the integration is performed on open surfaces.

\section{Stationary solutions in solid rotation}
\label{sec_alfven}

In the most general conditions, a 3D MHD solution cannot be stationary. For example it is not possible to obtain a steady solution combining differential rotation and nonaxisymmetric magnetic fields. In the particular case of a perfectly conducting solid rotator with a frozen-in nonaxisymmetric magnetic field, which is the usual assumption that we make to model our star, it is in principle possible to model the flow in the stellar surroundings as a steady one in the corotating frame of reference. In the laboratory frame of reference this solution is not stationary (i.e., at a given position quantities vary in time) but it simply behaves as a fixed configuration that rotates rigidly at the angular speed ot the central rotator $\Omega_\star$. In this particular case, the curl of the electric field of a stationary solution in the corotating frame of reference must be zero. 
\[
\nabla \times \vec{E} = 0 \qquad \rightarrow \qquad \vec{E} = \nabla f \qquad \rightarrow \qquad \vec{B} \cdot \nabla f  = 0
\]
Since the electric field in the rotating frame of reference must be equal to zero on the surface of the star (frozen-in condition, i.e., the anchoring points of the magnetic field lines do not move) and $f$ is constant along the field lines, then $\nabla f = 0$ everywhere. Therefore $\vec{E} = 0$ and $\vec{u} \parallel \vec{B}$ everywhere. The stationary conservation of mass, angular momentum and energy can be written as
\begin{equation}
\begin{aligned}
\rho \vec{u} \cdot \nabla\left( \frac{\rho u }{B}\right) & = 0 \\
\rho \vec{u} \cdot \nabla\left( r v_\phi - \frac{r B_\phi B}{4 \pi \rho u}\right) + \frac{\partial P_\mathrm{t}}{\partial \phi} & = 0 \\
\rho \vec{u} \cdot \nabla\left(h+\frac{\vec{v}\cdot\vec{v}}{2}+\Phi_\mathrm{g} - \frac{r \Omega_\star B_\phi B}{4 \pi \rho u}\right) + \Omega_\star \frac{\partial P_\mathrm{t}}{\partial \phi} & = 0 \; .
\end{aligned}
\end{equation}
We therefore find that some quantities are constant along magnetic field lines, analogously to an axisymmetric steady solution, the mass-to-magnetic flux ratio, $k$,
\begin{equation}
k = \frac{\rho u}{B} \; ,
\end{equation}
and the effective rotation rate of the magnetic field lines, $\Omega_\star$,
\begin{equation}
\Omega_\star = \frac{1}{r}\left(v_\phi-B_\phi\frac{k}{\rho}\right) \; .
\end{equation}
Contrary to axisymmetric stationary solutions, the specific angular momentum, $\Lambda$,
\begin{equation}
\Lambda = r\left(v_\phi-\frac{B_\phi}{4\pi k}\right)
\label{eq_L_app},
\end{equation} 
and the specific energy, $e$, given by the Bernoulli equation
\begin{equation}
e = h+\frac{\vec{v}\cdot\vec{v}}{2}+\Phi_\mathrm{g} - \frac{r \Omega_\star B_\phi}{4 \pi k},
\end{equation}
are not invariant along magnetic field lines. Angular momentum and energy can flow in the azimuthal direction from one magnetic flux tube to another due to the total (thermal plus magnetic) pressure gradient in the $\phi$ direction. 
On the other hand, in a steady situation, the angular momentum and the mass fluxes through an arbitrary closed surface containing the central star are the same independently of the shape and distance of the chosen surface.
Since the stellar torque, $\dot{J}$, expressed as
\begin{equation}
\dot{J} = \int_{S} \left(\Lambda \rho \vec{u} + r P_\mathrm{t} \hat{\phi}\right)\cdot \mathrm{d}\vec{S}
\label{eq_l_torque}
\end{equation} 
and the mass accretion and loss rate, $\dot{M}$, expressed as
\begin{equation}
\dot{M} = \int_{S} \rho \vec{u} \cdot \mathrm{d}\vec{S} \; ,
\end{equation}
do not depend on the chosen surface $\vec{S}$, it is possible to define unambiguously an average specific angular momentum as
\begin{equation}
\langle \Lambda \rangle = \frac{\dot{J}}{\dot{M}} \; .
\end{equation}
If we consider a trans-Alfv\'{e}nic outflow, as a stellar wind, and evaluate $\Lambda$ at the Alfv\'{e}n surface, where $u = B/\sqrt{4\pi \rho}$, we get that $\Lambda = r_A^2\Omega_\star$, where $r_A$ is the cylindrical distance from the rotation axis of the points on the Alfv\'{e}n surface. For a steady axisymmetric solution, since $\Lambda$ is invariant along magnetic field lines, the equivalence $\Lambda = r_A^2\Omega_\star$ is valid all along the line, while in a general 3D case this is valid at the Alfv\'{e}n point only, since $\Lambda$ is not constant along field lines. If we evaluate the integral Eq. (\ref{eq_l_torque}) at the Alfv\'{e}n surface $\vec{S}_A$:
\begin{equation}
\dot{J} = \int_{S_A} \left(r^2 \Omega_\star \rho \vec{u} + r P_\mathrm{t} \hat{\phi}\right)\cdot \mathrm{d}\vec{S}_A \,
\end{equation}
the average specific angular momentum can be written as 
\begin{equation}
\langle \Lambda \rangle = \langle r_A^2 \rangle \Omega_\star+ 
\int_{S_A} r P_\mathrm{t} \hat{\phi}\cdot \mathrm{d}\vec{S}_A/\dot{M} \, ,
\label{eq_alfv}
\end{equation}
where
\begin{equation}
\langle r_A^2 \rangle = \frac{\int_{S_A} r^2 \rho \vec{u} \cdot \mathrm{d}\vec{S}_A}{\int_{S} \rho \vec{u} \cdot \mathrm{d}\vec{S}}
\end{equation}
is the mass-loss weighted cylindrical radius of the Alfv\'{e}n surface. 
Therefore the usual equivalence
\begin{equation}
\langle \Lambda \rangle = \langle r_A^2 \rangle \Omega_\star
\label{eq_ral}
\end{equation}
is valid only if the second integral in Eq. (\ref{eq_alfv}) is
zero. This is the case of axisymmetric solutions, where
Eq. (\ref{eq_ral}) is employed customarily, but it is not necessarily
true for a more general 3D case with a nonaxisymmetric magnetic
field, in which the Alfv\'{e}n surface is not axisymmetric.  

\section{Spectropolarimetric observations}
\label{sec_spol}

In Table \ref{tab_spec_data}, we list the sample of CTTs-Class II
objects observed with the ESPaDOnS and/or SPIRou spectropolarimeters
at the CFHT, and, when possible, with VLTI/GRAVITY in interferometry,
that we used to test our findings in Sect. \ref{sec_spec}. In different columns we
list the stellar parameters: object name; stellar mass, $M_\star$;
stellar radius, $R_\star$; the photospheric effective temperature,
$T_\mathrm{eff}$, used to compute the Kelvin-Helmholtz contraction 
timescale, $\tau_\mathrm{KH}$ (Eq. \ref{eq_kh}), given in column 13; the
rotation period, $P_\star$; the mass accretion rate,
$\dot{M}_\mathrm{acc}$; the intensity of the dipolar component of the
magnetosphere, $B_{\star,\mathrm{dip}}$, and its misalignment with
respect to the rotation axis $\Theta$; the size of the Br$\gamma$
emitting region, $R_{\mathrm{Br\gamma}}$, derived from interferometry;
the corotation radius, $R_\mathrm{co}$. We then list the results of our
numerical modeling: the truncation radius, $R_\mathrm{t}$
(Eq. \ref{eq_rt_yaccf}), in units of the stellar and corotation radii;
the star-disk-interaction, $\dot{J}_\mathrm{SDI}/J_\star$
(Eq. \ref{eq_tsdi}), and the stellar wind torque,
$\dot{J}_\mathrm{SW}/J_\star$ (Eqs. \ref{eq_jsw}-\ref{eq_ysw}),
assuming a mass ejection efficiency, $\dot{M}_\mathrm{SW} = 1\%
\dot{M}_\mathrm{acc}$, both divided by the stellar angular
momentum, $J_\star$, to provide the inverse of the associated
spin-up/spin-down timescale; the stellar spin evolution timescale
$\tau_{\Omega}$, that takes into account both the external
stellar torques and contraction. 
Stars are listed according to decreasing $R_\mathrm{t}/R_\mathrm{co}$ ratio.

The quantity $\tau_{\Omega}$ was computed as follows. 
Assuming solid body rotation, the conservation of the angular momentum, 
$J_\star = k^2 R_\star^2 M_\star \Omega_\star$, of a star subject to 
an external torque, $\dot{J}$, conveys the stellar spin evolution 
\citep[see, e.g.,][]{Matt:2010aa} via
\begin{equation}
\frac{\dot{\Omega}_\star}{\Omega_\star} =
\frac{\dot{J}}{J_\star} - \frac{\dot{M}_\mathrm{acc}}{M_\star} -
2\frac{\dot{R}_\star}{R_\star} \; ,
\label{eq_spinev}
\end{equation}
where the evolution of the stellar radius can be expressed as
\begin{equation}
\frac{\dot{R}_\star}{R_\star} = 2\frac{\dot{M}_\mathrm{acc}}{M_\star}-
\frac{28\pi R_\star^3\sigma T_\mathrm{eff}^4}{3GM_\star^2} \; ,
\label{eq_rstarev}
\end{equation}
assuming that the gravitational potential energy released by stellar
contraction is emitted as blackbody radiation \citep{Collier-Cameron:1993aa}. 
Equations (\ref{eq_spinev}) and (\ref{eq_rstarev}) can be expressed
in terms of the following characteristic timescales, assuming positive (negative)
values if they contribute to stellar spin-up (spin-down): $\tau_\mathrm{J} = 
J_\star/\dot{J} = J_\star/(\dot{J}_\mathrm{SDI}+\dot{J}_\mathrm{SW})$, the 
timescale associated with the total external torque; $\tau_\mathrm{M} = 
-M_\star/\dot{M}_\mathrm{acc}$, the spin-down timescale associated with mass
accretion, that determines both an increase in the stellar moment of inertia and 
of its potential gravitational energy, thus slowing down stellar contraction; 
$\tau_{KH}$, the Kelvin-Helmholtz spin-up timescale associated with stellar contraction, 
already provided by Eq. (\ref{eq_kh}). Combining Eqs. (\ref{eq_spinev}) and (\ref{eq_rstarev}), 
we can express the spin evolution timescale 
$\tau_\Omega = \Omega_\star/\dot{\Omega}_\star$ as
\begin{equation}
\tau_\Omega = \frac{\tau_\mathrm{J}\tau_\mathrm{KH}\tau_\mathrm{M}}
{\tau_\mathrm{KH}\tau_\mathrm{M}+5\tau_\mathrm{J}\tau_\mathrm{KH}+2\tau_\mathrm{J}\tau_\mathrm{M}}
\approx
\frac{\tau_\mathrm{J}\tau_\mathrm{KH}}{\tau_\mathrm{KH}+2\tau_\mathrm{J}} \; ,
\end{equation}
where the last approximation was obtained by neglecting the spin-down due to mass 
accretion ($\lvert\tau_\mathrm{M}\rvert \gg \lvert\tau_\mathrm{J}\rvert$ and
$\lvert\tau_\mathrm{M}\rvert \gg \tau_\mathrm{KH}$) since, for a typical range  
$\dot{M}_\mathrm{acc} \sim 10^{-7}-10^{-10}\ \mathrm{M}_{\sun}\ \mathrm{yr}^{-1}$, 
$\tau_\mathrm{M}$ corresponds to very long timescales, $\sim 10^7 - 10^{10}$ years.

\begin{figure}[!t]
        \centering
        \includegraphics[width=\linewidth]{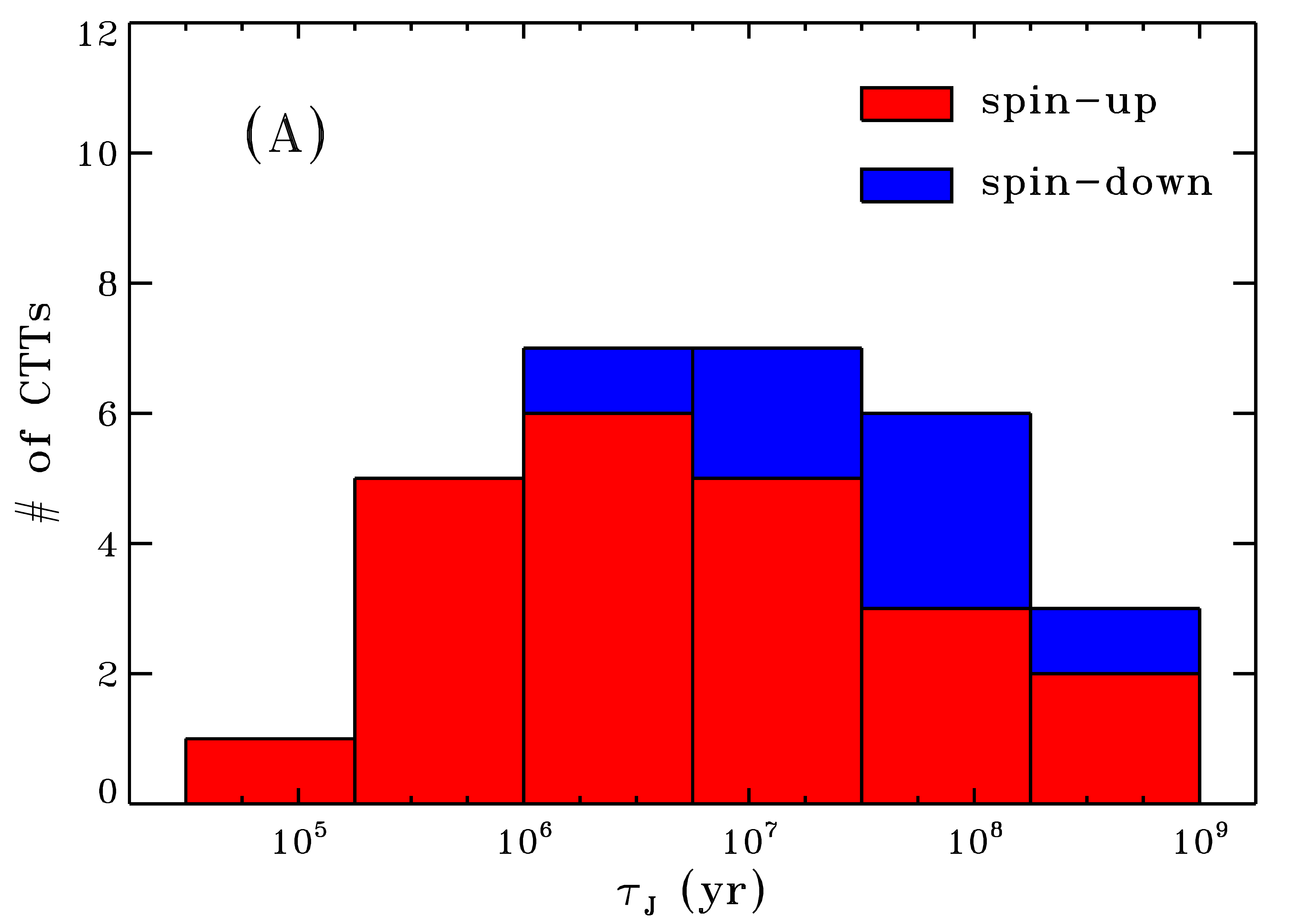}\\[1ex]
        \includegraphics[width=\linewidth]{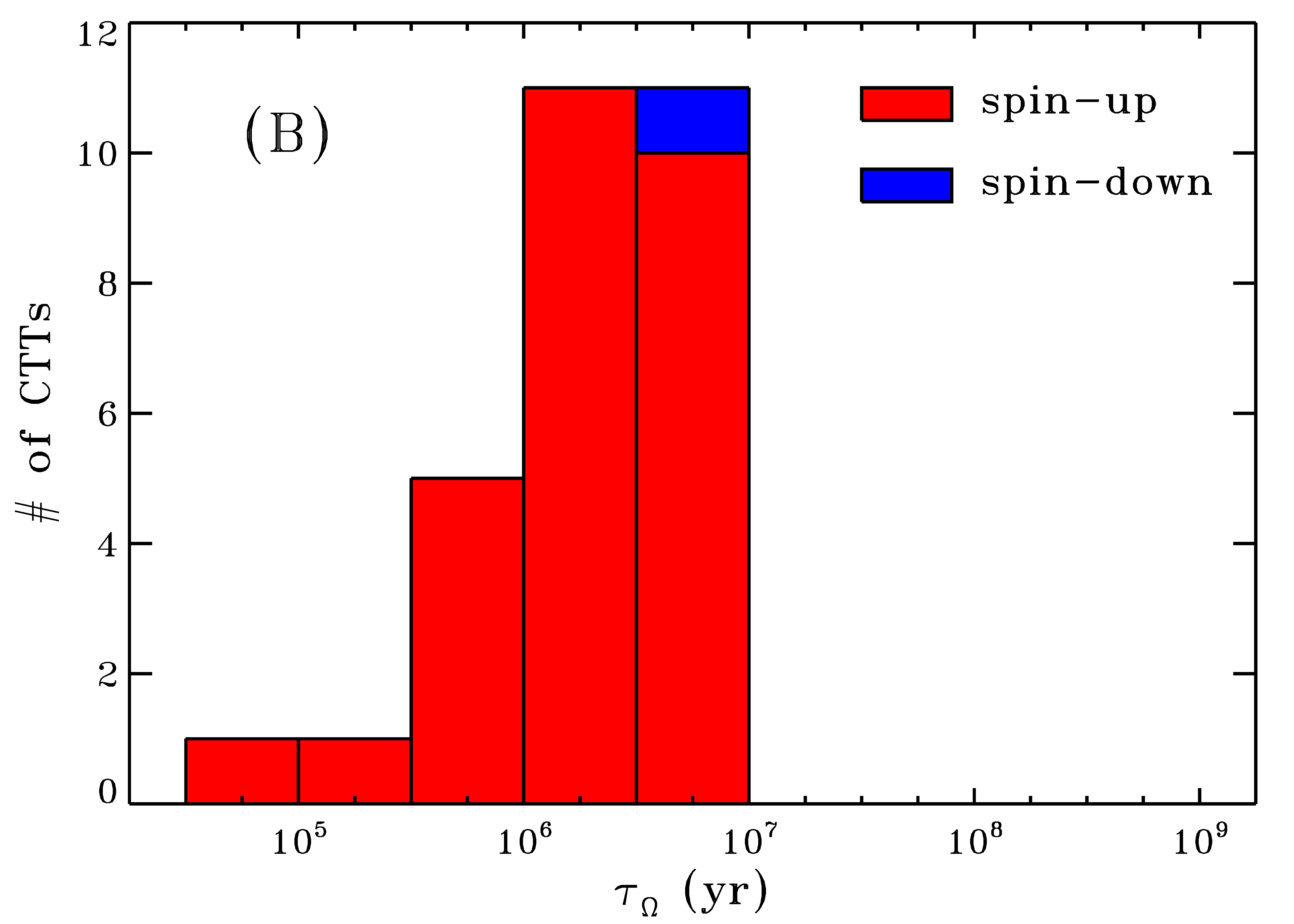}
        \caption{Distribution of the stellar torque timescale $\tau_\mathrm{J}$ (\textit{panel a}) and the 
                spin evolution timescale $\tau_\Omega$ (\textit{panel b}) in the stellar sample listed in Table 
                \ref{tab_spec_data}. Red (blue) bars identify spin-up (spin-down) timescales.} 
        \label{fig_hists}
\end{figure}

In Fig. \ref{fig_hists}, we plot the distributions in our stellar sample of the stellar 
total torque timescale $\tau_\mathrm{J}$ (panel a) and the 
stellar spin evolution timescale $\tau_\Omega$ (panel b). These histograms clearly 
show that stars subject to a spin-down torque not strong enough to oppose stellar contraction 
or to a weak spin-up torque characterized by $\tau_\mathrm{J} \gtrsim 10^7$ years, end up 
spinning up on a typical Kelvin-Helmholtz contraction timescale $\tau_\mathrm{KH} < 10^7$ years. 
On the other hand, stars characterized by short spin-up evolution timescales 
$\tau_\Omega < 10^6$ years are likely subject to a strong accretion torque.

   \begin{sidewaystable*}
        \caption{Stellar sample and model results}
        \label{tab_spec_data}
        \centering
        \begin{tabular}{c c c c c c c c c c c c c c c c c}
           \hline\hline\\[-2.ex]
       Star & $M_\star$ & $R_\star$ & $T_\mathrm{eff}$ & $P_\star$ &
       $\log_{10}\dot{M}_\mathrm{acc}$ & $B_{\star,\mathrm{dip}}$ & $\Theta$ & 
       $R_{\mathrm{Br\gamma}}$ & $R_\mathrm{co}$ & $R_\mathrm{t}$ & $R_\mathrm{t}$ &
       $\tau_\mathrm{KH}$ & $\dot{J}_\mathrm{SDI}/J_\star$ & 
       $\dot{J}_\mathrm{SW}/J_\star$ & $\tau_{\Omega}$ & Refs. \\       
       & ($M_\sun$) & ($R_\sun$) & (K) & (days) & ($M_\sun$ yr$^-1$) &
       (G) & (${}^\circ$) & ($R_\star$) & ($R_\star$) & ($R_\star$) & ($R_\mathrm{co}$) & ($10^6$ yr) & ($10^{-6}$ yr$^{-1}$) & ($10^{-6}$ yr$^{-1}$) & ($10^6$ yr) & \\ 
       \\[-2.ex]\hline\\[-2.ex]
           IRAS04125+2902 & \multirow{2}*{0.70} & \multirow{2}*{1.45} & \multirow{2}*{3889} & \multirow{2}*{11.35} & \multirow{2}*{-11} & \multirow{2}*{950} & \multirow{2}*{15} & \multirow{2}*{---} &\multirow{2}*{13.0} & \multirow{2}*{12.1} & \multirow{2}*{0.93} & \multirow{2}*{10.5} & \multirow{2}*{-0.018} & \multirow{2}*{-0.037} & \multirow{2}*{7.42} & \multirow{2}*{1}\\
           (2025) &&&&&&&&&&&&&&&& \\
           IRAS04125+2902 & \multirow{2}*{0.70} & \multirow{2}*{1.45} & \multirow{2}*{3889} & \multirow{2}*{11.35} & \multirow{2}*{-11} & \multirow{2}*{800} & \multirow{2}*{5} & \multirow{2}*{---} & \multirow{2}*{13.0} & \multirow{2}*{11.8} & \multirow{2}*{0.91} & \multirow{2}*{10.5} & \multirow{2}*{-0.011} & \multirow{2}*{-0.029} & \multirow{2}*{6.62} & \multirow{2}*{1}\\
           (2024) &&&&&&&&&&&&&&&& \\
           AA Tau & 0.70 & 2.00 & 4000 & 8.20 & -9.2 & 1720 & 10 & --- &  7.59 & 6.85 & 0.90 & 3.88 & -0.259 & -0.492 & -5.07 & 2, 3\\
           V2247 Oph & 0.36 & 2.00 & 3500 & 3.50 & -9.8 & 110 & 40 & --- & 3.45 & 2.85 & 0.83 & 1.61 & 0.001 & -0.011 & 0.814 & 2, 4\\
           PDS 70 (2024) & 0.875 & 1.30 & 4140 & 3.01 & -10 & 420 & 31 & --- & 6.45 & 5.17 & 0.80 & 17.7 & 0.007 & -0.011 & 9.27 & 5\\
           V2129 Oph (2009) & 1.35 & 2.00 & 4500 & 6.53 & -9.2 & 970 & 10 & --- & 8.12 & 6.49 & 0.80 & 8.30 & 0.051 & -0.082 & 4.80 & 2, 6\\
           LkCa 15 & 1.25 & 1.6 & 4500 & 5.70 & -9.2 & 1350 & 20 & --- & 9.04 & 7.11 & 0.79 & 13.9 & 0.081 & -0.098 & 8.05 & 7\\
           DN Tau (2010) & 0.65 & 1.90 & 3950 & 6.32 & -9.2 & 530 & 25 & --- & 6.56 & 5.14 & 0.78 & 3.78 & 0.081 & -0.079 & 1.90 & 8\\
           BP Tau (Feb 2006) & 0.70 & 1.95 & 4000 & 7.60 & -8.6 & 1220 & 10 & --- & 7.40 & 5.77 & 0.78 & 3.86 & 0.405 & -0.376 & 1.89 & 2, 9\\
           CR Cha & 1.90 & 2.50 & 4900 & 2.30 & -9.0 & 220 & 70 & --- & 3.63 & 2.81 & 0.77 & 5.99 & 0.013 & -0.007 & 2.97 & 2, 10\\
           JH 223 & 0.4 & 1.1 & 3528 & 3.31 & -10.2 & 250 & 28 & --- & 6.26 & 4.84 & 0.77 & 11.6 & 0.013 & -0.010 & 5.72 & 11\\
           BP Tau (Dec 2006) & 0.70 & 1.95 & 4000 & 7.60 & -8.6 & 960 & 30 & --- & 7.40 & 5.57 & 0.75 & 3.86 & 0.476 & -0.259 & 1.39 & 2, 9\\
           DoAr 44 & 1.20 & 2.00 & 4600 & 2.96 & -8.2 & 800 & 20 & 5.0 & 4.61 & 3.48 & 0.75 & 6.01 & 0.257 & -0.109 & 2.04 & 12, 13\\
           GQ Lup (2009) & 1.05 & 1.70 & 4300 & 8.40 & -9.0 & 1070 & 30 & --- & 10.4 & 7.54 & 0.72 & 9.81 & 0.274 & -0.106 & 2.72 & 14\\
           DN Tau (2012) & 0.65 & 1.90 & 3950 & 6.32 & -9.2 & 300 & 30 & --- & 6.57 & 4.73 & 0.72 & 3.78 & 0.110 & -0.033 & 1.66 & 8\\
           PDS 70 (2022) & 0.875 & 1.30 & 4140 & 3.01 & -10 & 200 & 37 & --- & 6.45 & 4.65 & 0.72 & 17.7 & 0.012 & -0.004 & 8.25 & 5\\
           GQ Lup (2011) & 1.05 & 1.70 & 4300 & 8.40 & -9.0 & 900 & 30 & --- & 10.39 & 7.35 & 0.71 & 9.81 & 0.281 & -0.081 & 2.50 & 14\\
           TW Hya (2020) & 0.80 & 1.16 & 4050 & 3.61 & -8.7 & 1190 & 23 & 3.5 & 7.92 & 5.59 & 0.71 & 22.8 & 0.406 & -0.104 & 2.64 & 15, 16\\
           GM Aur & 0.95 & 2.02 & 4287 & 6.04 & -8.3 & 730 & 15 & --- & 6.79 & 4.75 & 0.70 & 4.84 & 0.665 & -0.138 & 1.09 & 17\\
           V2129 Oph (2005) & 1.35 & 2.00 & 4500 & 6.53 & -9.2 & 280 & 20 & --- & 8.12 & 5.43 & 0.67 & 8.30 & 0.086 & -0.012 & 3.20 & 2, 18\\
           CI Tau (2016) & 0.90 & 2.00 & 4200 & 9.01 & -7.6 & 1700 & 20 & 4.8 & 8.79 & 5.82 & 0.66 & 4.86 & 6.018 & -0.783 & 0.181 & 19, 20\\
           CI Tau (2019) & 0.90 & 2.00 & 4200 & 9.01 & -8.0 & 780 & 11 & 4.8 & 8.79 & 5.56 & 0.63 & 4.86 & 2.387 & -0.190 & 0.392 & 20, 21\\
           TW Hya (2008) & 0.80 & 1.16 & 4050 & 3.61 & -8.9 & 370 & 40 & 3.5 & 7.92 & 4.87 & 0.62 & 22.8 & 0.272 & -0.015 & 2.97 & 2, 16, 22\\
           DO Tau (2025) & 0.54 & 1.90 & 3450 & 5.128 & -7.7 & 320 & 66 & 1.3 & 5.36 & 3.22 & 0.60 & 4.49 & 2.90 & -0.098 & 0.326 & 23, 24\\
           S CrA N & 0.80 & 2.30 & 4300 & 7.30 & -7.0 & 816 & 34 & 5.0 & 6.39 & 3.82 & 0.60 & 2.30 & 13.93 & -0.521 & 0.073 & 25, 26\\
           V4046 Sgr A & 0.95 & 1.12 & 4250 & 2.42 & -9.3 & 100 & 60 & --- & 6.66 & 3.73 & 0.56 & 29.4 & 0.062 & -0.001 & 7.92 & 2, 27\\
           DO Tau (2021) & 0.54 & 1.90 & 3450 & 5.128 & -7.7 & -190 & 50 & 1.3 & 5.36 & 2.98 & 0.56 & 4.49 & 2.81 & -0.044 & 0.330 & 23, 24\\
           V4046 Sgr B & 0.85 & 1.04 & 4250 & 2.42 & -9.3 & 80 & 80 & --- & 6.91 & 3.68 & 0.53 & 29.4 & 0.073 & -0.001 & 7.29 & 2, 27\\
           CV Cha & 2.00 & 2.50 & 5500 & 4.40 & -7.5 & 140 & 60 & --- & 5.69 & 2.90 & 0.51 & 4.18 & 1.302 & -0.008 & 0.591 & 2, 10\\
           
       \hline\hline\\[-2.ex]
        \end{tabular}  
        \begin{minipage}[]{\textwidth}
                \footnotesize{{\bf References.} (1)
                  \citet{Donati:2025aa}; (2) \citet{Johnstone:2014ab}};
                (3) \citet{Donati:2010ab}; (4) \citet{Donati:2010aa};
                (5) \citet{Donati:2024ad}; (6) \citet{Donati:2011ab};
                (7) \citet{Donati:2019aa}; (8) \citet{Donati:2013aa};
                (9) \citet{Donati:2008ab}; (10) \citet{Hussain:2009aa}; (11) \citet{Freitas:2026aa};
                (12) \citet{Bouvier:2020aa}; (13) \citet{Bouvier:2020ab}; (14) \citet{Donati:2012aa};
                (15) \citet{Donati:2024ab}; (16) \citet{Gravity:2020aa}; (17) \citet{Zaire:2024aa};
                (18) \citet{Donati:2007aa}; (19) \citet{Donati:2020aa}; (20) \citet{Gravity:2023aa};
                (21) \citet{Donati:2024aa}; (22) \citet{Donati:2011aa}; 
                (23) \citet{Donati:2026aa}; (24) \citet{Gravity:2026aa};
                (25) \citet{Nowacki:2023aa}; (26) \citet{Gravity:2024aa}; (27) \citet{Donati:2011ac} 
        \end{minipage}   
      \end{sidewaystable*}

\end{appendix}

\end{document}